\documentclass[aps,amsmath,amssymb,prd,reprint,groupedaddress,showkeys]{revtex4-2}

\usepackage{CJK}
\usepackage{upgreek}
\usepackage{booktabs}
\usepackage{graphicx}
\usepackage{hyperref}

\hypersetup{
  bookmarksopen=true,
  pdfstartview=FitH,
  pdftitle={Gluon Condensation Signature in the GeV Gamma-Ray Spectra of Pulsars},
  pdfauthor={Zou, Ze-Cheng; Huang, Yong-Feng; Li, Cheng-Ming; Zheng, He-Rui; Zhu, Wei},
  pdfsubject={Gluon Condensation Signature in Gamma-Ray Pulsars},
  pdfkeywords={Gamma-ray astronomy; Nuclear physics; Pulsars},
  colorlinks=true,
  citecolor=blue,
  linkcolor=blue,
  urlcolor=blue,}

\begin{document}
\begin{CJK*}{UTF8}{gbsn}

\title{Gluon Condensation Signature in the GeV Gamma-Ray Spectra of Pulsars}

\author{Ze-Cheng Zou (邹泽城)}
\affiliation{School of Astronomy and Space Science, Nanjing University, Nanjing 210023, China}
\affiliation{Key Laboratory of Modern Astronomy and Astrophysics (Nanjing University), Ministry of Education, Nanjing 210023, China}

\author{Yong-Feng Huang (黄永锋)}
\affiliation{School of Astronomy and Space Science, Nanjing University, Nanjing 210023, China}
\affiliation{Key Laboratory of Modern Astronomy and Astrophysics (Nanjing University), Ministry of Education, Nanjing 210023, China}

\author{Cheng-Ming Li (李程明)}
\affiliation{School of Physics and Microelectronics, Zhengzhou University, Zhengzhou 450001, China}

\author{He-Rui Zheng (郑合瑞)}
\affiliation{School of Physics and Microelectronics, Zhengzhou University, Zhengzhou 450001, China}

\author{Wei Zhu (朱伟)}
\email[Author (WZ) to whom correspondence should be addressed. ]{wzhu@phy.ecnu.edu.cn}
\affiliation{Department of Physics, East China Normal University, Shanghai 200241, China}

\date{\today}

\begin{abstract}
The accumulation of gluons inside nucleons, i.e., the gluon condensation, may lead to a characteristic broken power-law gamma-ray spectrum in high-energy nucleon collisions. Here we show that the observed spectra of at least 25 sources in the second \textit{Fermi} Large Area Telescope Catalog of Gamma-ray Pulsars can be well fitted by such a broken power-law function that has only four free parameters. It strongly indicates that the gamma-ray emission from these pulsars is of hadronic origin, but with gluon condensation inside hadrons. It is well known that the quark-gluon distribution in a free nucleon is different from that in a bound nucleon. This work exposes the nuclear $A$ dependence of the gluon condensation effect, where $A$ refers to the baryon number. Our study reveals the gluon condensation under the condition of $A\to\infty$, which may open a new window for eavesdropping on the structure of compact stars on the subnuclear level.
\end{abstract}

\maketitle
\end{CJK*}

\section{Introduction}

Nucleons (protons and neutrons) are important building blocks of materials. Any new discoveries about the properties of nucleons will deepen our understanding of the Universe. In the infinite momentum frame, a nucleon consists of partons, i.e., quarks and gluons. The gluon distribution inside nucleons dominates high-energy hadronic processes. In general, the number of gluons in a nucleon grows with gluon splitting. However, the increase of gluon number cannot be too fast due to the restriction of the unitarity. Quantum chromodynamics (QCD) research based on the Jalilian-Marian-Iancu-McLerran-Weigert-Leonidov-Kovner (JIMWLK) nonlinear evolution equations predicts that at a critical value of the gluon density, the color-charged gluons may reach an equilibrium state between splitting and fusion of gluons, which is called the color glass condensation (CGC; for a review, see \cite{2011PThPS.187...17M}). An advanced QCD evolution equation proposed by Zhu \emph{et al.\ }\cite{1999NuPhB.551..245W,2008ChPhL..25.3605Z,2016NuPhB.911....1Z,2017NuPhB.916..647Z,2022NuPhB.98415961Z} shows that the Balitsky-Fadin-Kuraev-Lipatov (BFKL) singularities in the nonlinear QCD evolution equations may continually evolve the CGC solution to the smaller-$x$ range and arise a chaotic solution. The dramatic chaotic oscillations will produce strong shadowing and antishadowing effects, forcing the gluons to gather in a state at a critical momentum of $(x_c,k_c)$, where $x_c$ is the fraction of the proton's longitudinal momentum carried by the condensed gluons and $k_c$ is their transverse momentum. In other words, a realistic gluon condensation (GC) presents.

Such a GC will induce various effects in high-energy $\gamma$-rays produced during nucleon collisions if the critical gluon momentum $(x_c,k_c)$ is in the appropriate energy range. Unfortunately, the exact values of $(x_c,k_c)$ cannot be precisely determined theoretically due to the uncertainties in the evolution equations. Phenomena associated with the GC effect have not been observed in various experiments carried out at the largest hadron collider---Large Hadron Collider (LHC). It implies that the GC-threshold may be beyond the current maximum energy of LHC. However, high-energy protons associated with various compact stars in the Universe may exceed the GC-threshold and cause the GC effect in their collisions with other particles.

Specifically, cosmic $\gamma$-rays can be generated in the hadronic scenario of $p+p\to\pi^0\to2\gamma$ \cite{2004vhec.book.....A}. In this study, the GC effect will be considered in such a hadronic scenario, which is called the GC-model. The sharp peak in the momentum distribution of gluons in the GC-model can tremendously increase the cross section of the hadron-hadron collisions. Consequently, the GC-model predicts a broken power-law (BPL) spectrum with an exponential cutoff in the high-energy $\gamma$-rays associated with such collisions \cite{2018IJMPE..2750073Z,2018ApJ...868....2F,2020ApJ...889..127Z,2021JCAP...01..038Z,2021JCAP...08..065R}, i.e.,
\begin{equation}
  \Phi^\mathrm{GC}_\gamma(E_\gamma)=\begin{cases}
    \Phi_0\left(\frac{E_\gamma}{E_\pi^\mathrm{GC}}\right)^{-\Gamma_1},&\\
    &\hspace{-2.9cm}\text{if }E_{\gamma}\leqslant E_{\pi}^\mathrm{GC};\\
    \Phi_0\left(\frac{E_\gamma}{E_\pi^\mathrm{GC}}\right)^{-\Gamma_2},&\\
    &\hspace{-2.9cm}\text{if }E_{\pi}^\mathrm{GC}<E_{\gamma}<E_{\pi}^\mathrm{cut};\\
    \Phi_0\left(\frac{E_\gamma}{E_\pi^\mathrm{GC}}\right)^{-\Gamma_2}\exp\left(-\frac{E_\gamma}{E_\pi^\mathrm{cut}}+1\right),&\\
    &\hspace{-2.9cm}\text{if }E_{\gamma}\geqslant E_{\pi}^\mathrm{cut};
  \end{cases}\label{eq:1.1}
\end{equation}
where $E_\gamma$ is the energy of $\gamma$-ray photons, $E_\pi^\mathrm{GC}$ is the break energy at the GC-threshold determined by the critical momentum $(x_c,k_c)$, $\Gamma_1$ and $\Gamma_2$ are power-law indices, $E_\pi^\mathrm{cut}$ is the cutoff energy, and $\Phi_0$ is the normalization constant.

The BPL function is widely used to describe the very-high-energy (VHE) $\gamma$-ray spectra in astrophysics. The GC-model gives it an intuitive physical explanation. It has been shown that the VHE $\gamma$-ray emissions from supernova remnants (SNRs) and active galactic nuclei can be well explained by the  GC-model \cite{2018IJMPE..2750073Z,2018ApJ...868....2F,2020ApJ...889..127Z,2021JCAP...01..038Z,2021JCAP...08..065R}. The parameter $E_\pi^\mathrm{GC}$ is found to be larger than $100\,\mathrm{GeV}$ for the $pA$ or $AA$ collisions in those circumstances, where $A$ is the baryon number of a normal nucleus. In this study, we will adopt the GC-model to study the GeV $\gamma$-ray spectra of pulsars. A pulsar, i.e., namely a neutron star, is composed of a huge number of neutrons ($A\gg300$), which makes it an ideal object to test the GC-model.

This paper is organized as follows. First, the basic idea of the GC-model is briefly summarized in Sec.~\ref{sec:gcmod} for the sake of completeness. Then, in Sec.~\ref{sec:1420}, taking PSR J1420-6048 as a typical example, we use the GC-model to analyze its GeV $\gamma$-ray spectrum in detail. In Sec.~\ref{sec:2pc}, more $\gamma$-ray pulsars (24 sources) taken from the second \textit{Fermi} Large Area Telescope (\textit{Fermi}-LAT) Catalog of Gamma-ray Pulsars (2PC; \cite{2013ApJS..208...17A}) are further analyzed. Finally, Sec.~\ref{sec:con} presents our discussions and conclusion.

\section{The GC-model\label{sec:gcmod}}

According to the hadronic scenario of high-energy radiation \cite{2004vhec.book.....A}, about half of the energies of the parent protons are taken away by valence quarks, which form the leading particles. The remaining energies are transformed into the secondary hadrons (mainly pions) in the central region through gluons. GeV $\gamma$-ray photons are then produced through $\pi_0\to2\gamma$ process, whose spectrum is described in the laboratory frame as
\begin{equation}\begin{split}
  \Phi_\gamma(E_\gamma) = C_\gamma&\left(\frac{E_\gamma}{\mathrm{GeV}}\right)^{-\beta_\gamma}
  \int_{E_\pi^\mathrm{min}}^{E_\pi^\mathrm{cut}}\mathrm{d}E_\pi\left(\frac{E_p}{\mathrm{GeV}}\right)^{-\beta_p}\\
  &\times N_\pi(E_p,E_\pi)\frac{\mathrm{d}\omega_{\pi-\gamma}(E_\pi,E_\gamma)}{\mathrm{d}E_\gamma},\end{split}\label{eq:2.1}
\end{equation}
where the spectral index $\beta_\gamma$ incorporates the energy loss caused by the absorption of pions by the medium, and $N_\pi$ is the distribution function of pions. $C_\gamma$ is a normalization constant that incorporates the kinematic factor and the flux dimension. As usual, the accelerated protons are assumed to follow a simple power-law form of $N_p \propto E_p^{-\beta_p}$ in the source frame.

Neglecting the harmonization mechanism, $N_\pi$ should be proportional to the cross section of gluon minijet production. Note that $N_\pi$ is a complex quantity, which involves unknown nonperturbative QCD effects. Usually, a multi-parameter empirical formulation is used to describe it. In the presence of gluon condensation, the GC-effect leads to a huge amount of condensed gluons at the threshold $x_c$. These gluons should participate in the $pp$ collisions intensively, which inevitably leads to a large number of secondary mesons. As a result, the pion production reaches a maximum value at the corresponding energy. For simplicity, let us assume that all the kinetic energy at the center-of-mass frame is consumed in creating pions during the $pp$ collision. Taking this approximation, the complicated hadronization mechanism can be overcome. We can then use the relativistic forms of momentum and energy conservation to analytically obtain the solution of $N_\pi$ as \cite{2018IJMPE..2750073Z,2021JCAP...08..065R,2022NuPhB.98415961Z}
\begin{equation}\begin{split}
  &\ln N_\pi=0.5\ln(E_p/\mathrm{GeV})+a,\\&\ln N_\pi=\ln(E_\pi/\mathrm{GeV})+b,\\
  &\phantom{log}\text{with }E_\pi\in\left[E_\pi^\mathrm{GC},E_\pi^\mathrm{cut}\right],
  \end{split}\label{eq:2.2}
\end{equation}
where $a = 0.5\ln(2m_p/\mathrm{GeV})-\ln(m_\pi/\mathrm{GeV})+\ln K$ and $b = \ln(2m_p/\mathrm{GeV})-2\ln(m_\pi/\mathrm{GeV})+\ln K$. Here $m_p$ is the proton mass, $m_\pi$ is the pion mass, and $K$ is the inelasticity which can be taken as $K\simeq1/2$. We see that Equation~\ref{eq:2.2} has a typical power-law form.

Considering the standard $\pi^0\to2\gamma$ process and substituting Equation~\ref{eq:2.2} into Equation~\ref{eq:2.1}, one can straightforwardly get the GC-characteristic spectrum as
\begin{equation}
  E_\gamma^2\Phi^\mathrm{GC}_\gamma(E_\gamma)\simeq\begin{cases}
    \frac{2e^bC_\gamma}{2\beta_p-1}(E_\pi^\mathrm{GC})^3\left(\frac{E_\gamma}{E_\pi^\mathrm{GC}}\right)^{-\beta_\gamma+2},\hspace{-2.9cm}&\\
    &\text{if }E_\gamma\leqslant E_\pi^\mathrm{GC};\\
    \frac{2e^bC_\gamma}{2\beta_p-1}(E_\pi^\mathrm{GC})^3\left(\frac{E_\gamma}{E_\pi^\mathrm{GC}}\right)^{-\beta_\gamma-2\beta_p+3},\hspace{-2.9cm}&\\
    &\text{if }E_\pi^\mathrm{GC}<E_\gamma<E_\pi^\mathrm{cut};\\
    \frac{2e^bC_\gamma}{2\beta_p-1}(E_\pi^\mathrm{GC})^3\left(\frac{E_\gamma}{E_\pi^\mathrm{GC}}\right)^{-\beta_\gamma-2\beta_p+3}\hspace{-2.9cm}&\\
    \phantom{\frac{2e^b}{2\beta_p}}\times\exp\left(-\frac{E_\gamma}{E_\pi^\mathrm{cut}}+1\right),&\text{if }E_\gamma\geqslant E_\pi^\mathrm{cut}.
  \end{cases}\label{eq:2.3}
\end{equation}
It is the detailed version of Equation~\ref{eq:1.1} in the framework of the GC-model. Note that the two parameters of $E_\pi^\mathrm{GC}$ and $E_\pi^\mathrm{cut}$ are connected with each other through
\begin{equation}
  E_\pi^\mathrm{cut}=\mathrm{e}^{b-a}\sqrt{\frac{2m_p}{k_c^2}}\left(E_\pi^\mathrm{GC}\right)^2,\label{eq:2.4}
\end{equation}
where all energies take the GeV unit. Additionally, another useful formula of the GC-model is \cite{2022NuPhB.98415961Z}
\begin{equation}
  E_p=\frac{2m_p}{m^2_{\pi}}E_{\pi}^2,\label{eq:2.5}
\end{equation}
which comes directly from Equation~\ref{eq:2.2} and gives the relation between the energy of the incident proton and $E_{\pi}$ in the laboratory frame.

Equation~\ref{eq:2.3} is the $\gamma$-ray energy spectrum predicted by the GC-model. Note that this BPL distribution at $E_\gamma\leqslant E_\pi^\mathrm{cut}$ is an analytic solution of the GC-model, rather than a phenomenological parameterized formula. The exponential cutoff in Equation~\ref{eq:2.3} comes from the prominent suppression of $\gamma$-rays at $E_\gamma>E_\pi^\mathrm{cut}$. The reason is that the gluons condensate at $x_c$ and very few gluons exist at $x<x_c$. As a result, almost no gluons can participate in the $pA$ interaction at a higher energy range. Consequently, the number of photons with energy larger $E_\pi^\mathrm{cut}$ decreases sharply, leading to an exponential cutoff.

The exact value of the threshold energy ($E_\pi^\mathrm{GC}$) depends on the baryon number ($A$) of the target nucleus for the $pA$ (or $AA$) collisions. In other words, it is $A$-dependent. For example, $E_\pi^\mathrm{GC}$ is ${\sim}20\,\mathrm{TeV}$ for the $pp$ collision, and it is $100\,\mathrm{GeV}$ for the $p$--heavy nucleus collisions \cite{2022NuPhB.98415961Z}. Generally speaking, $E_\pi^\mathrm{GC}$ decreases monotonously when $A$ increases. When a proton collides with a very heavy clump composed of a huge number ($A^*$) of neutrons, e.g.\ $A^*\gg300$, then $E_\pi^\mathrm{GC}$ will be very small and could fall in the GeV range. Neutron stars provide a natural environment in which heavy clumps of neutrons can emerge. We will show below that the observed GeV $\gamma$-rays from many pulsars could be produced in this way.

\section{The GC-model for PSR J1420-6048\label{sec:1420}}

The matter in the outer crust of a neutron star is in a lattice structure. The density of materials deep inside the neutron star is much larger than that in the crust. The superfluid neutrons in the interior can penetrate through the lattice space and escape to the surface since there is enough empty space between two adjacent atoms in the crust. Consequently, neutron clusters with a large baryon number ($A^*$) can form on the stellar surface. Note that the inner core of the neutron star can be regarded as a big nucleus whose baryon number $A^{**}$ is effectively infinity. So we have $A^{**}\gg A^*\gg A$.

On the other hand, electrons can be accelerated to TeV energy by shock waves in the pulsar wind nebulae (PWNe) \cite{2019ApJ...876L...8B}. Since the proton mass is much larger than the electron mass, protons can be accelerated to even much higher energies due to a negligible radiation loss. Once these energetic protons enter the strong magnetic field of the pulsar, they will move along the magnetic field lines. Note that the moving direction of the proton is determined by its initial velocity rather than by its charge. They will finally bombard the two poles of the pulsar, and collide with neutron clusters (with the baryon number $A^*$) or nuclei (baryon number $A$) to produce GeV $\gamma$-rays. Such a $\gamma$-ray emission should be modulated by the rotation of the neutron star, showing a pulsating behavior. The value of $A^*$ is determined by the lattice size, which further constrains the GC-threshold $E_\pi^\mathrm{GC}$ to be in the GeV range.

\begin{figure}
  \includegraphics[width=8.6cm]{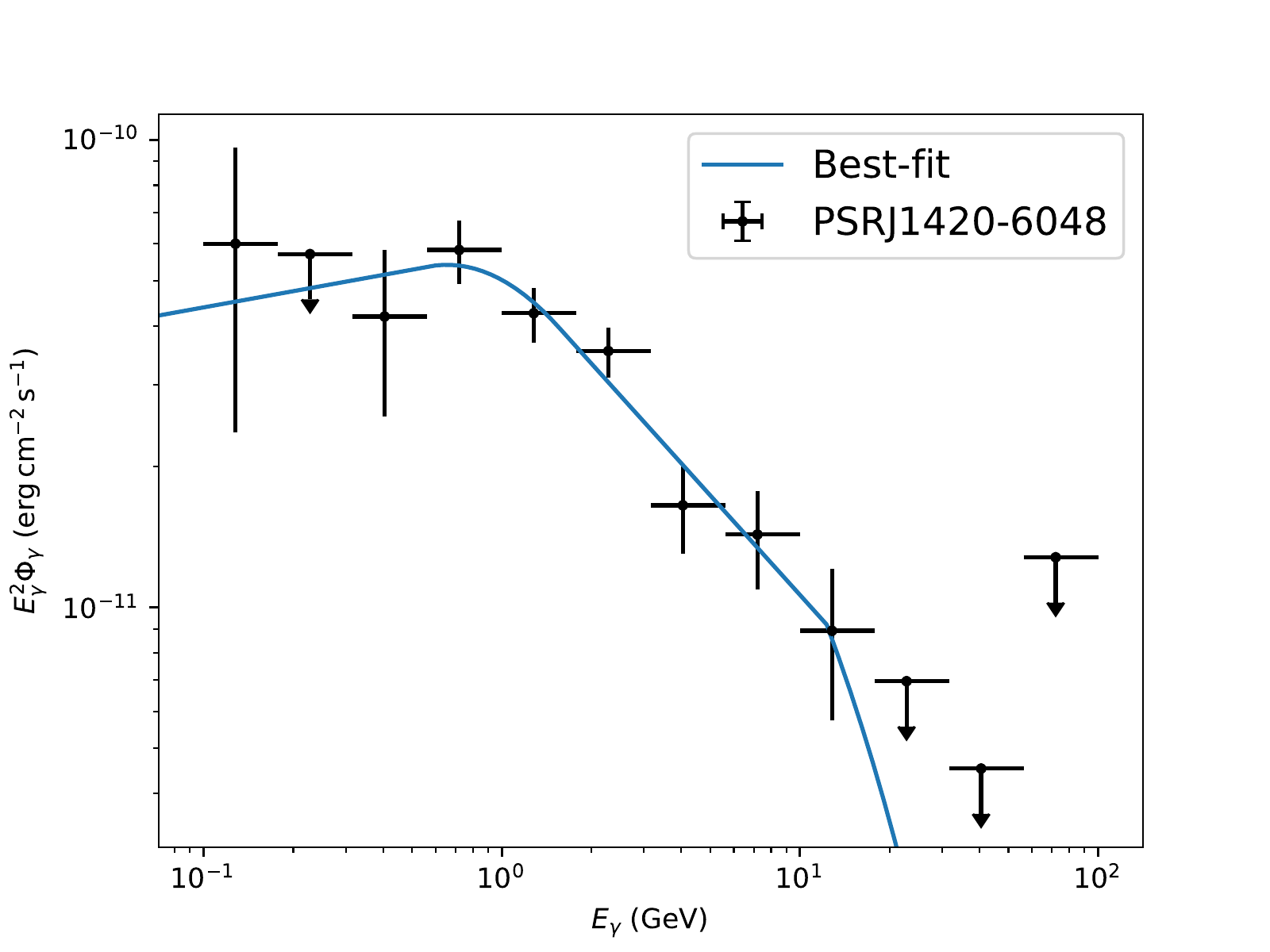}
  \caption{Observed spectrum of PSR J1420-6048 and our best fit by using the GC-model. The observational data are taken from the 2PC \cite{2013ApJS..208...17A}. Note that the second broken position is predicted by Equation~\ref{eq:2.4}.\label{Fig.1}}
\end{figure}

PSR J1420-6048 is a young pulsar \cite{2013ApJS..208...17A} and is associated with a complex extended radio nebula. Its $\gamma$-ray spectrum in the GeV-energy band is shown in Fig.~\ref{Fig.1}. We have used our model (i.e., Equation~\ref{eq:2.3}) to fit the observed spectrum of PSR J1420-6048. The best-fit results are illustrated in Fig.~\ref{Fig.1}, and the corresponding model parameters are presented in Table~\ref{tab:1}. From Fig.~\ref{Fig.1}, we see that our model can explain the observed $\gamma$-ray spectrum of PSR J1420-6048 quite well. Note that in our model, the cutoff energy (i.e., the position of the second break) is not a free parameter, but is connected with the first break through Equation~\ref{eq:2.4}. Here, we further discuss some details of our model.

(1) Nuclei $A$ and neutron clusters $A^*$ coexist in the surface layer of a neutron star. From Equation~\ref{eq:2.5}, we see that in the $pA^*$ collisions, to produce significant ${\sim}1\,\mathrm{GeV}$ $\gamma$-ray photons, the energy of the incident protons should be $E_p^\mathrm{cut}=10\,\mathrm{TeV}$. By contrast, for $pA$ collisions, the incident protons must be accelerated to $E_p^\mathrm{cut}=10^{10}\,\mathrm{GeV}$ to generate high-energy $\gamma$-rays through the GC-effect at $E_\pi^\mathrm{GC}\geqslant100\,\mathrm{GeV}$. Obviously, the latter energy is beyond the range of most PWNe, and the GC-spectra associated with $pA$ collisions are then very rare in pulsars. Only very few pulsars may show such a GC feature.

(2) Where are the excess photons around $E_\gamma\sim1\,\mathrm{GeV}$ in the $pA^*$ collisions according to the standard hadronic mechanism without the GC-effect? Usually, a significant portion of the kinetic energy in hadron collisions is consumed to heat up secondary particles in the central region of the collisions. Consequently, we have $N_\pi\sim\ln\sqrt{s/m_\pi}$ when the GC-effect is not included, where $\sqrt{s}$ is the interaction energy at the center-of-mass frame. However, when the GC-effect is considered, almost all the kinetic energy will be utilized to produce new particles, which leads to $N_\pi\sim\sqrt{s/m_\pi}$. Taking $E_p=10\,\mathrm{TeV}$ as an example, we find that the increase in the number of particles is nearly 100 times. Therefore, the possible contributions of the $pA^*$ collisions without the GC effect to $\Phi_\gamma$ are much weaker than that of the $pA^*$ collisions with the GC effect. This example also shows that the hadronic collision with the GC-effect is an efficient mechanism to convert kinetic energy to high-energy radiation. It is thus not surprising that although the strength of VHE protons in the pulsar environment might be weak, the GC-effect can still lead to significant GeV emission.

(3) Curvature radiation is generally considered to be a viable mechanism for the GeV $\gamma$-ray emission of pulsars. The spectrum of a single source in this model is usually well described by a power-law function with an exponential cutoff
\begin{equation}
  \Phi_\gamma=\Phi_0\left(\frac{E_\gamma}{E_0}\right)^{-\Gamma}\exp\left(-\frac{E_\gamma}{E_\mathrm{cut}}\right).\label{eq:3.1}
\end{equation}
Note that photons emitted through the curvature radiation are usually distributed in a relatively narrow energy range. To fit the observed spectrum in a wide range, the superposition of multiple components is necessary. In fact, the radiation of leptons is closely related to their motions and trajectories. We need to sum up all the possible contributions from these radiations. In this case, the number of free parameters increases correspondingly, which reduces the reliability of the model due to its complexity.

Usually, the emission from electrons moving in a strong magnetic field contains two components, i.e., curvature radiation and synchrotron radiation. GeV $\gamma$-rays are believed to be due to curvature radiation, while keV emission comes from synchrotron radiation. These two components should be connected with each other since they are produced by the same group of electrons. A model that combines the synchrotron radiation and the curvature radiation together is called the synchrotron-curvature (SC) model, which uses the correlation between these two components to effectively reduce the number of free parameters. Previously, PSR J1420-6048 has been used as an example to test the SC model \cite{2019MNRAS.489.5494T}. It is argued that the SC model, which has only four free parameters, can fit the observational data well. However, we note that the observed X-ray spectrum, which is explained as the synchrotron component, covers only a narrow energy range in the keV band. It is still unclear whether the SC model is consistent with observations or not when a wider energy band is involved. Therefore, more X-ray observations on PSR J1420-6048 are needed in the future. Anyway, as a promising new mechanism, our GC-model remains to be a competitive choice.

\section{Comparison with More Gamma-ray Pulsars\label{sec:2pc}}

\begin{figure*}
  \includegraphics[width=0.368\textwidth]{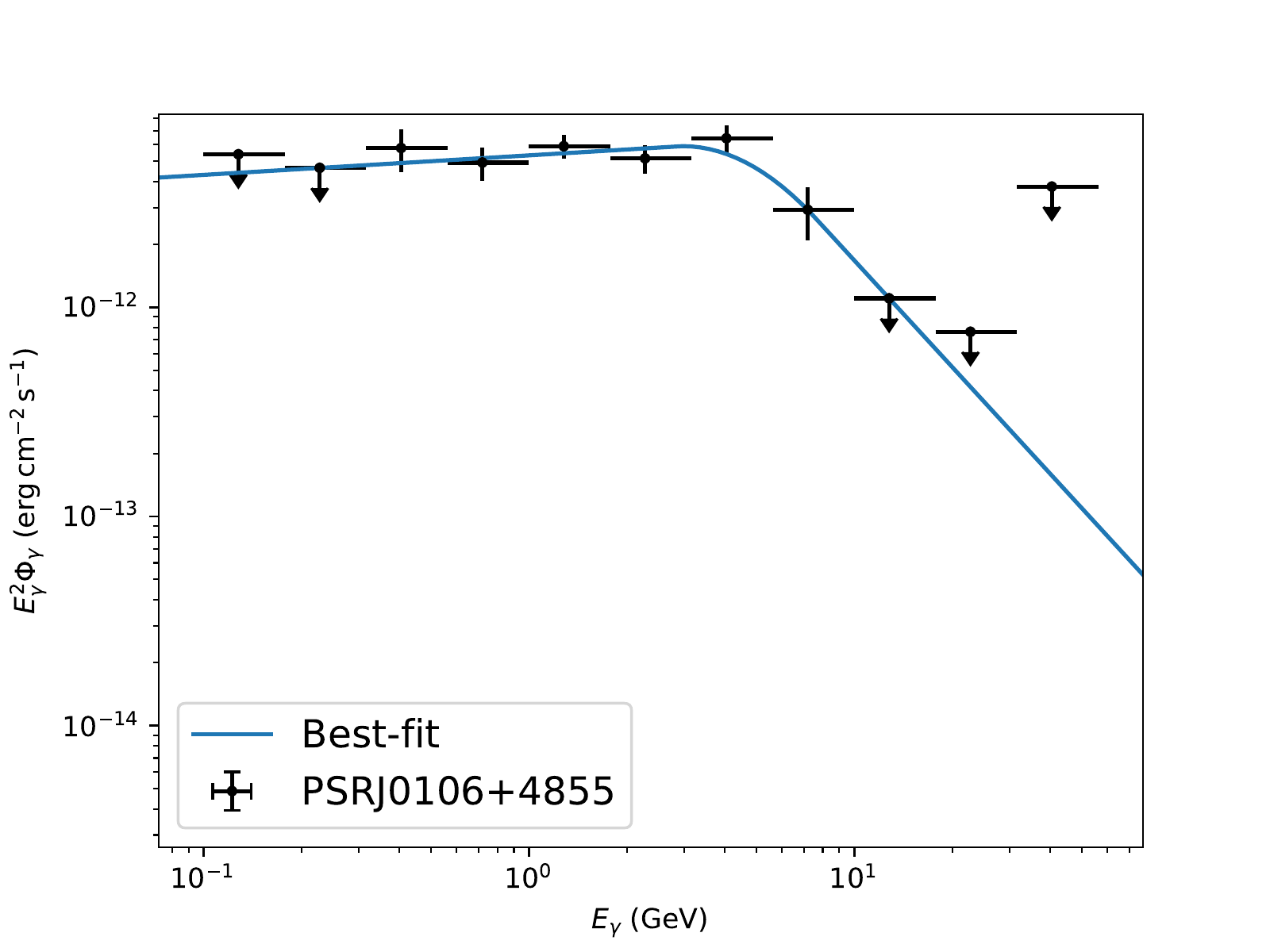}\!\!\!\!\!\!\!\!\!\!\!%
  \includegraphics[width=0.368\textwidth]{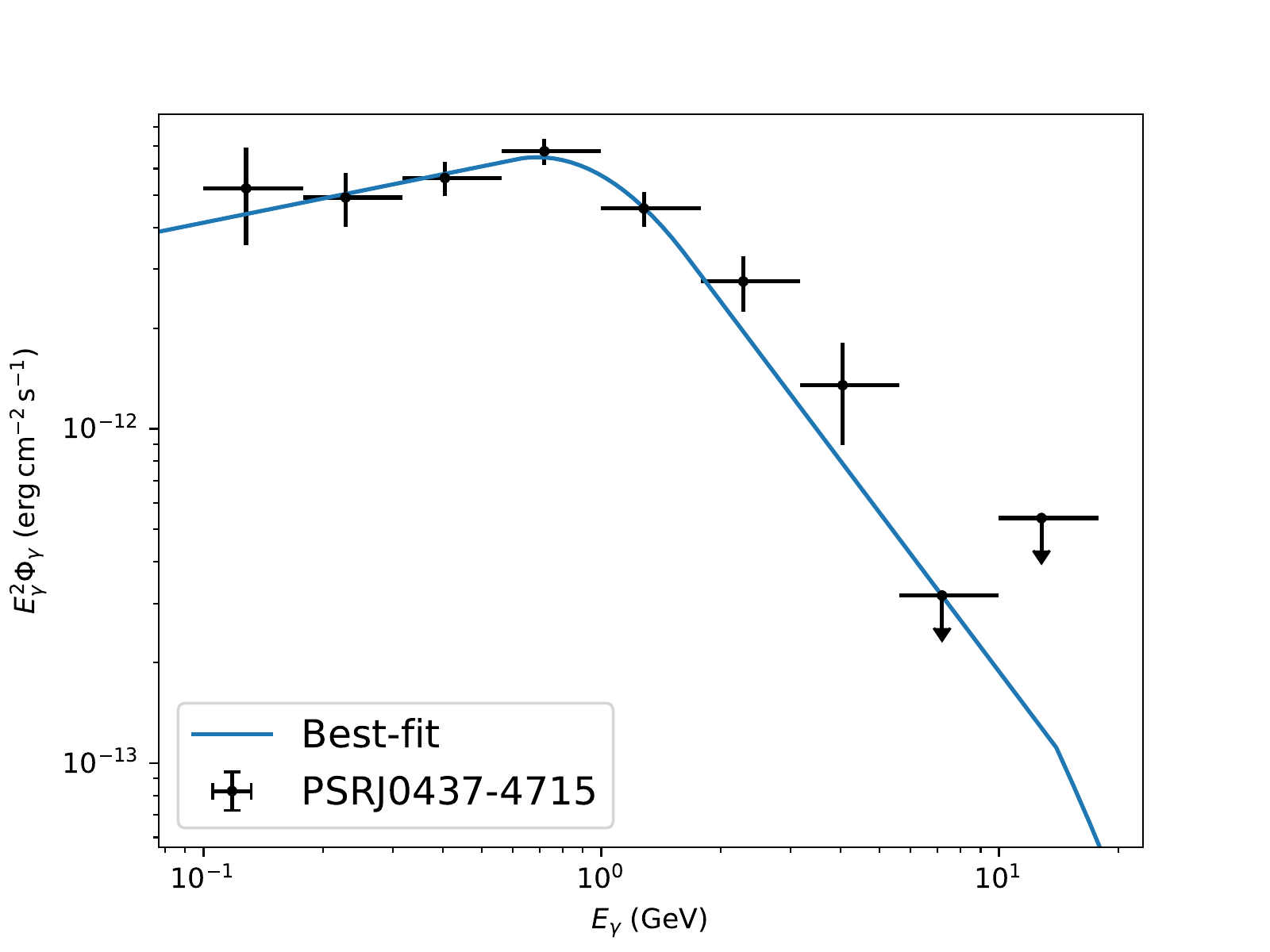}\!\!\!\!\!\!\!\!\!\!\!%
  \includegraphics[width=0.368\textwidth]{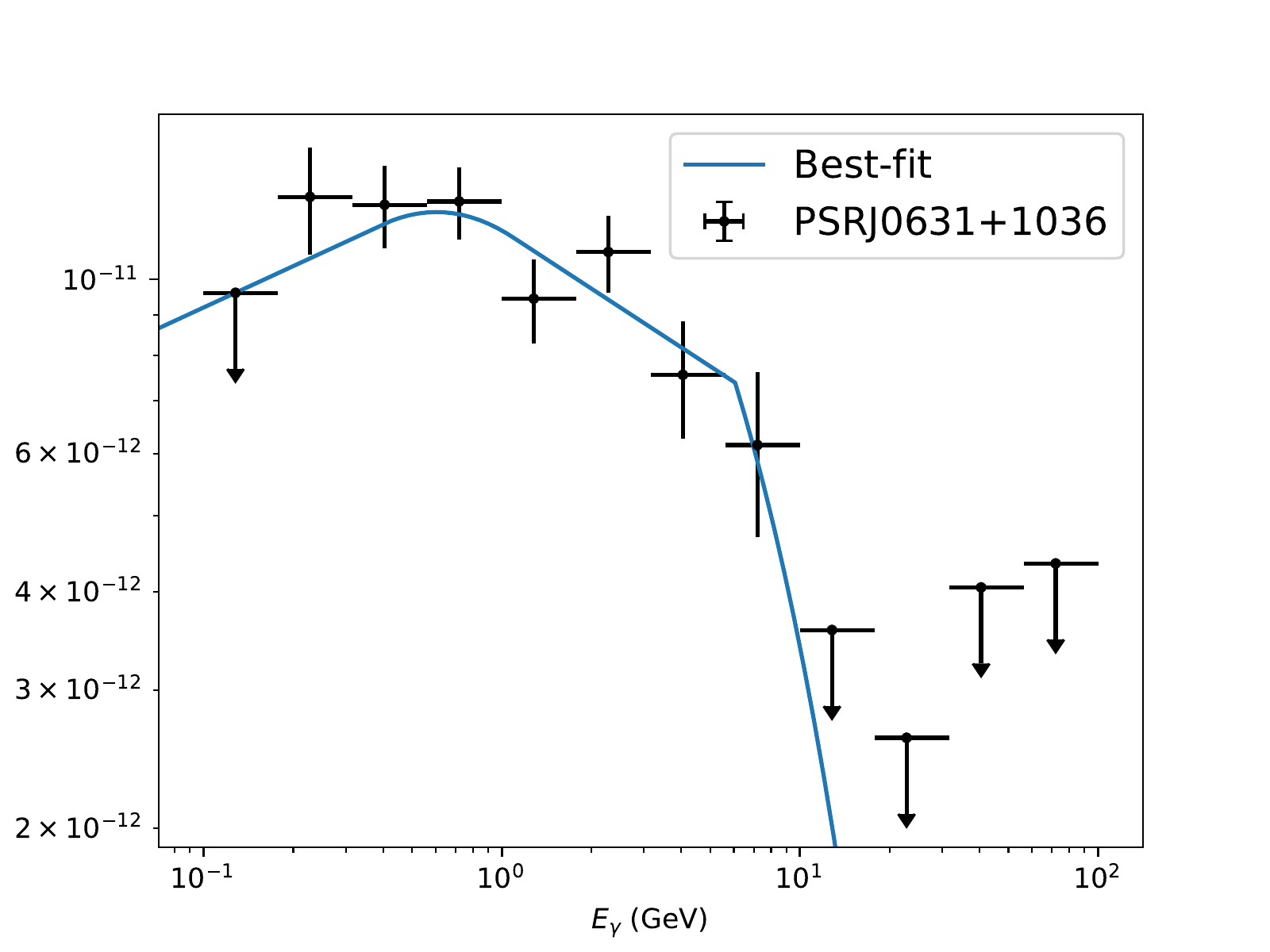}
  \includegraphics[width=0.368\textwidth]{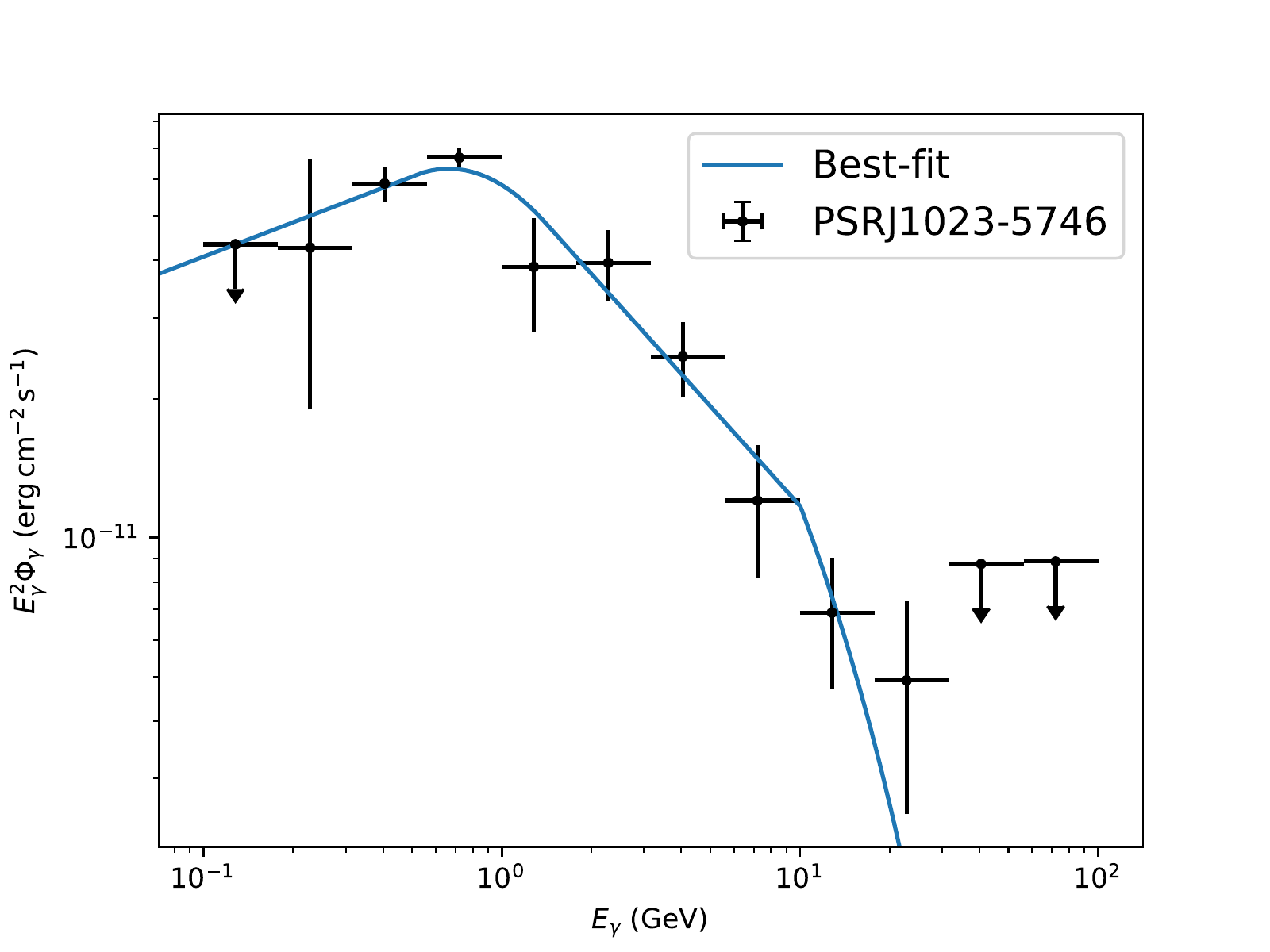}\!\!\!\!\!\!\!\!\!\!\!%
  \includegraphics[width=0.368\textwidth]{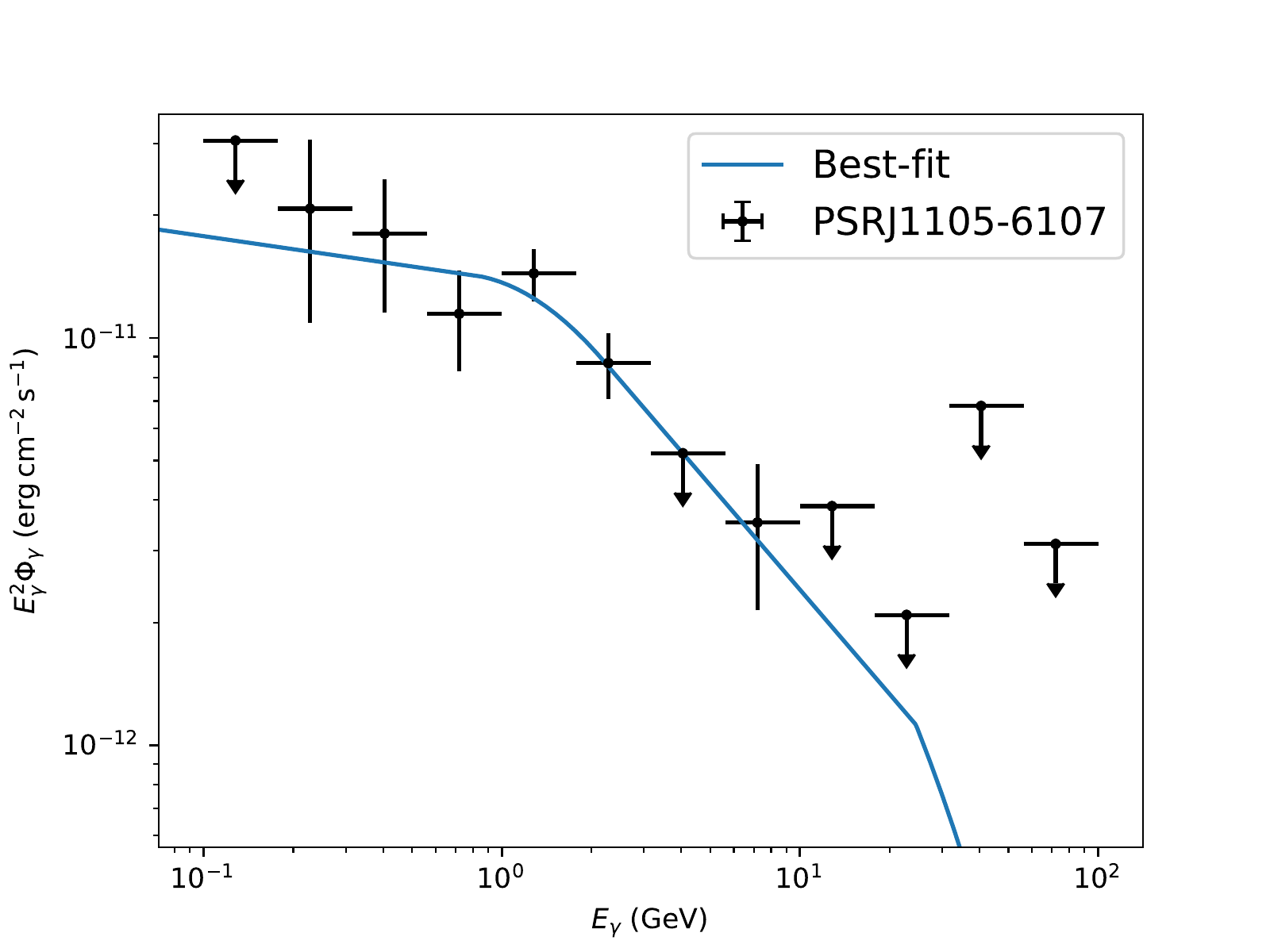}\!\!\!\!\!\!\!\!\!\!\!%
  \includegraphics[width=0.368\textwidth]{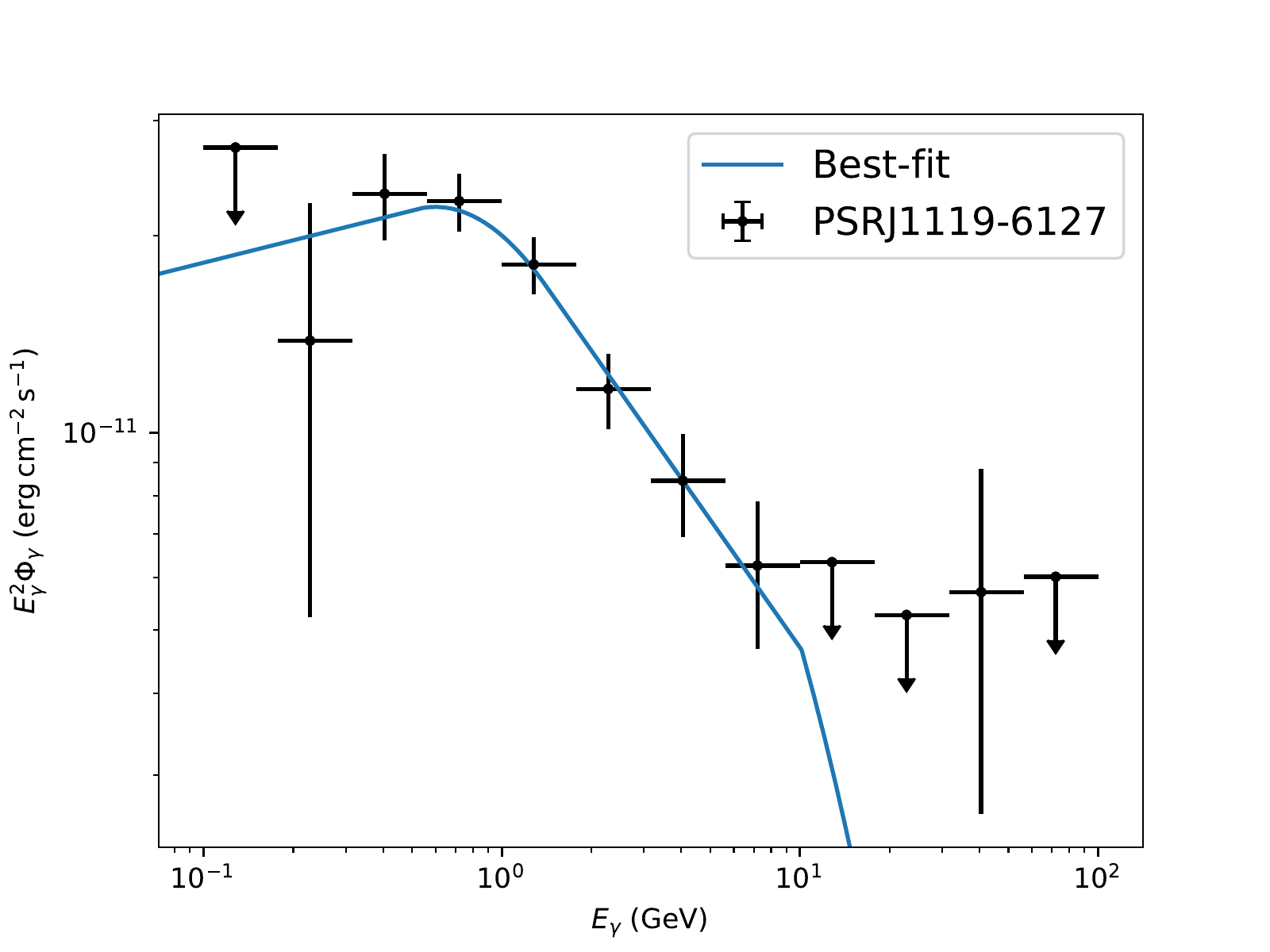}
  \includegraphics[width=0.368\textwidth]{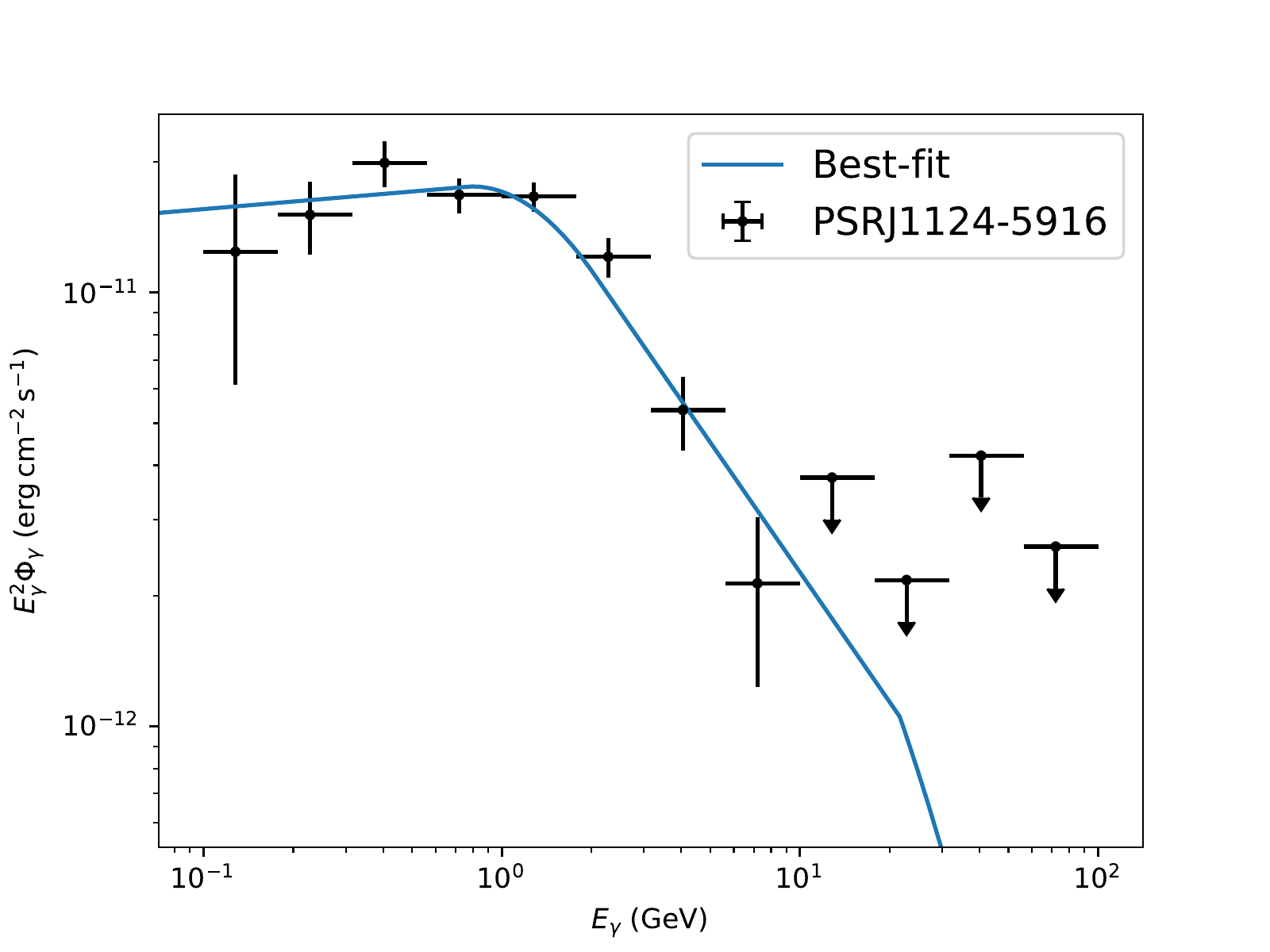}\!\!\!\!\!\!\!\!\!\!\!%
  \includegraphics[width=0.368\textwidth]{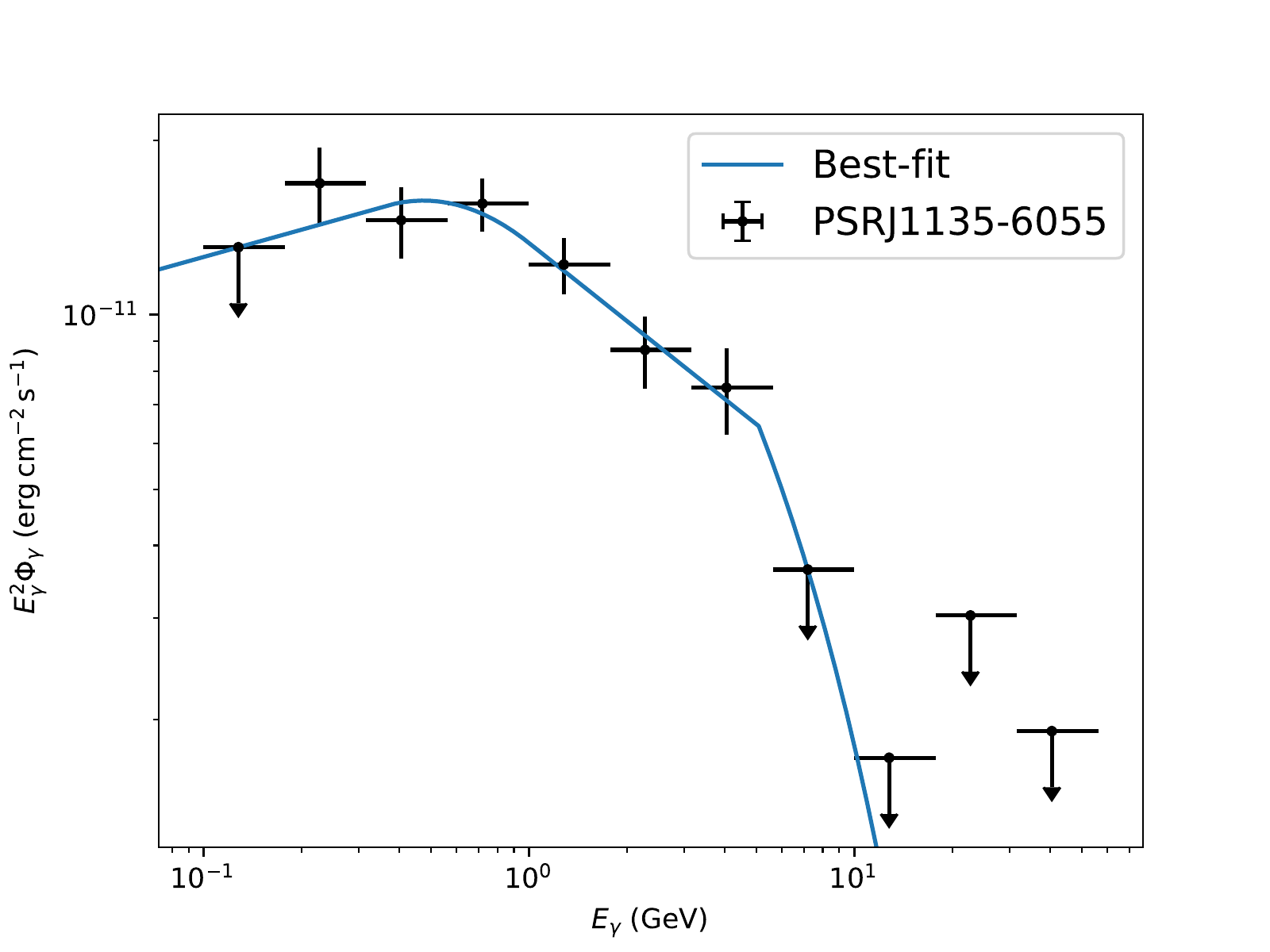}\!\!\!\!\!\!\!\!\!\!\!%
  \includegraphics[width=0.368\textwidth]{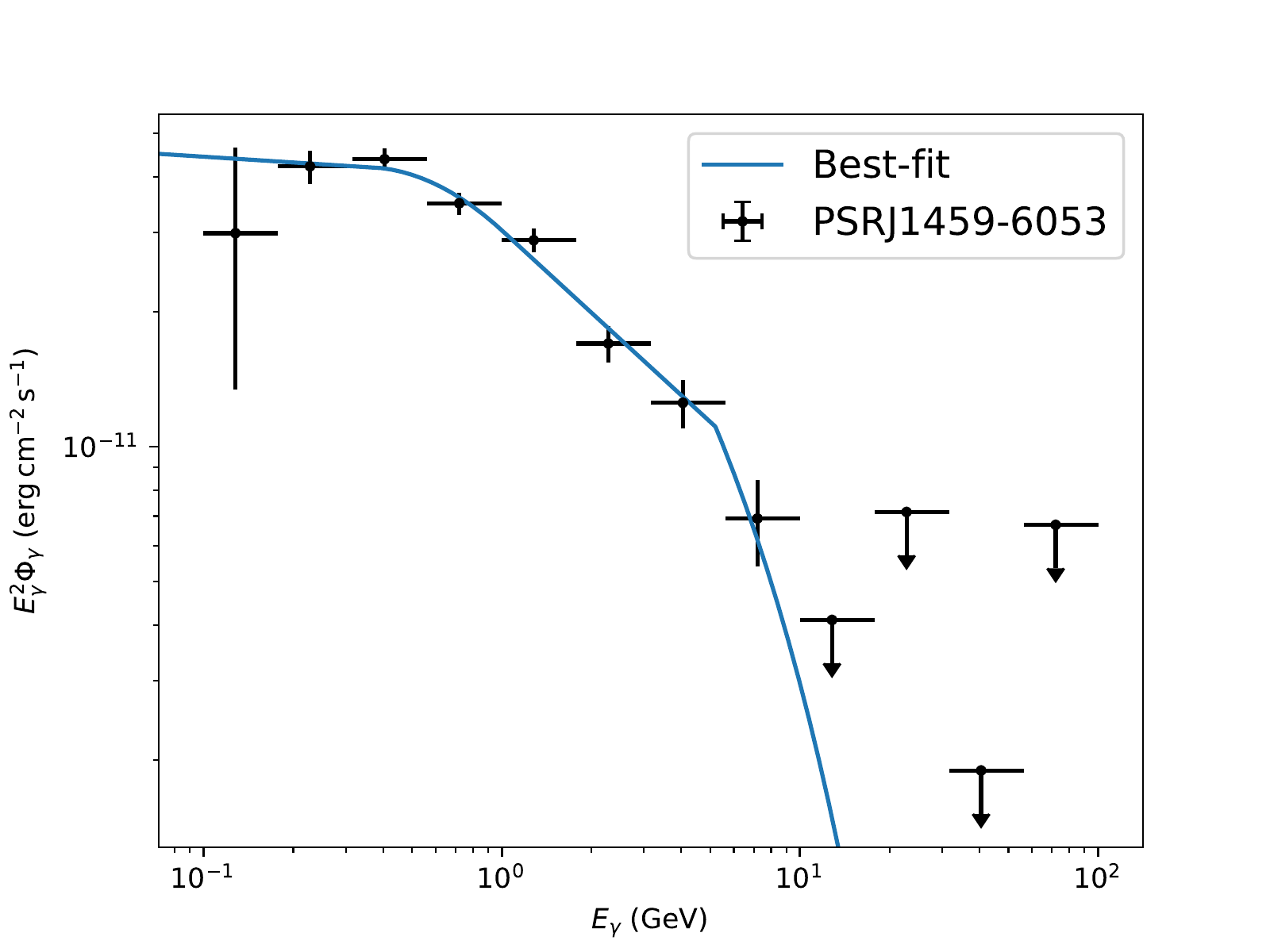}
  \includegraphics[width=0.368\textwidth]{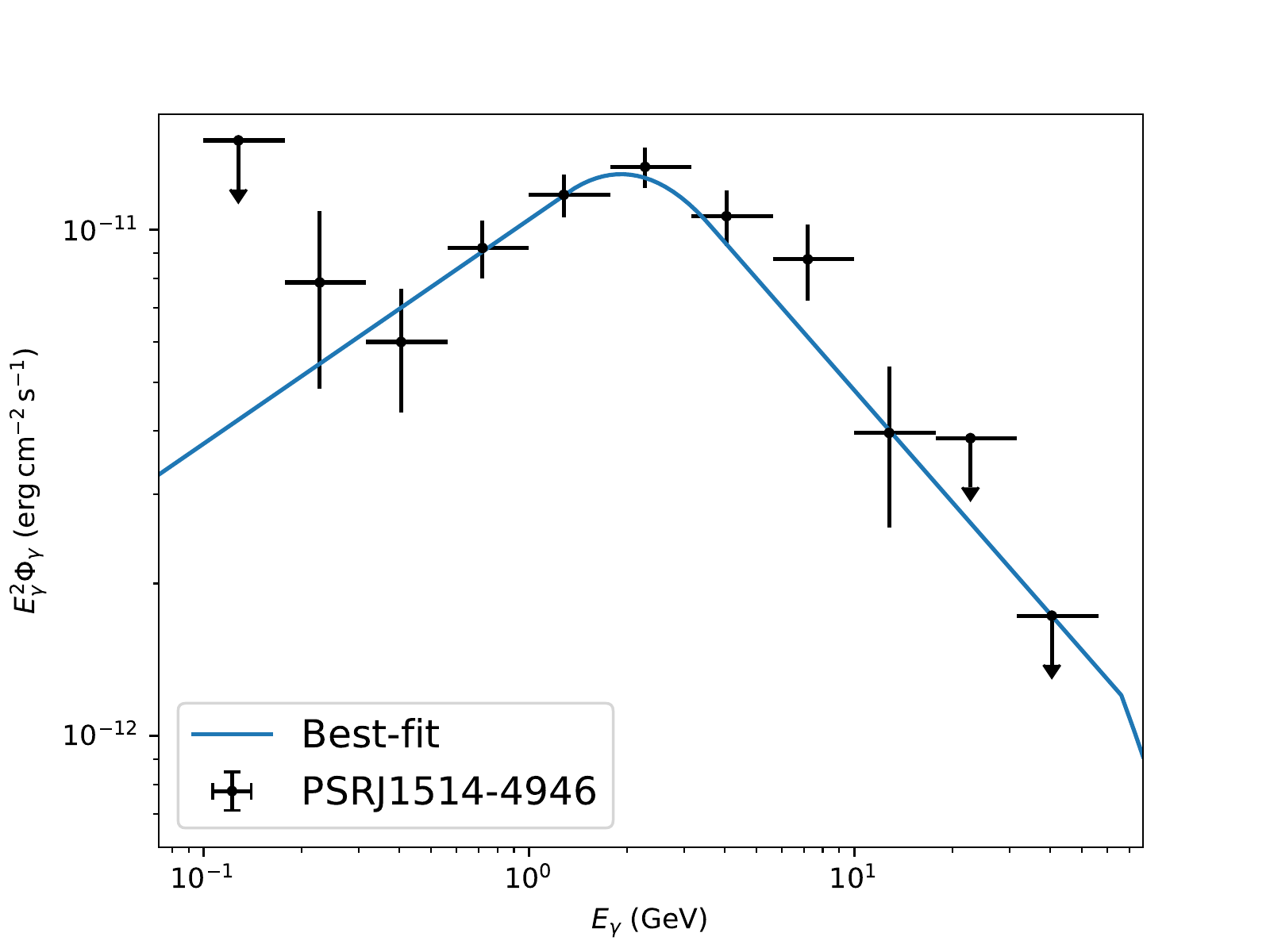}\!\!\!\!\!\!\!\!\!\!\!%
  \includegraphics[width=0.368\textwidth]{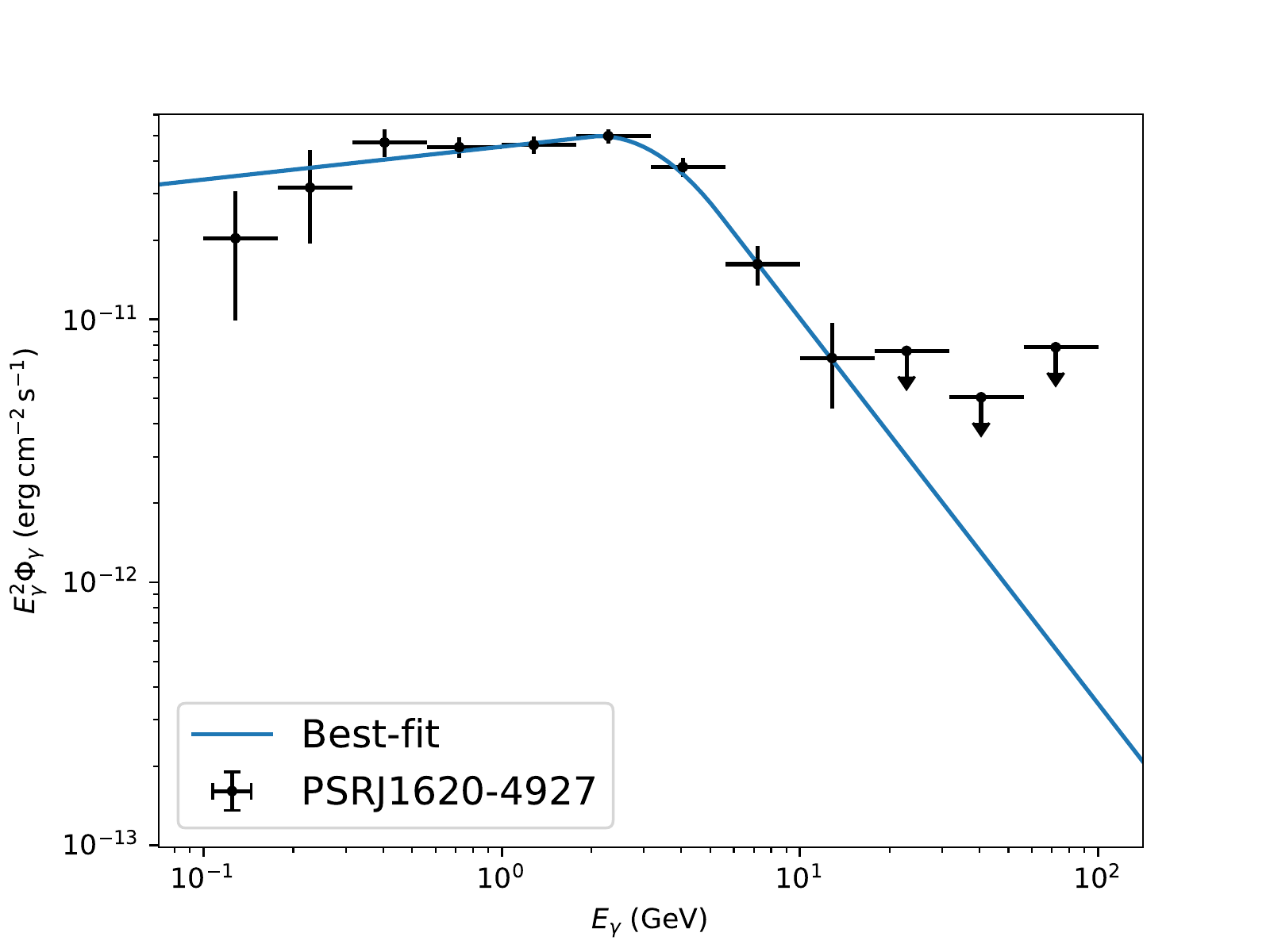}\!\!\!\!\!\!\!\!\!\!\!%
  \includegraphics[width=0.368\textwidth]{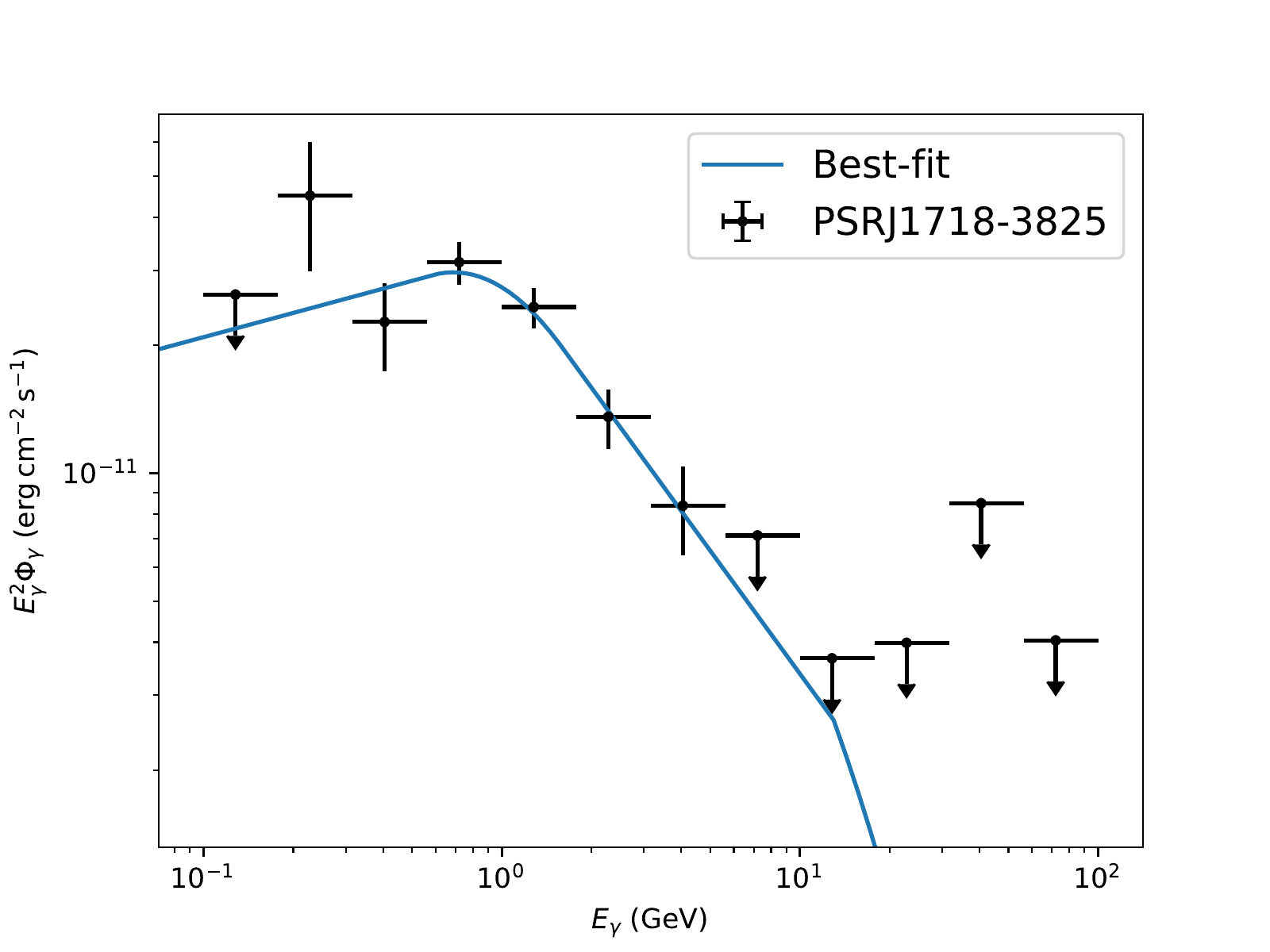}
  \caption{Phase-averaged spectra of 24 $\gamma$-ray pulsars in the 2PC catalog, and the best-fit results by using the GC-model. The four parameters of $E_\pi^\mathrm{GC}$, $C_\gamma$, $\beta_p$, and $\beta_\gamma$ derived from our best fit are presented in Table~\ref{tab:1}. Note that in the spectra of some pulsars when a second spectral break could be seen, the position of the second break is connected with that of the first break by Equation~\ref{eq:2.4}. The observational data are taken from the 2PC catalog \cite{2013ApJS..208...17A}.\label{Fig.2}}
\end{figure*}
\begin{figure*}
  \setcounter{figure}{1}
  \includegraphics[width=0.368\textwidth]{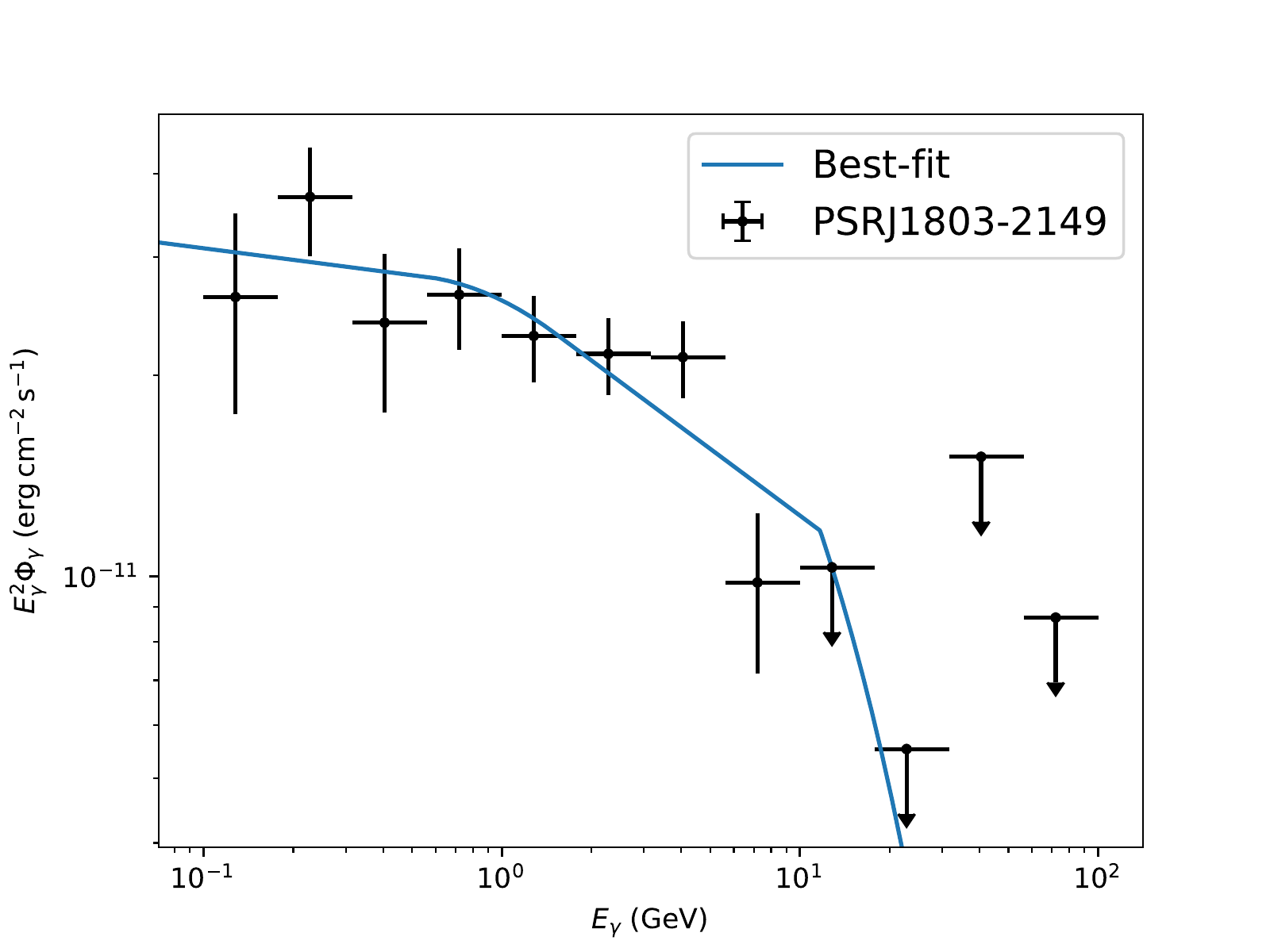}\!\!\!\!\!\!\!\!\!\!\!%
  \includegraphics[width=0.368\textwidth]{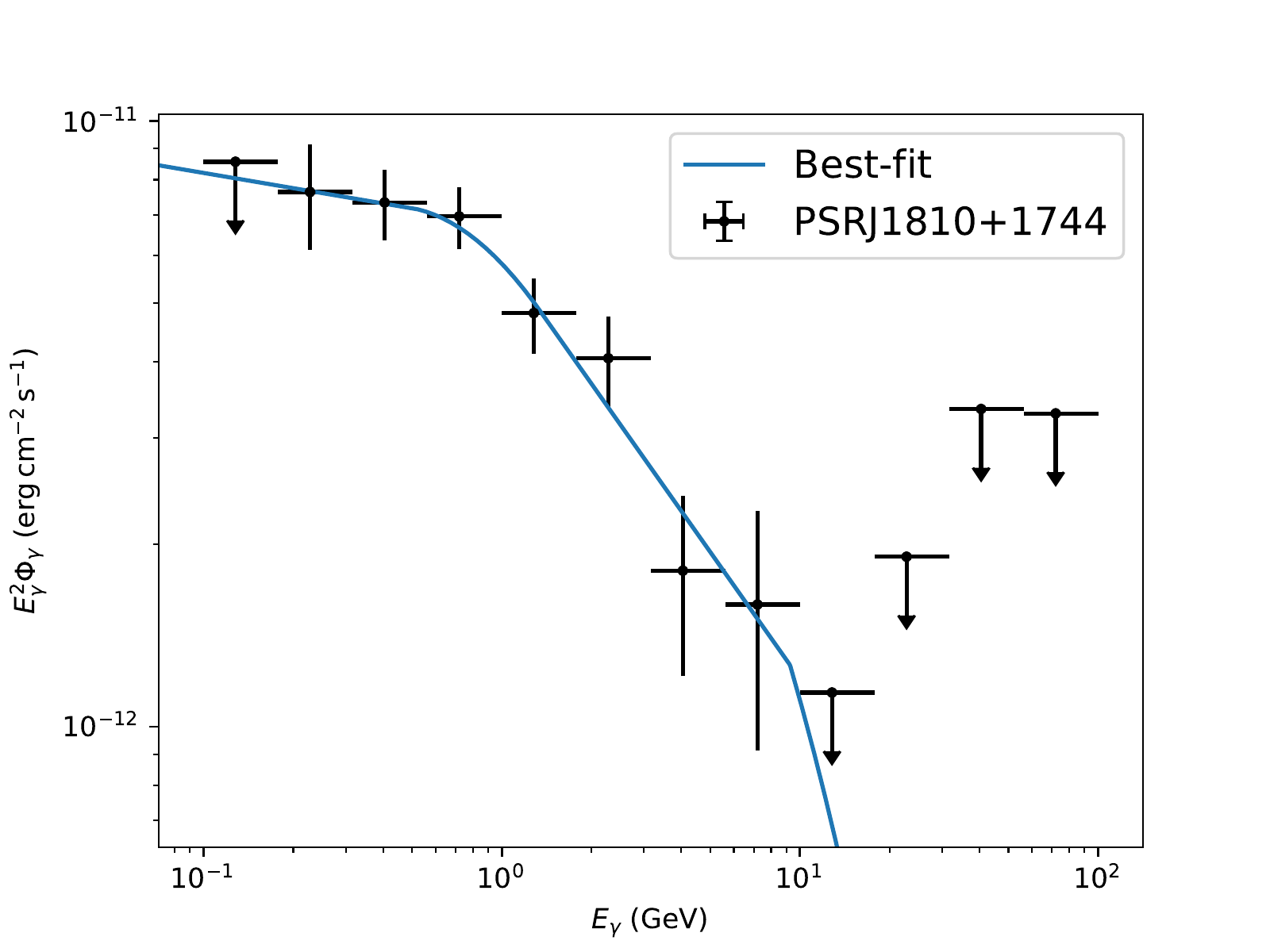}\!\!\!\!\!\!\!\!\!\!\!%
  \includegraphics[width=0.368\textwidth]{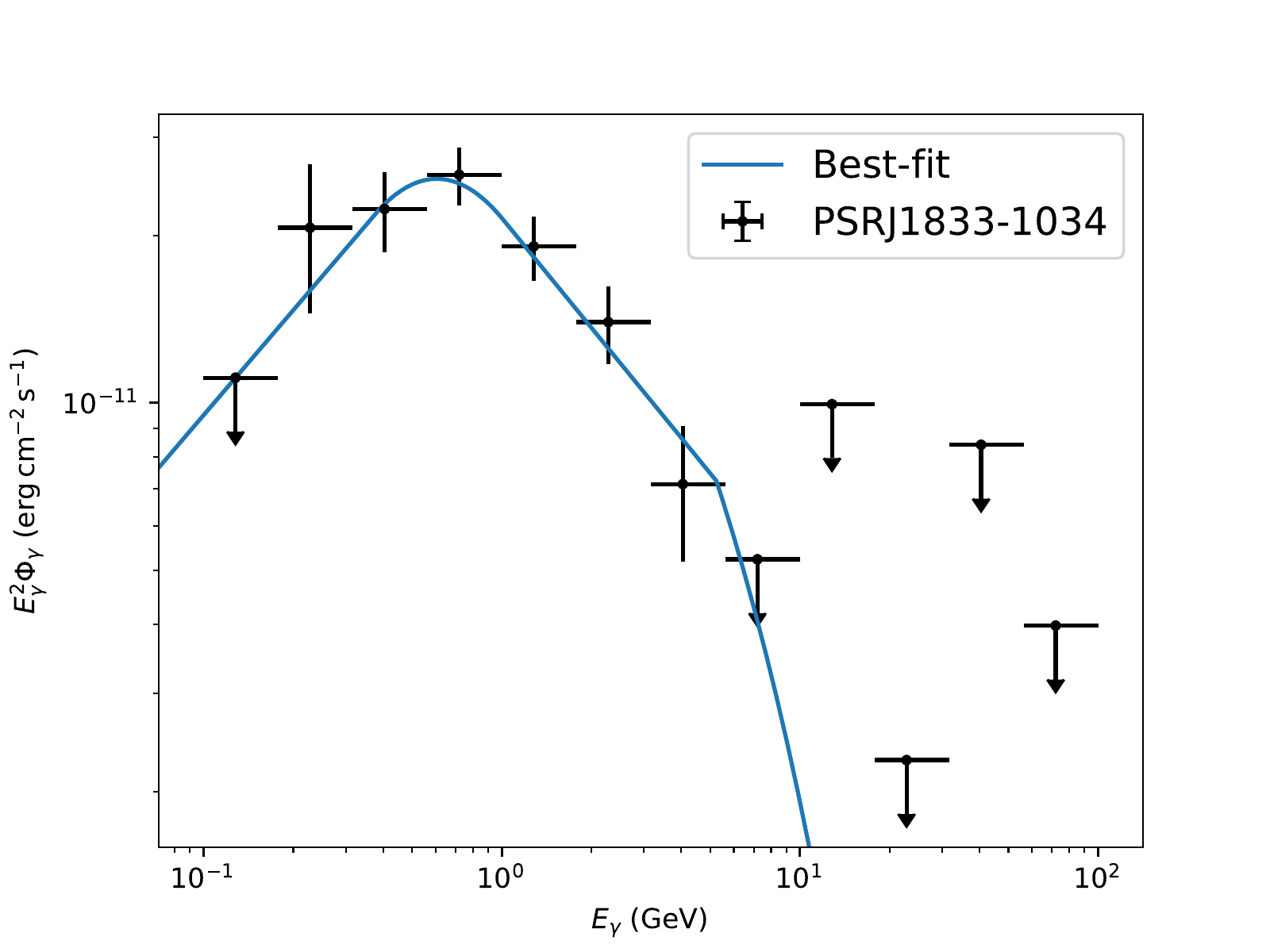}
  \includegraphics[width=0.368\textwidth]{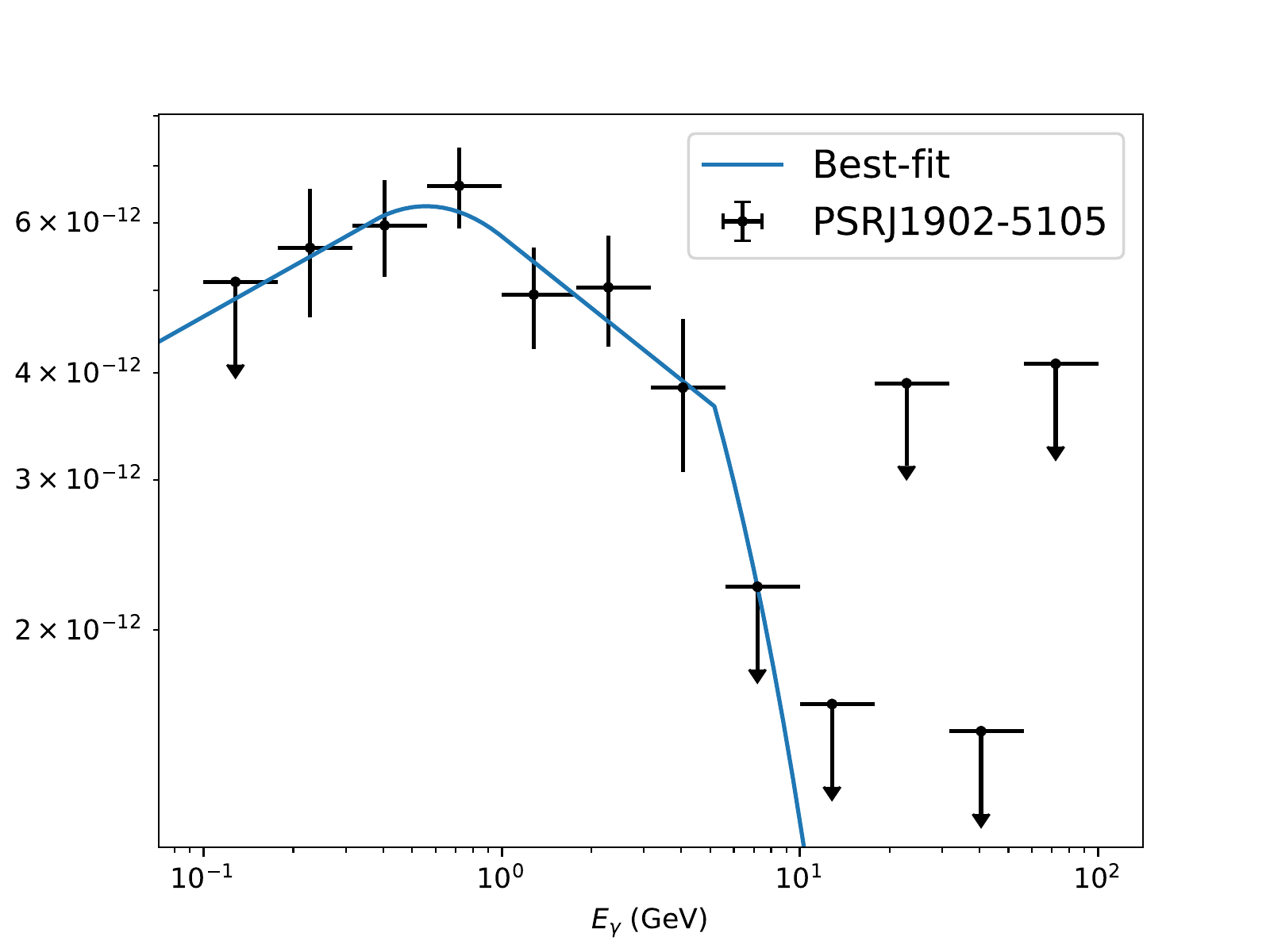}\!\!\!\!\!\!\!\!\!\!\!%
  \includegraphics[width=0.368\textwidth]{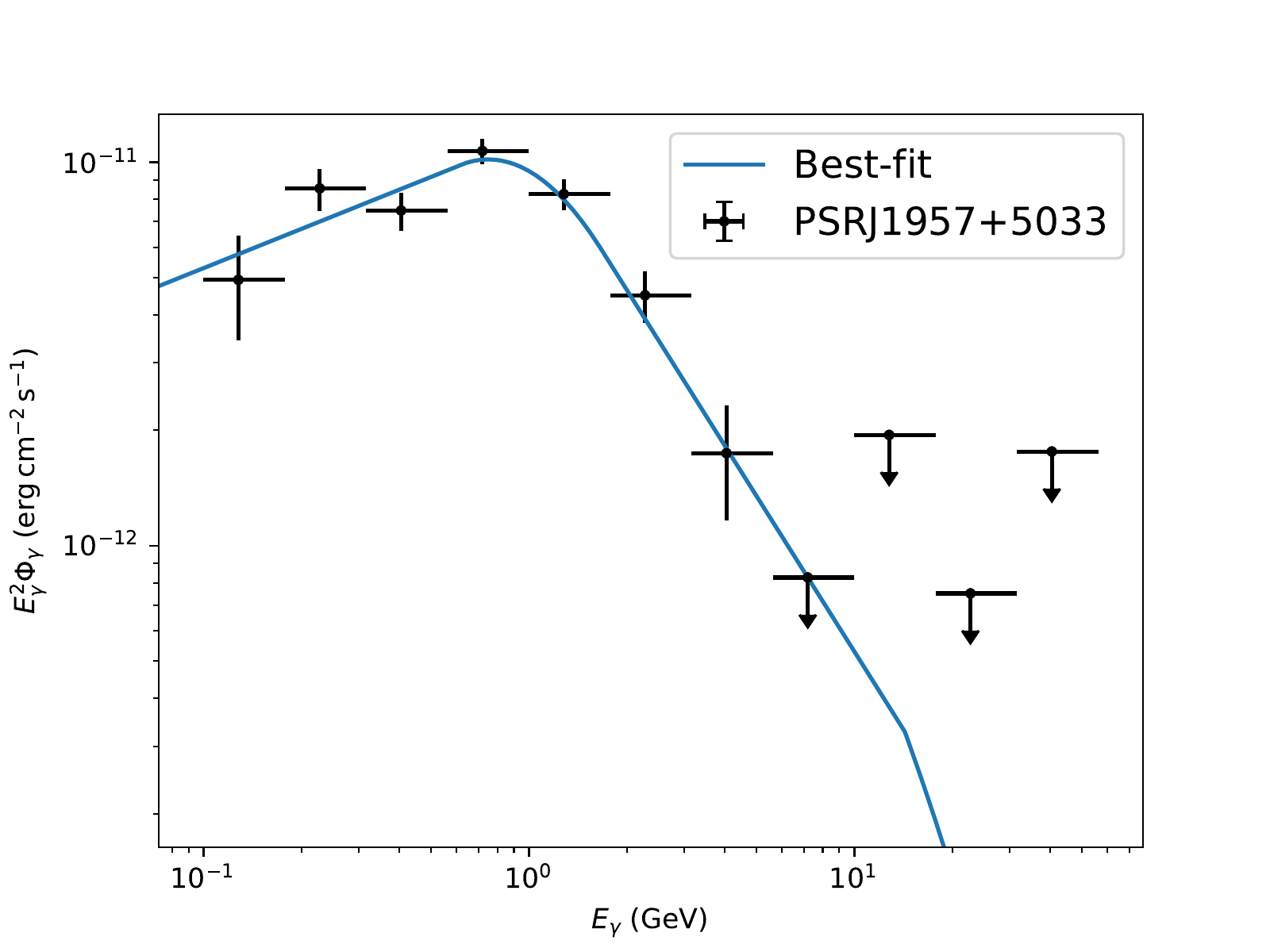}\!\!\!\!\!\!\!\!\!\!\!%
  \includegraphics[width=0.368\textwidth]{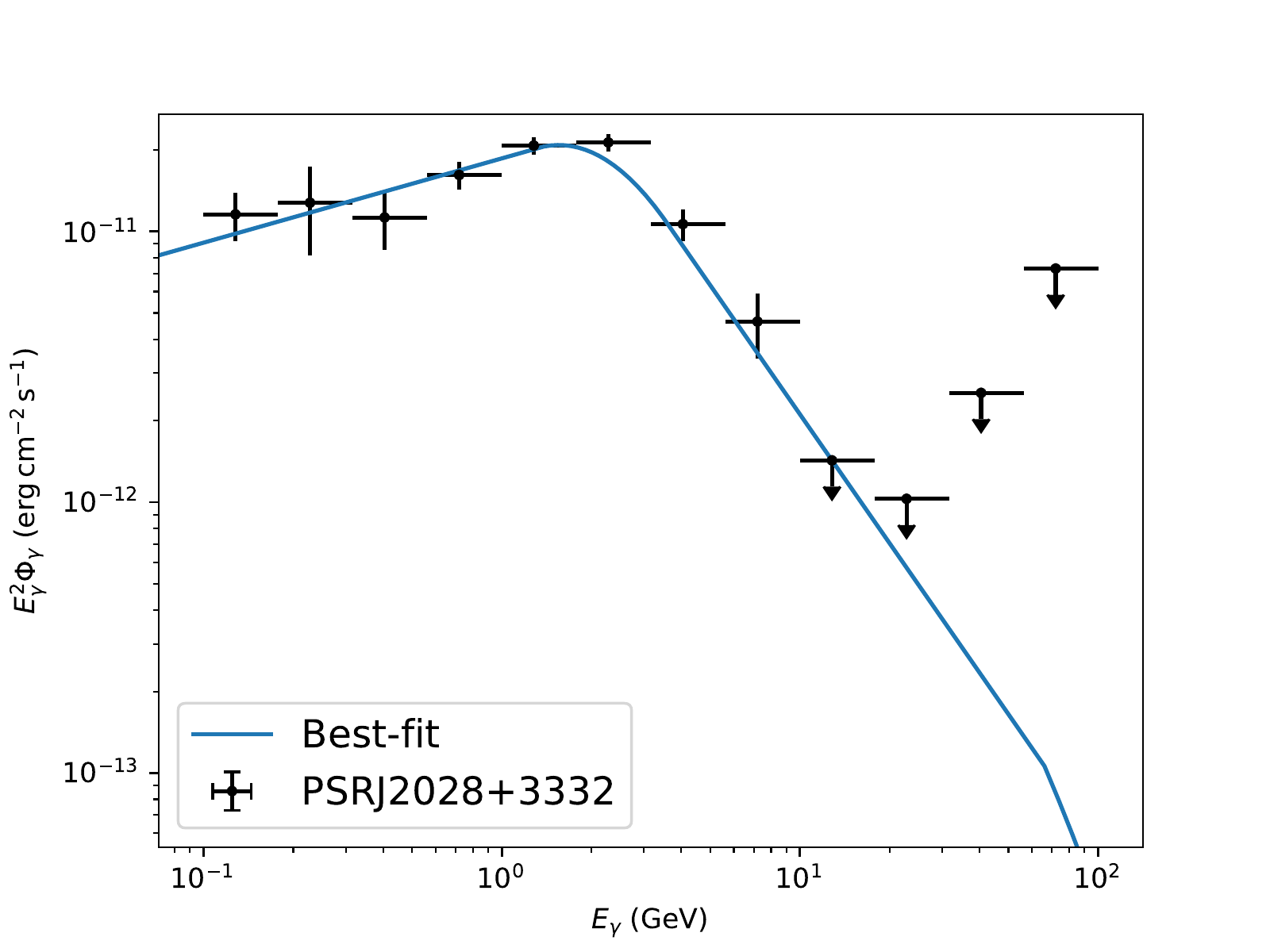}
  \includegraphics[width=0.368\textwidth]{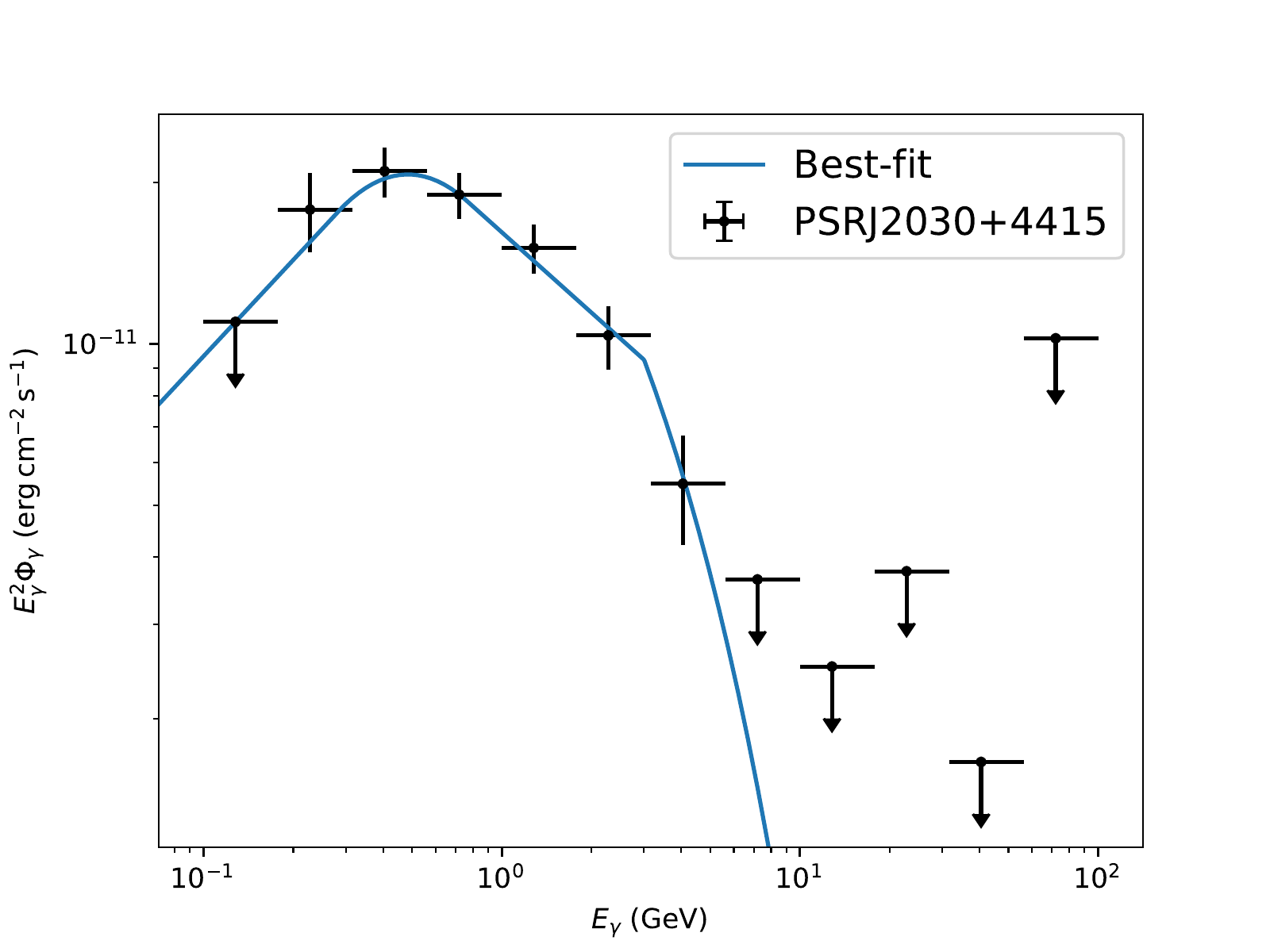}\!\!\!\!\!\!\!\!\!\!\!%
  \includegraphics[width=0.368\textwidth]{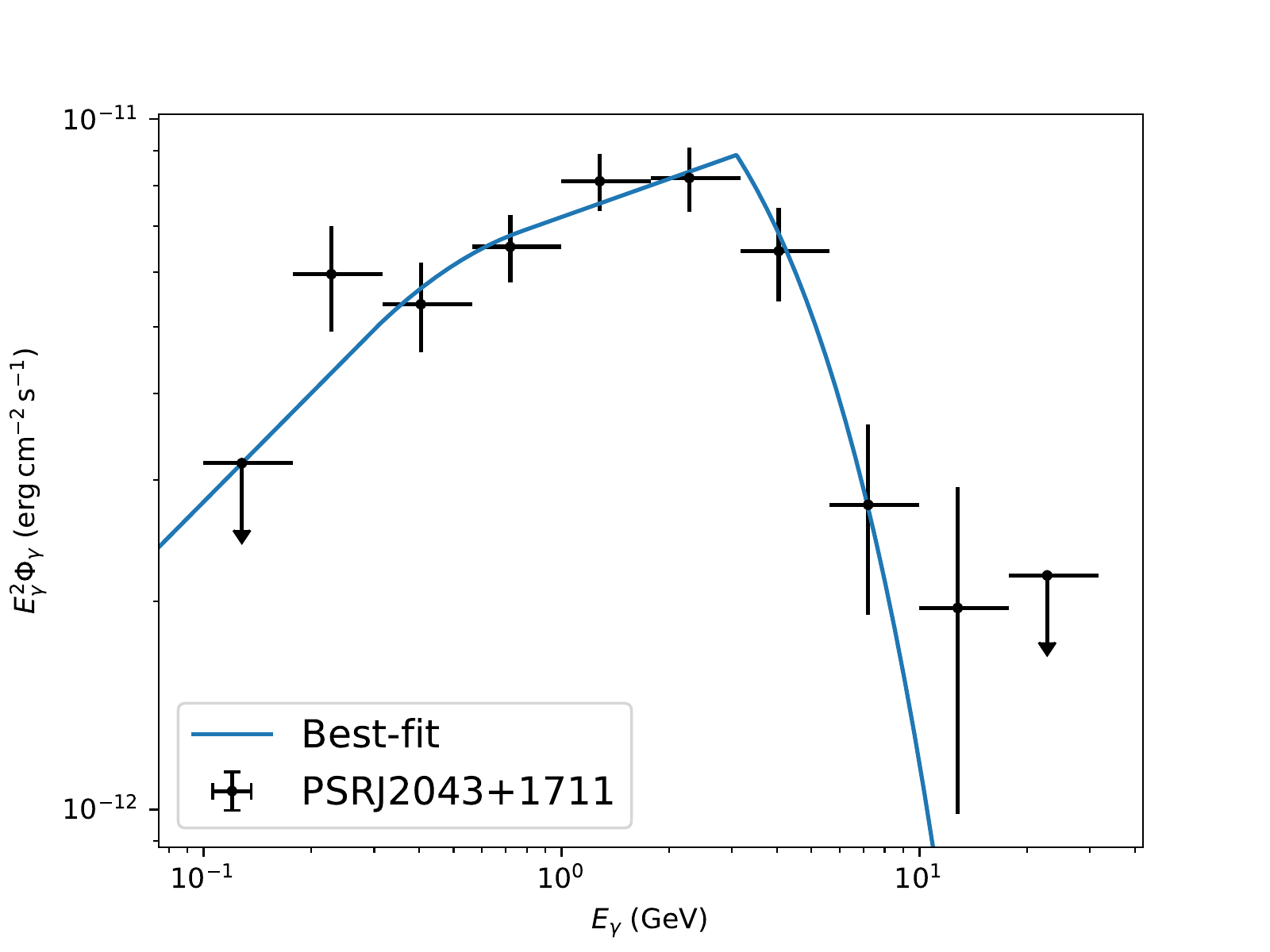}\!\!\!\!\!\!\!\!\!\!\!%
  \includegraphics[width=0.368\textwidth]{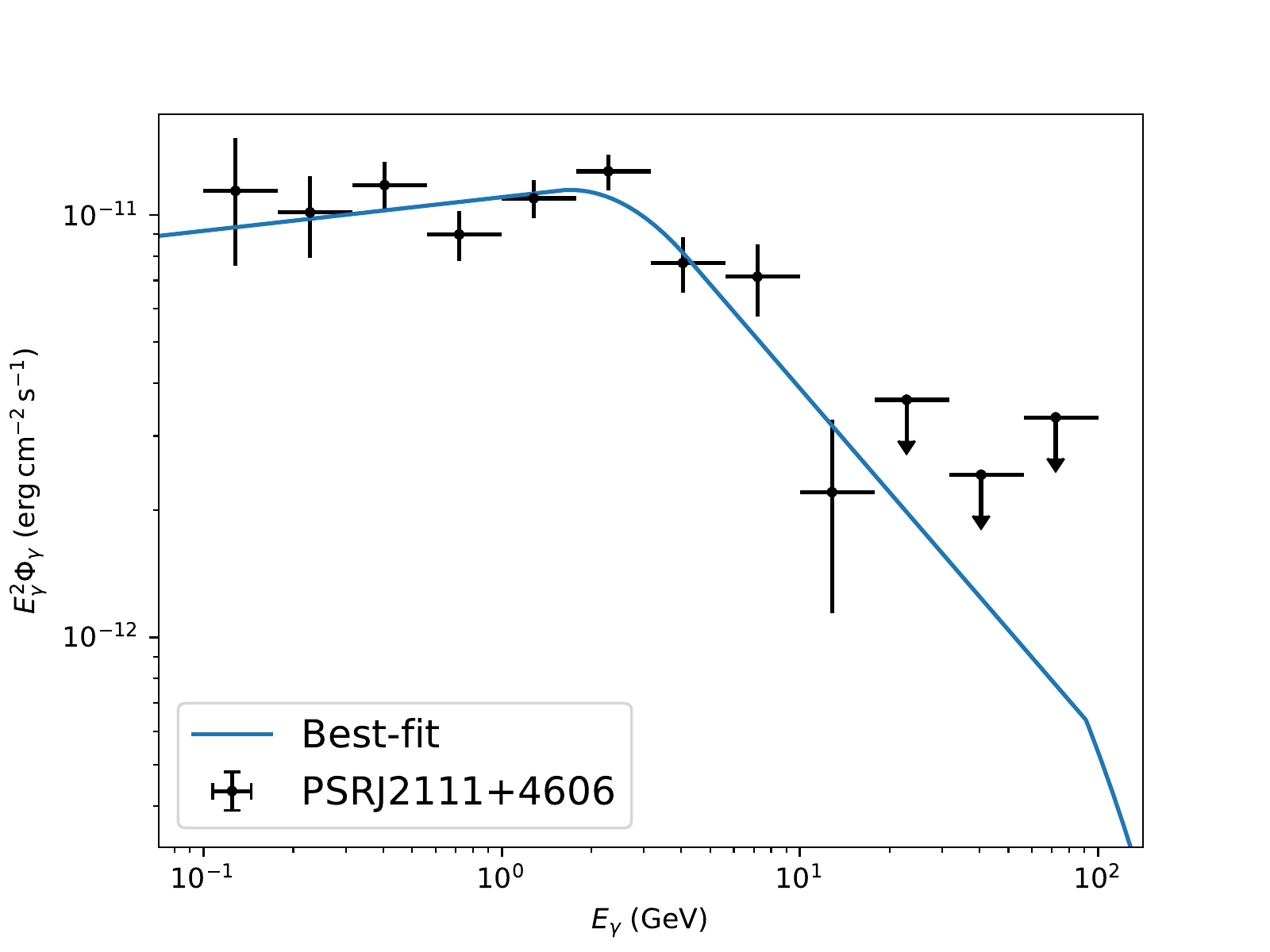}
  \includegraphics[width=0.368\textwidth]{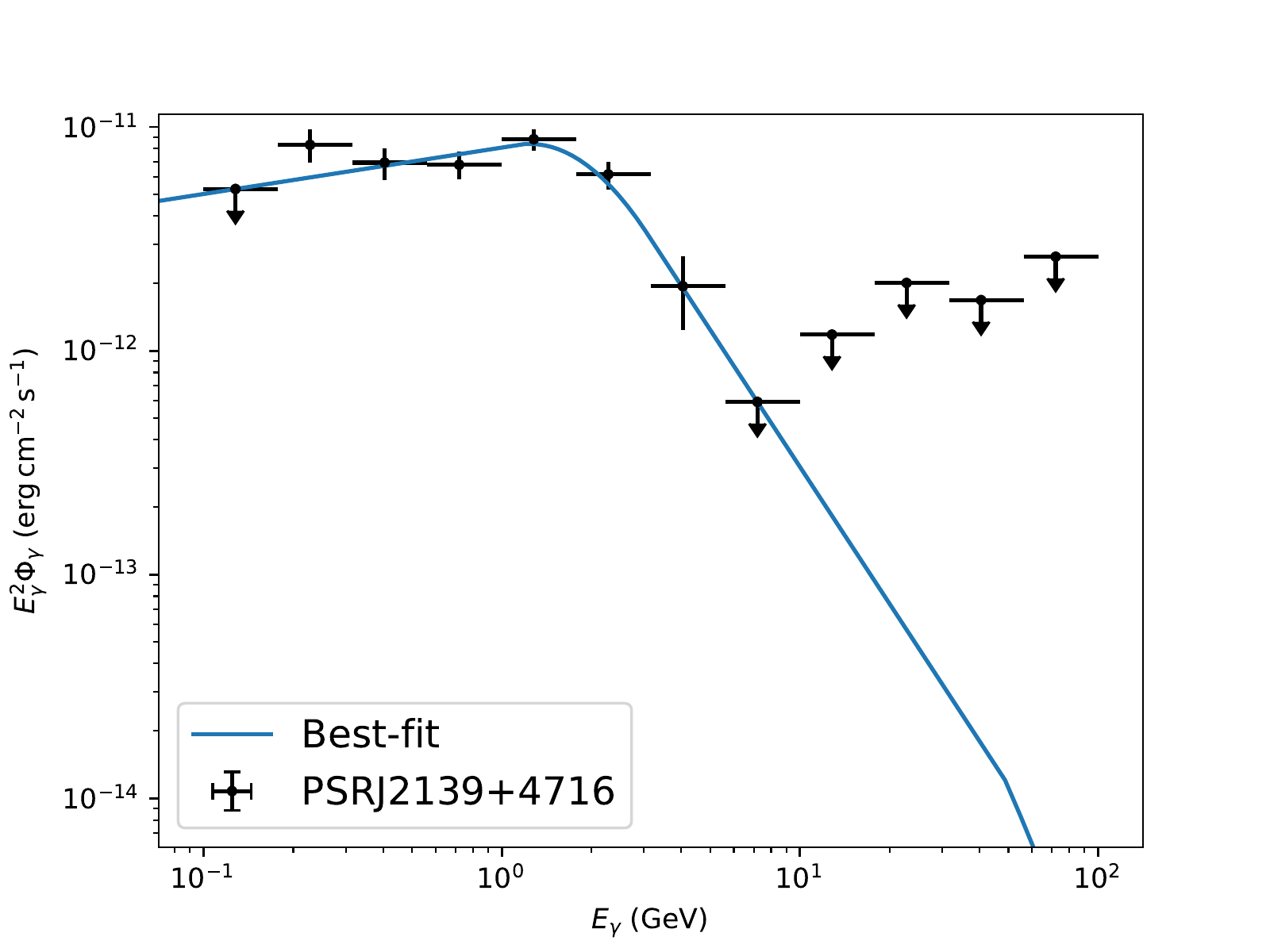}\!\!\!\!\!\!\!\!\!\!\!%
  \includegraphics[width=0.368\textwidth]{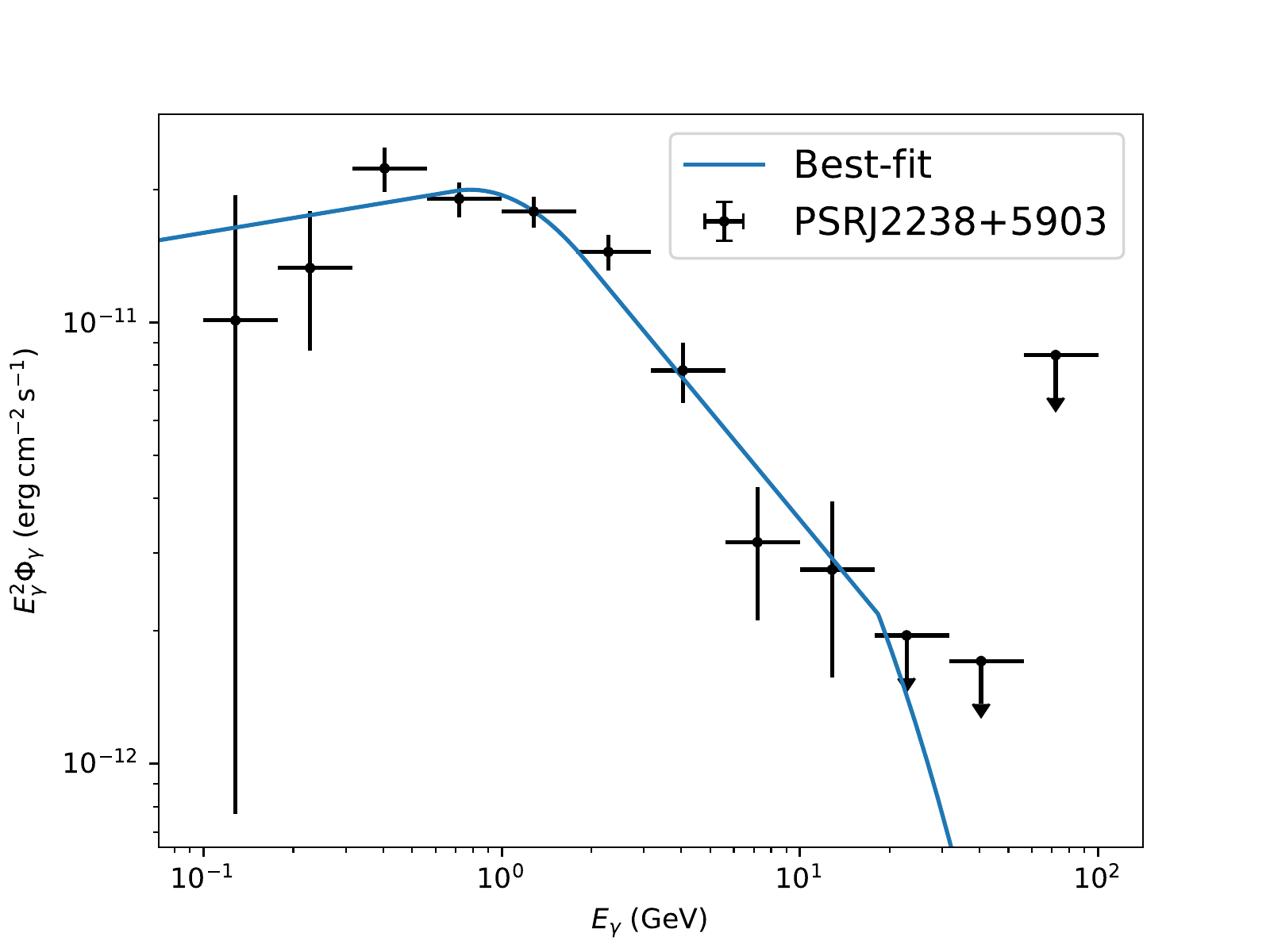}\!\!\!\!\!\!\!\!\!\!\!%
  \includegraphics[width=0.368\textwidth]{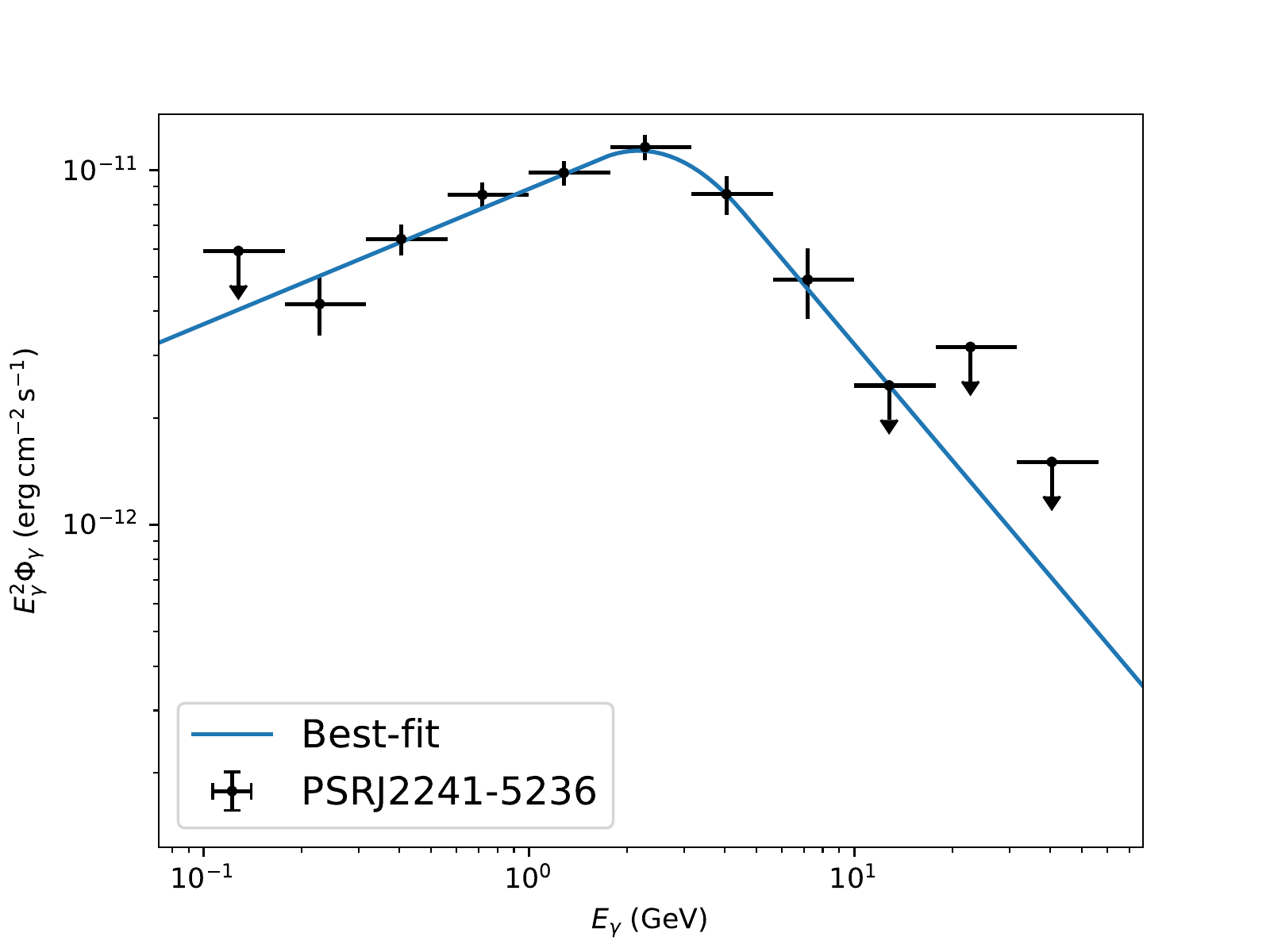}
  \caption{(Continued).}
\end{figure*}

\begin{table*}
  \begin{ruledtabular}
    \caption{\label{tab:1}Best-fit parameters of our GC-model for 25 $\gamma$-ray pulsars. The fitting results are correspondingly illustrated in Figs.~\ref{Fig.1} and~\ref{Fig.2}.}
    \begin{tabular}{l*5{c}}
    Pulsar & $E_\pi^\mathrm{GC}$ & $C_\gamma$ & $\beta_p$ & $\beta_\gamma$ & $\chi^2/\mathrm{d.o.f.}$\\
    & $\left(\mathrm{GeV}\right)$ & $\left(\mathrm{GeV^{-2}\,cm^{-2}\,s^{-1}}\right)$ & & & \\
    \colrule
    PSR J0106$+$4855 & 4.64 & 6.70E-13 & 1.39 & 1.91 & $1.53/2=0.77$ \\
    PSR J0437$-$4715 & 1.00 & 8.02E-11 & 1.41 & 1.76 & $4.59/3=1.53$ \\
    PSR J0631$+$1036 & 0.66 & 1.16E-10 & 0.71 & 1.82 & $4.41/3=1.47$ \\
    PSR J1023$-$5746 & 0.85 & 6.72E-10 & 0.98 & 1.75 & $4.70/5=0.94$ \\
    PSR J1105$-$6107 & 1.32 & 2.65E-11 & 0.87 & 2.11 & $1.46/2=0.73$ \\
    PSR J1119$-$6127 & 0.85 & 1.75E-10 & 0.88 & 1.89 & $4.34/4=1.08$ \\
    PSR J1124$-$5916 & 1.24 & 6.02E-11 & 1.03 & 1.94 & $7.08/4=1.77$ \\
    PSR J1135$-$6055 & 0.60 & 2.79E-10 & 0.80 & 1.84 & $1.58/2=0.79$ \\
    PSR J1420$-$6048 & 0.94 & 3.45E-10 & 0.91 & 1.88 & $3.37/4=0.84$ \\
    PSR J1459$-$6053 & 0.61 & 6.26E-10 & 0.78 & 2.04 & $6.12/4=1.53$ \\
    PSR J1514$-$4946 & 2.17 & 1.03E-11 & 1.09 & 1.56 & $5.58/4=1.40$ \\
    PSR J1620$-$4927 & 3.24 & 1.50E-11 & 1.30 & 1.87 & $3.80/5=0.76$ \\
    PSR J1718$-$3825 & 0.96 & 2.52E-10 & 1.08 & 1.81 & $2.71/2=1.36$ \\
    PSR J1803$-$2149 & 0.91 & 5.95E-11 & 0.64 & 2.06 & $7.44/4=1.86$ \\
    PSR J1810$+$1744 & 0.81 & 4.78E-11 & 0.81 & 2.08 & $1.63/3=0.54$ \\
    PSR J1833$-$1034 & 0.61 & 1.01E-09 & 1.14 & 1.37 & $1.78/2=0.89$ \\
    PSR J1902$-$5105 & 0.61 & 8.56E-11 & 0.74 & 1.80 & $1.05/2=0.52$ \\
    PSR J1957$+$5033 & 1.01 & 1.15E-10 & 1.35 & 1.66 & $4.88/3=1.63$ \\
    PSR J2028$+$3332 & 2.17 & 2.67E-11 & 1.45 & 1.69 & $4.42/4=1.11$ \\
    PSR J2030$+$4415 & 0.46 & 1.59E-09 & 1.05 & 1.40 & $1.09/2=0.54$ \\
    PSR J2043$+$1711 & 0.47 & 1.25E-10 & 0.67 & 1.48 & $6.00/4=1.50$ \\
    PSR J2111$+$4606 & 2.55 & 3.90E-12 & 0.95 & 1.92 & $6.98/5=1.40$ \\
    PSR J2139$+$4716 & 1.86 & 1.94E-11 & 1.62 & 1.79 & $3.62/2=1.81$ \\
    PSR J2238$+$5903 & 1.14 & 7.91E-11 & 0.96 & 1.89 & $9.96/5=1.99$ \\
    PSR J2241$-$5236 & 2.76 & 5.51E-12 & 1.23 & 1.62 & $2.55/3=0.85$
    \end{tabular}
  \end{ruledtabular}
\end{table*}

There are 117 $\gamma$-ray pulsars reported in the 2PC catalog \cite{2013ApJS..208...17A}. We have systematically searched through the catalog to find out how many pulsars are compatible with the GC-model. In our study, the uncertainty ($\sigma_i$) of the observed flux ($E^2_\gamma\Phi_{\gamma,i}^\mathrm{obs}$) and the corresponding central energy ($E_{\gamma,i}$) of the spectral bin are taken from \cite{2013ApJS..208...17A}. The predicted $E^2_\gamma\Phi_{\gamma,i}^\mathrm{pre}$ by the GC-model is calculated by using Equation~\ref{eq:2.3}. Each normal data point $i$ contributes a likelihood $L_i$ as
\begin{equation}
  \ln L_i=-\frac12\frac{\left(E^2_\gamma\Phi_{\gamma,i}^\mathrm{pre}-E^2_\gamma\Phi_{\gamma,i}^\mathrm{obs}\right)^2}{\sigma_i^2}-\frac12\ln\left(2\uppi\sigma_i^2\right).\label{eq:4.1}
\end{equation}
For those data points that only present an upper limit for the flux, $L_i$ is calculated as \cite{2022arXiv220508566M}
\begin{equation}
  \ln L_i=-\frac12\frac{\left\{\max\left[\left(E^2_\gamma\Phi_{\gamma,i}^\mathrm{pre}-E^2_\gamma\Phi_{\gamma,i}^\mathrm{obs}\right),0\right]\right\}^2}{\left(0.01\times E^2_\gamma\Phi_{\gamma,i}^\mathrm{obs}\right)^2}.\label{eq:4.2}
\end{equation}
This equation gives an appropriate penalty to the model when the predicted flux is higher than the observational upper limit. The best-fit parameters of the GC-model are then obtained by maximizing the total likelihood, $\sum_i\ln L_i$.

\begin{figure}
  \includegraphics[width=0.535\textwidth]{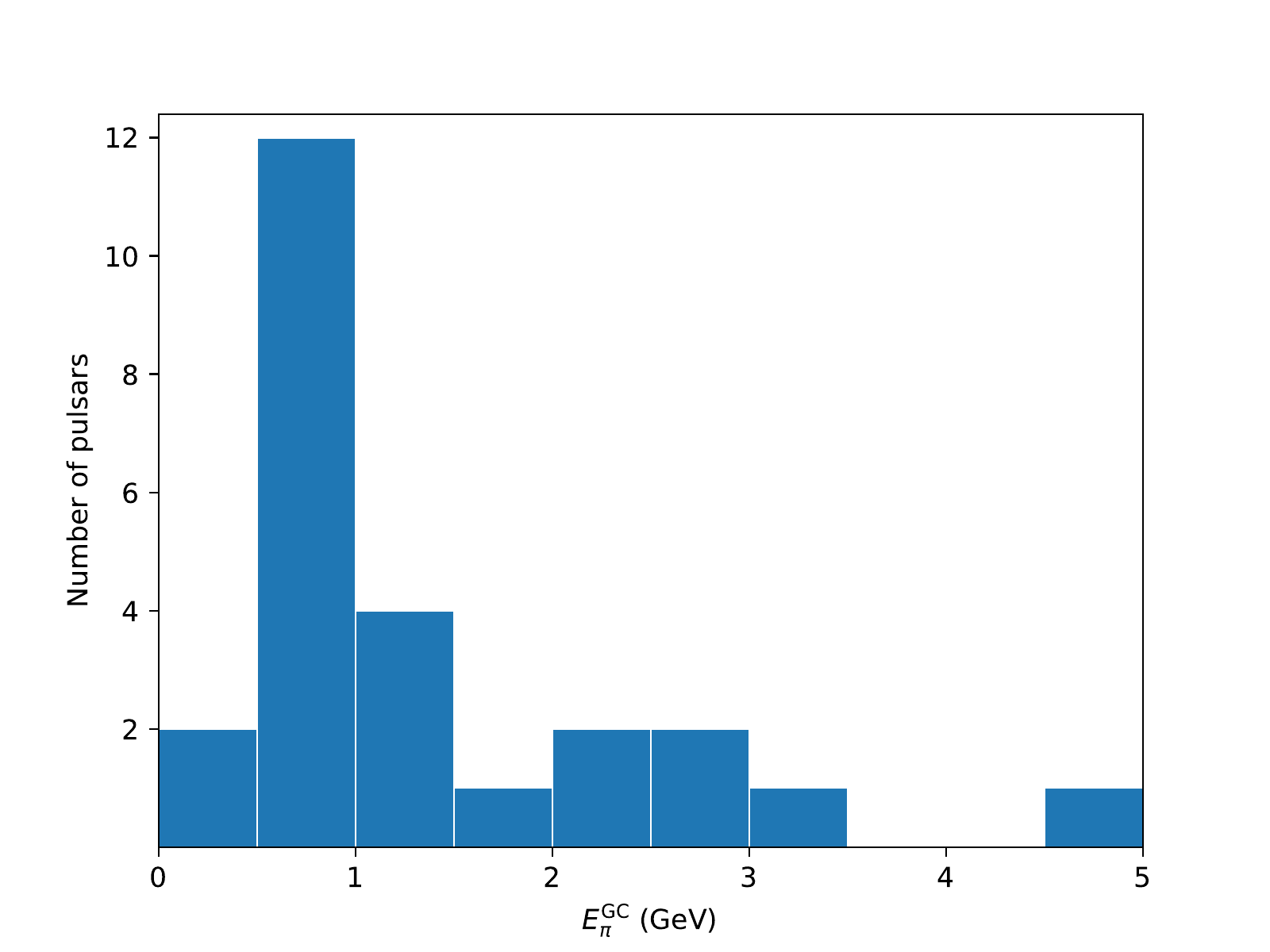}
  \caption{Distribution of $E_\pi^\mathrm{GC}$ for all the 25 pulsars listed in Table~\ref{tab:1}.\label{Fig.3}}
\end{figure}

To be more specific, the goodness of fit is assessed by utilizing the reduced-$\chi^2$
\begin{equation}
  \chi^2/\mathrm{d.o.f.}=\sum_i\frac{\left(E^2_\gamma\Phi_{\gamma,i}^\mathrm{pre}-E^2_\gamma\Phi_{\gamma,i}^\mathrm{obs}\right)^2}{\sigma^2_i}\times\frac{1}{\mathrm{d.o.f.}},\label{eq:4.3}
\end{equation}
where the degree of freedom ($\mathrm{d.o.f.}$) is defined as the number of observational data points subtracted by the number of model parameters. Note that when calculating the degree of freedom, only those data points representing a clear positive detection are counted, while the data points corresponding to an upper limit are omitted. This is reasonable since these upper limits usually present poor constraints on the theoretical model.

For the 117 $\gamma$-ray pulsars reported in the 2PC catalog, 25 sources have poor spectrum data with large uncertainties \cite{2013ApJS..208...17A}. For the rest pulsars, we have compared their spectra with the GC-model. It is found that at least 25 $\gamma$-ray pulsars are well fitted by our model, with $0.5<\chi^2/\mathrm{d.o.f.}<2$. With PSR J1420-6048 individually discussed in Sect.~\ref{sec:1420}, the rest results are illustrated in Fig.~\ref{Fig.2}. From this figure, we can clearly see that the spectra of these pulsars can all be described by a BPL function, and thus are compatible with our theoretical prediction \cite{2017NuPhB.916..647Z}.

According to the GC-model (see Equation~\ref{eq:2.3}), the $\gamma$-ray spectrum is determined by a set of four free parameters, $(E_\pi^\mathrm{GC}, C_\gamma, \beta_p,$ and $\beta_\gamma)$. The parameters derived from our best-fit results are listed in Table~\ref{tab:1}. Note that $\beta_p$ is the power-law index of the distribution of the protons accelerated in the environment around the pulsar. Usually, $\beta_p\sim2.7$ for protons diffusing and propagating in an environment with chaotic magnetic fields. However, in Table~\ref{tab:1}, we can see that $\beta_p$ is significantly less than $2.7$, which is special in the GC-model. On the other hand, $\beta_\gamma$ is affected by the medium absorption of pions so that a small value of $\beta_\gamma\in(1,2)$ is expected in our GC-model. For the parameter of $E_\pi^\mathrm{GC}$, it is mainly in a range of $(0.3\,\mathrm{GeV},4.6\,\mathrm{GeV})$ as shown in Table~\ref{tab:1} and Fig.~\ref{Fig.3}. This is also consistent with our expectations. According to Equation~\ref{eq:2.4}, $E_\pi^\mathrm{cut}$ could range from 1\,GeV to 300\,GeV. In the future, more spectrum data would be helpful to further test our model.

Finally, we discuss some of the pulsars listed in Table~\ref{tab:1}.

(i) We noticed that PSR J0659+1414, J1420-6048, and J1718-3825 have previously been explained by adopting the SC model in \cite{2019MNRAS.489.5494T}. However, no $\chi^2/\mathrm{d.o.f.}$ is provided in that study. Therefore, a direct quantitative comparison between the SC model and the GC-model is not available. It is even possible that both models may take effect together. More observations are needed to further clarify the issue.

(ii) For PSR J0437-4715, it has been argued that electrons in this pulsar can be accelerated up to TeV energy in the bow shock wind nebula \cite{2019ApJ...876L...8B}. Therefore, one can expect that protons can also be accelerated to $E_p\sim20\,\mathrm{TeV}$ and reach the GC-threshold $E_\pi^\mathrm{GC}\sim1\,\mathrm{GeV}$, since the cooling due to synchrotron loss is less efficient for protons.

(iii) For PSR J1420-6048, a candidate PWN, HESS J1420-607, is found to be associated with this compact star. It is a very hard $\gamma$-ray source extending to $20\,\mathrm{TeV}$ \cite{2006A&A...456..245A}. According to the leptonic model, HESS J1420-607 involves a VHE acceleration mechanism for electrons, which can also accelerate protons to TeV energy. These protons may arrive at PRS J1420-6048 and bombard neutron cluster $A^*$ on the stellar surface. Therefore, the acceleration of high-energy protons seems to be quite common in PWN-pulsar interactions.

\begin{figure}
  \includegraphics[width=0.535\textwidth]{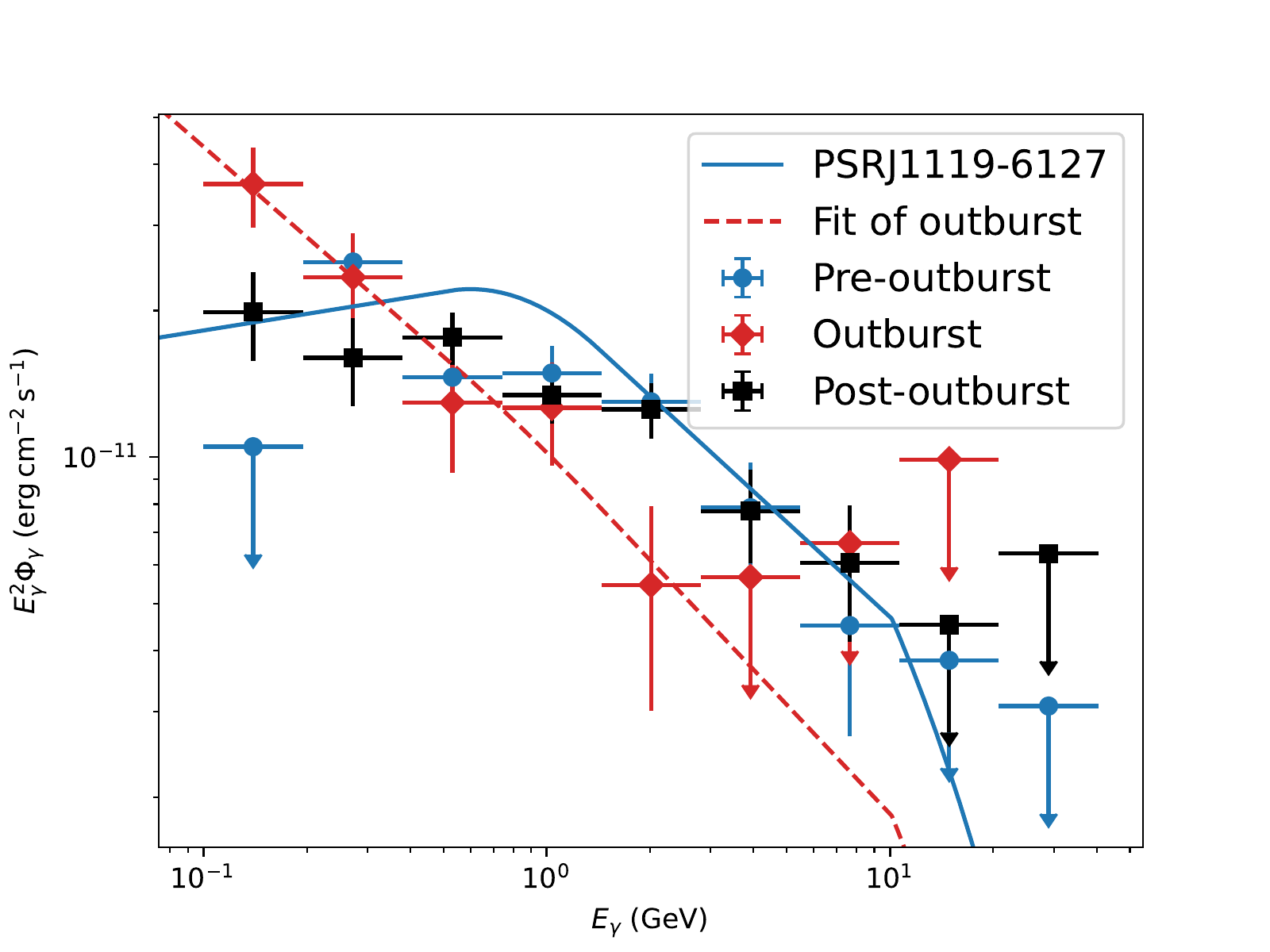}
  \caption{Spectra of PSR J1119-6127 in the normal state (circles for pre-outburst stage and squares for post-outburst stage) and the outburst state (diamonds). During the outburst, the parameters changed to $C_\gamma=1.49\times10^{-11}\,\mathrm{GeV^{-2}\,cm^{-2}\,s^{-1}}$, $\beta_p=0.57$, and $\beta_\gamma=2.62$, but the $E_\pi^\mathrm{GC}$ parameter does not vary and is still $E_\pi^\mathrm{GC}=0.85\mathrm{GeV}$. The solid curve is our best fit to the pulsar during its normal state, while the dashed curve is the one during the outburst state.\label{Fig.4}}
\end{figure}

(iv) PSR J1119-6127 is a highly magnetized pulsar, from which an outburst was observed in July 2016 \cite{2020ApJ...902...96W}. Six months later, the $\gamma$-ray spectrum recovers its normal state. The spectra of the curvature radiation are sensitive to the configuration of the magnetic field. The above result means that the GeV emission of PSR J1119-6127 origins from the outer magnetosphere, since its magnetosphere has been reconfigured by the outburst. We point out that the GC-model may provide a new understanding. The $pA^*$ collisions occur on the stellar surface. The strong perturbation caused by the outburst in the magnetosphere may affect the propagations of protons and $\gamma$-rays, but it does not influence the $pA^*$ collisions. On the other hand, the influence of the magnetic reconfiguration on the shock acceleration mechanisms is weak. Thus, the two parameters of $\beta_p$ and $\beta_\gamma$ will be changed during the outburst and they restore their original values when the outburst is finished. In Fig.~\ref{Fig.4}, the spectra of PSR J1119-6127 in different states are shown and are compared with our GC-model. We see that the model can explain the observations satisfactorily.

\section{Discussions and Conclusion\label{sec:con}}

In this study, possible astrophysical phenomena connected to the GC in nucleons are studied. Superfluid neutrons inside a neutron star may penetrate to the stellar surface and form neutron cluster $A^*$. $\gamma$-rays emitted during the $pA^*$ hadronic collisions then follow a special broken power-law spectrum. It is found that the observed GeV $\gamma$-ray spectra of at least 25 pulsars in the 2PC catalog can be well explained by such a GC-model. The results indicate that a hadronic model with the GC effect is a possible mechanism for GeV $\gamma$-ray pulsars.

\begin{figure}
  \includegraphics[width=0.535\textwidth]{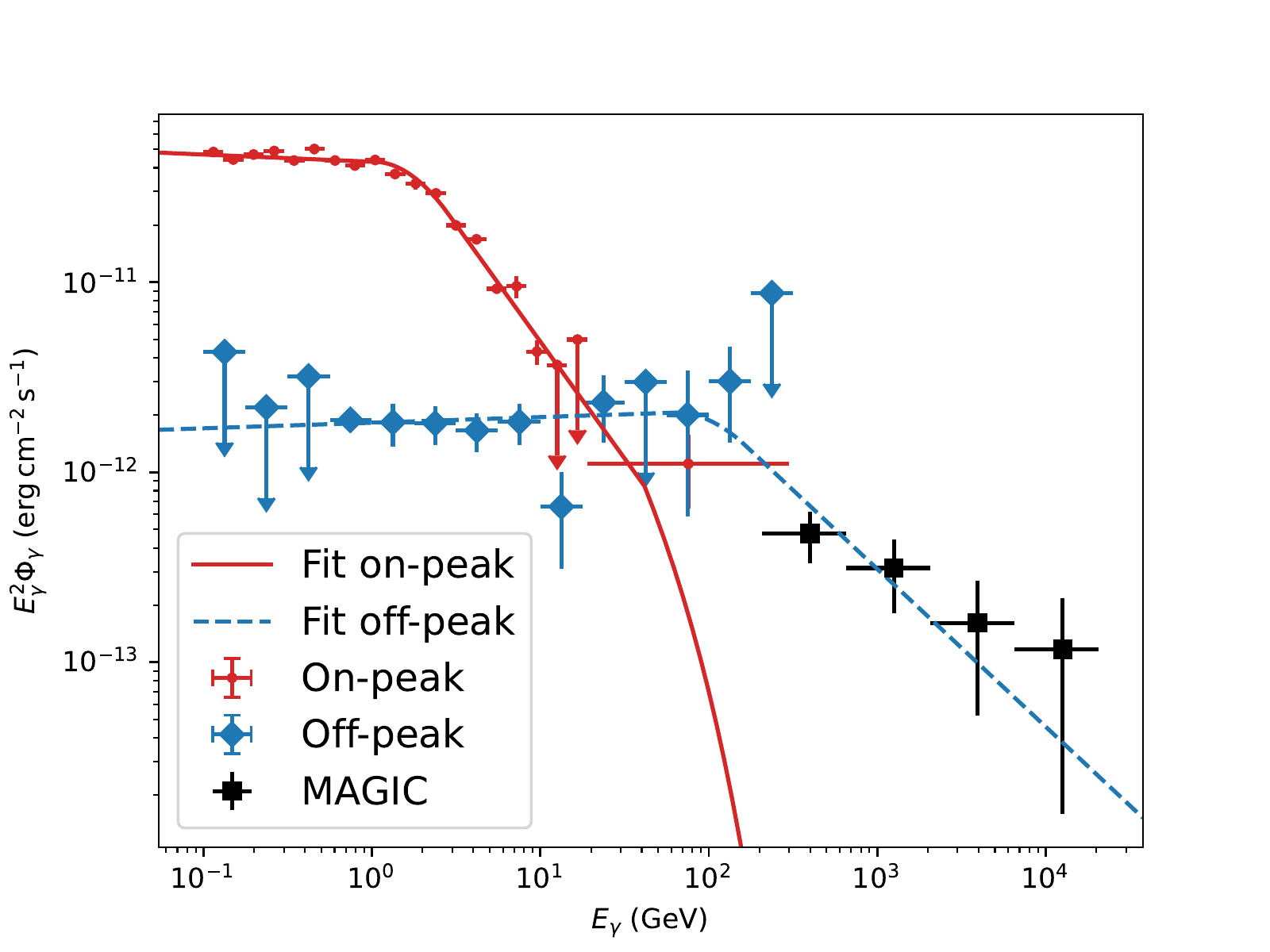}
  \caption{Observed $\gamma$-ray spectra of the pulsar/PWN complex PSR J0205+6449/3C58, and our best-fit results by using the GC-model. The observational data are taken from \cite{2018ApJ...858...84L} and \cite{2021JCAP...08..065R}. For the on-peak spectrum (dots), the best-fit (solid curve) parameters are $E_\pi^\mathrm{GC}=1.72\,\mathrm{GeV}$, $\beta_p=1.10$, $\beta_\gamma=2.04$, and $C_\gamma=5.99\times10^{-11}\,\mathrm{GeV^{-2}\,cm^{-2}\,s^{-1}}$. For the off-peak spectrum (diamonds for \textit{Fermi}-LAT and squares for MAGIC), the best-fit (dashed curve) parameters are $E_\pi^\mathrm{GC}=100\,\mathrm{GeV}$, $\beta_p=0.93$, $\beta_\gamma=1.97$, and $C_\gamma=1.09\times10^{-17}\,\mathrm{GeV^{-2}\,cm^{-2}\,s^{-1}}$.\label{Fig.5}}
\end{figure}

PSR J0205+6449/3C58 is an interesting pulsar/PWN complex. Classified as a PWN, 3C58 is an extended flat-spectrum radio source, and PSR J0205+6449 is a pulsar located in 3C58. Fig.~\ref{Fig.5} shows the on-peak and off-peak spectra of this complex observed by \textit{Fermi}-LAT and MAGIC. The observed spectra are usually explained in the framework of the leptonic model: the off-peak emission (including the MAGIC data) is believed to originate from inverse-Compton (IC) scattering, while the on-peak emission is due to synchrotron radiation, which provides soft seed photons for the IC process \cite{2018ApJ...858...84L}. Different from the usual leptonic model, \cite{2021JCAP...08..065R} have argued that the broken power-law spectrum of the off-peak emission at around ${\sim}100\,\mathrm{GeV}$ could be explained by the GC-model. In this study, we have fitted both the on-peak and off-peak spectra of the PSR J0205+6449/3C58 complex with the GC-model. The results are presented in Fig.~\ref{Fig.5}. We see that the on-peak spectrum can be well fitted by $pA^*$ collisions (i.e., the solid curve in Fig.~\ref{Fig.5}), which has a typical break energy of $E_\pi^\mathrm{GC} = 1.72\,\mathrm{GeV}$. However, the off-peak spectrum is much harder. It has a break energy of $E_\pi^\mathrm{GC}=100\,\mathrm{GeV}$, which indicates that it should be due to $pA$ collisions. Anyway, note that the parameters of $\beta_\gamma$ and $\beta_p$ are similar for the two emission components, which is reasonable in the GC-model.

Due to the lack of knowledge of the internal structure of nucleons, the energy scale of the GC effects is still an open issue. Here we present some discussions on this point. Taking $E_\pi^\mathrm{GC} = 1\,\mathrm{GeV}$ and $5\,\mathrm{GeV}$ as two examples, then from Equations~\ref{eq:2.4} and~\ref{eq:2.5}, we can calculate $E_\pi^\mathrm{cut}$ as $E_\pi^\mathrm{cut} =14\,\mathrm{GeV}$ and $350\,\mathrm{GeV}$, respectively. The corresponding energy of protons should be $E_p^\mathrm{GC}=100\,\mathrm{GeV}$ and $2.5\,\mathrm{TeV}$, with $E_p^\mathrm{cut}=20\,\mathrm{TeV}$ and $10\,\mathrm{PeV}$, respectively. On the other hand, various observations have shown that protons can be easily accelerated up to ${\sim}100\,\mathrm{TeV}$ by PWNe or SNRs in our galaxy. Therefore, there is no problem to apply the GC-model if $E_\pi^\mathrm{GC}\sim1\,\mathrm{GeV}$. However, the requirement of $E_p=10\,\mathrm{PeV}$ seems to exceed the usual maximum energy ranges of most Galactic sources (usually less than $100\,\mathrm{TeV}$). It is interesting to note that Galactic $\gamma$-rays significantly higher than $100\,\mathrm{TeV}$, and even up to $1.4\,\mathrm{PeV}$ were recently reported \cite{2021Natur.594...33C,2021NatAs...5..460T}, which implies that PeV accelerators (PeVatrons) should exist in our Galaxy. Furthermore, since $E_p=100\,\mathrm{TeV}$ is larger than $E_p^\mathrm{GC}=2.5\,\mathrm{TeV}$, we should still be able to observe a significant portion of the GC spectrum at this energy scale. Therefore, the applications of the GC-model for the 25 pulsars (in Table~\ref{tab:1}) are granted.

The above estimations relate to the acceleration mechanism of protons in pulsars. The $\gamma$-ray spectrum of a pulsar typically has a complex structure, where different energy bands have corresponding radiation models and possibly different particle acceleration mechanisms. In this paper, we only focus on the generation mechanism of the energy spectra. As for the proton acceleration process in the pulsar environment, like the conventional hadron model, the accelerated proton distribution is usually described by a power-law distribution. Therefore, the proton acceleration mechanism is a topic to be studied.

Now we try to make some speculations on the possible proton acceleration mechanisms in pulsars. As mentioned above, the existence of an acceleration mechanism in the Milky Way that accelerates cosmic rays (mainly protons) to TeV/PeV is an observed fact \cite{2021Sci...373..425L}. SNRs are the main sources of high-energy cosmic rays in our galaxy, where such accelerators should be present. For example, variant electromagnetic fields at large scales \cite{1972ARA&A..10..427R}, plasma turbulence acceleration of plasma \cite{1973ARA&A..11..363T}, shock wave based on the Fermi acceleration mechanism \cite{1990cup..book.....G} or even some unknown physical mechanisms could be candidates for the above-mentioned accelerators.

We imagine that the charged particles are thrown by the strong electromagnetic field of a fast-rotating neutron star and pulsate the PWN. They will form shock waves in the magnetosphere. Theoretically, the Fermi mechanism can accelerate protons to above 100 TeV in the interstellar magnetic field \cite{1990cup..book.....G}. Thus the produced VHE charged particles are ejected outward, in which electrons/positrons cross the magnetic dipole field lines and produce a large number of soft X-rays by synchrotron radiation.

At the same time, a part of particles also change their direction of motion in the magnetosphere due to collisions. These relativistic particles move along the curved magnetic lines in the dipole magnetic field of a neutron star. Because the magnetic field is so strong, any movement across the magnetic lines is practically impossible \cite{1998scichina..book.....Y}. These particles can only move along or inverse magnetic lines, which depends on the orientation of the particles. Due to the focusing effect of the magnetic field, they fly toward the magnetic poles along the direction of magnetic line gathering. The electrons/positrons release most of their energy through curvature radiation and soon become non-relativistic hot electrons. This is the SC model \cite{2019MNRAS.489.5494T}. The $\gamma$-ray spectrum of curvature radiation can be simulated by Equation~\ref{eq:3.1} after proper superposition.

On the other hand, protons lose very little energy due to their large mass. They accelerate along magnetic dipole-type lines and eventually hit the nucleus $A$ or neutron cluster $A^*$ of the pulsar dipole. If $E_p>20\,\mathrm{TeV}$, the $pA^*$ collisions may produce a BPL spectrum near 1 GeV (see Equation~\ref{eq:2.3}).

The proton density in the magnetosphere of neutron stars may be much lower than the electron density. Thus, the ``$\pi$ bump'' in the $pA$ collisions without the GC effect only provides a small contribution to the GeV $\gamma$-ray spectra \cite{2020ApJ...896...76A}. However, the highly efficient kinetic energy-photon conversion in the GC model can greatly increase the photon yield, which is sufficient to produce a significant observable gamma spectrum. For example, an electron with 10 GeV may emit at most 10 photons with 1 GeV, while a proton with 20 TeV can produce about $10^4$ such photons in the GC model.

Of course, not all GeV $\gamma$-ray spectra of pulsars are governed by the GC model. If the proton flow is too weak, or if the proton energy is below the GC-threshold, the curvature radiation of the leptons or other radiation mechanisms will dominate the GeV $\gamma$-ray spectrum, and the spectrum will then show a deviation from Equation~\ref{eq:2.3}. Thus, according to the shape of the resulting $\gamma$-ray spectrum, we can determine whether the GC model or the SC model dominates the GeV $\gamma$-ray spectrum of a pulsar.

It is well known that the quark-gluon distributions in a free nucleon are different from that in a bound nucleon, which is reflected in the shadowing--antishadowing effect (i.e., the EMC effect; for a review, see \cite{1994PhR...240..301A}) and the CGC effect. In this study, the baryon number dependence of the GC-effect is investigated by effectively considering the case of $A\to\infty$, which may open a new window to probe the quark-gluon distribution in the nucleon. It may also shed new light on the structure of compact stars.

Conventional models of the GeV $\gamma$-ray spectra of pulsars do not account for the structure of the nucleon. This paper proposes a hadronic model with the GC effect, which attempts to naturally explain the above observed BPL in terms of a lower GC-threshold in a large neutron cluster $A^*$. This example provides a new approach to exploring pulsars from the perspective of particle physics. We aim to apply this method to further pulsar phenomena.

In summary, because of the GC in nucleons, $\gamma$-ray spectra emitted in the hadronic collisions may present a typical broken power law, which can be used to explain some of the observed GeV $\gamma$-ray spectra from pulsars. These results indicate that the standard hadronic model but with the GC effect is a possible mechanism of pulsar GeV $\gamma$-rays except leptonic mechanisms. We also show that the nuclear $A$ dependence of the GC-effect when $A\to\infty$, which may open a new window for eavesdropping on the structure of compact stars on the subnuclear level.

\begin{acknowledgments}
We thank the anonymous referee for helpful comments and suggestions. This work is supported by the National Natural Science Foundation of China (Nos.\ 11851303, 12233002, 12041306, 12147103, and 12005192).
\end{acknowledgments}


\begin{thebibliography}{28}%
\makeatletter
\providecommand \@ifxundefined [1]{%
 \@ifx{#1\undefined}
}%
\providecommand \@ifnum [1]{%
 \ifnum #1\expandafter \@firstoftwo
 \else \expandafter \@secondoftwo
 \fi
}%
\providecommand \@ifx [1]{%
 \ifx #1\expandafter \@firstoftwo
 \else \expandafter \@secondoftwo
 \fi
}%
\providecommand \natexlab [1]{#1}%
\providecommand \enquote  [1]{``#1''}%
\providecommand \bibnamefont  [1]{#1}%
\providecommand \bibfnamefont [1]{#1}%
\providecommand \citenamefont [1]{#1}%
\providecommand \href@noop [0]{\@secondoftwo}%
\providecommand \href [0]{\begingroup \@sanitize@url \@href}%
\providecommand \@href[1]{\@@startlink{#1}\@@href}%
\providecommand \@@href[1]{\endgroup#1\@@endlink}%
\providecommand \@sanitize@url [0]{\catcode `\\12\catcode `\$12\catcode
  `\&12\catcode `\#12\catcode `\^12\catcode `\_12\catcode `\%12\relax}%
\providecommand \@@startlink[1]{}%
\providecommand \@@endlink[0]{}%
\providecommand \url  [0]{\begingroup\@sanitize@url \@url }%
\providecommand \@url [1]{\endgroup\@href {#1}{\urlprefix }}%
\providecommand \urlprefix  [0]{URL }%
\providecommand \Eprint [0]{\href }%
\providecommand \doibase [0]{https://doi.org/}%
\providecommand \selectlanguage [0]{\@gobble}%
\providecommand \bibinfo  [0]{\@secondoftwo}%
\providecommand \bibfield  [0]{\@secondoftwo}%
\providecommand \translation [1]{[#1]}%
\providecommand \BibitemOpen [0]{}%
\providecommand \bibitemStop [0]{}%
\providecommand \bibitemNoStop [0]{.\EOS\space}%
\providecommand \EOS [0]{\spacefactor3000\relax}%
\providecommand \BibitemShut  [1]{\csname bibitem#1\endcsname}%
\let\auto@bib@innerbib\@empty
\bibitem [{\citenamefont {{McLerran}}(2011)}]{2011PThPS.187...17M}%
  \BibitemOpen
  \bibfield  {author} {\bibinfo {author} {\bibfnamefont {L.}~\bibnamefont
  {{McLerran}}},\ }\bibfield  {title} {\bibinfo {title} {{\it The CGC and the
  Glasma: Two Lectures at the Yukawa Institute}},\ }\href
  {https://doi.org/10.1143/PTPS.187.17} {\bibfield  {journal} {\bibinfo
  {journal} {Progress of Theoretical Physics Supplement}\ }\textbf {\bibinfo
  {volume} {187}},\ \bibinfo {pages} {17} (\bibinfo {year} {2011})},\ \Eprint
  {https://arxiv.org/abs/1011.3204} {arXiv:1011.3204 [hep-ph]} \BibitemShut
  {NoStop}%
\bibitem [{\citenamefont {{Zhu}}(1999)}]{1999NuPhB.551..245W}%
  \BibitemOpen
  \bibfield  {author} {\bibinfo {author} {\bibfnamefont {W.}~\bibnamefont
  {{Zhu}}},\ }\bibfield  {title} {\bibinfo {title} {{\it A new approach to parton
  recombination in the QCD evolution equations}},\ }\href
  {https://doi.org/10.1016/S0550-3213(99)00237-0} {\bibfield  {journal}
  {\bibinfo  {journal} {NuPhB}\ }\textbf {\bibinfo {volume} {551}},\ \bibinfo
  {pages} {245} (\bibinfo {year} {1999})},\ \Eprint
  {https://arxiv.org/abs/hep-ph/9809391} {arXiv:hep-ph/9809391 [hep-ph]}
  \BibitemShut {NoStop}%
\bibitem [{\citenamefont {{Zhu}}\ \emph {et~al.}(2008)\citenamefont {{Zhu}},
  \citenamefont {{Shen}},\ and\ \citenamefont {{Ruan}}}]{2008ChPhL..25.3605Z}%
  \BibitemOpen
  \bibfield  {author} {\bibinfo {author} {\bibfnamefont {W.}~\bibnamefont
  {{Zhu}}}, \bibinfo {author} {\bibfnamefont {Z.-Q.}\ \bibnamefont {{Shen}}},\
  and\ \bibinfo {author} {\bibfnamefont {J.-H.}\ \bibnamefont {{Ruan}}},\
  }\bibfield  {title} {\bibinfo {title} {{\it Can a Chaotic Solution in the QCD
  Evolution Equation Restrain High-Energy Collider Physics?}},\ }\href
  {https://doi.org/10.1088/0256-307X/25/10/023} {\bibfield  {journal} {\bibinfo
   {journal} {ChPhL}\ }\textbf {\bibinfo {volume} {25}},\ \bibinfo {pages}
  {3605} (\bibinfo {year} {2008})},\ \Eprint {https://arxiv.org/abs/0809.0609}
  {arXiv:0809.0609 [hep-ph]} \BibitemShut {NoStop}%
\bibitem [{\citenamefont {{Zhu}}\ \emph {et~al.}(2016)\citenamefont {{Zhu}},
  \citenamefont {{Shen}},\ and\ \citenamefont {{Ruan}}}]{2016NuPhB.911....1Z}%
  \BibitemOpen
  \bibfield  {author} {\bibinfo {author} {\bibfnamefont {W.}~\bibnamefont
  {{Zhu}}}, \bibinfo {author} {\bibfnamefont {Z.}~\bibnamefont {{Shen}}},\ and\
  \bibinfo {author} {\bibfnamefont {J.}~\bibnamefont {{Ruan}}},\ }\bibfield
  {title} {\bibinfo {title} {{\it The chaotic effects in a nonlinear QCD evolution
  equation}},\ }\href {https://doi.org/10.1016/j.nuclphysb.2016.06.031}
  {\bibfield  {journal} {\bibinfo  {journal} {NuPhB}\ }\textbf {\bibinfo
  {volume} {911}},\ \bibinfo {pages} {1} (\bibinfo {year} {2016})},\ \Eprint
  {https://arxiv.org/abs/1603.04158} {arXiv:1603.04158 [hep-ph]} \BibitemShut
  {NoStop}%
\bibitem [{\citenamefont {{Zhu}}\ and\ \citenamefont
  {{Lan}}(2017)}]{2017NuPhB.916..647Z}%
  \BibitemOpen
  \bibfield  {author} {\bibinfo {author} {\bibfnamefont {W.}~\bibnamefont
  {{Zhu}}}\ and\ \bibinfo {author} {\bibfnamefont {J.}~\bibnamefont {{Lan}}},\
  }\bibfield  {title} {\bibinfo {title} {{\it The gluon condensation at high energy
  hadron collisions}},\ }\href
  {https://doi.org/10.1016/j.nuclphysb.2017.01.021} {\bibfield  {journal}
  {\bibinfo  {journal} {NuPhB}\ }\textbf {\bibinfo {volume} {916}},\ \bibinfo
  {pages} {647} (\bibinfo {year} {2017})},\ \Eprint
  {https://arxiv.org/abs/1702.02249} {arXiv:1702.02249 [hep-ph]} \BibitemShut
  {NoStop}%
\bibitem [{\citenamefont {{Zhu}}\ \emph {et~al.}(2022)\citenamefont {{Zhu}},
  \citenamefont {{Chen}}, \citenamefont {{Cui}},\ and\ \citenamefont
  {{Ruan}}}]{2022NuPhB.98415961Z}%
  \BibitemOpen
  \bibfield  {author} {\bibinfo {author} {\bibfnamefont {W.}~\bibnamefont
  {{Zhu}}}, \bibinfo {author} {\bibfnamefont {Q.}~\bibnamefont {{Chen}}},
  \bibinfo {author} {\bibfnamefont {Z.}~\bibnamefont {{Cui}}},\ and\ \bibinfo
  {author} {\bibfnamefont {J.}~\bibnamefont {{Ruan}}},\ }\bibfield  {title}
  {\bibinfo {title} {{\it The gluon condensation in hadron collisions}},\ }\href
  {https://doi.org/10.1016/j.nuclphysb.2022.115961} {\bibfield  {journal}
  {\bibinfo  {journal} {NuPhB}\ }\textbf {\bibinfo {volume} {984}},\ \bibinfo
  {eid} {115961} (\bibinfo {year} {2022})},\ \Eprint
  {https://arxiv.org/abs/2208.14219} {arXiv:2208.14219 [hep-ph]} \BibitemShut
  {NoStop}%
\bibitem [{\citenamefont {{Aharonian}}(2004)}]{2004vhec.book.....A}%
  \BibitemOpen
  \bibfield  {author} {\bibinfo {author} {\bibfnamefont {F.~A.}\ \bibnamefont
  {{Aharonian}}},\ }\href {https://doi.org/10.1142/4657} {\emph {\bibinfo
  {title} {{\it Very High Energy Cosmic Gamma Radiation: A Crucial Window on the
  Extreme Universe}}}}\ (\bibinfo  {publisher} {{World Scientific
  Publishing}},\ \bibinfo {address} {{Singapore}},\ \bibinfo {year}
  {2004})\BibitemShut {NoStop}%
\bibitem [{\citenamefont {{Zhu}}\ \emph {et~al.}(2018)\citenamefont {{Zhu}},
  \citenamefont {{Lan}},\ and\ \citenamefont {{Ruan}}}]{2018IJMPE..2750073Z}%
  \BibitemOpen
  \bibfield  {author} {\bibinfo {author} {\bibfnamefont {W.}~\bibnamefont
  {{Zhu}}}, \bibinfo {author} {\bibfnamefont {J.}~\bibnamefont {{Lan}}},\ and\
  \bibinfo {author} {\bibfnamefont {J.}~\bibnamefont {{Ruan}}},\ }\bibfield
  {title} {\bibinfo {title} {{\it The gluon condensation in high energy cosmic
  rays}},\ }\href {https://doi.org/10.1142/S0218301318500738} {\bibfield
  {journal} {\bibinfo  {journal} {IJMPE}\ }\textbf {\bibinfo {volume} {27}},\
  \bibinfo {eid} {1850073} (\bibinfo {year} {2018})},\ \Eprint
  {https://arxiv.org/abs/1709.03897} {arXiv:1709.03897 [hep-ph]} \BibitemShut
  {NoStop}%
\bibitem [{\citenamefont {{Feng}}\ \emph {et~al.}(2018)\citenamefont {{Feng}},
  \citenamefont {{Ruan}}, \citenamefont {{Wang}},\ and\ \citenamefont
  {{Zhu}}}]{2018ApJ...868....2F}%
  \BibitemOpen
  \bibfield  {author} {\bibinfo {author} {\bibfnamefont {L.}~\bibnamefont
  {{Feng}}}, \bibinfo {author} {\bibfnamefont {J.}~\bibnamefont {{Ruan}}},
  \bibinfo {author} {\bibfnamefont {F.}~\bibnamefont {{Wang}}},\ and\ \bibinfo
  {author} {\bibfnamefont {W.}~\bibnamefont {{Zhu}}},\ }\bibfield  {title}
  {\bibinfo {title} {{\it Looking for the Gluon Condensation Signature in Protons
  Using the Earth-limb Gamma-Ray Spectra}},\ }\href
  {https://doi.org/10.3847/1538-4357/aae781} {\bibfield  {journal} {\bibinfo
  {journal} {ApJ}\ }\textbf {\bibinfo {volume} {868}},\ \bibinfo {eid} {2}
  (\bibinfo {year} {2018})},\ \Eprint {https://arxiv.org/abs/1805.10618}
  {arXiv:1805.10618 [hep-ph]} \BibitemShut {NoStop}%
\bibitem [{\citenamefont {{Zhu}}\ \emph {et~al.}(2020)\citenamefont {{Zhu}},
  \citenamefont {{Liu}}, \citenamefont {{Ruan}},\ and\ \citenamefont
  {{Wang}}}]{2020ApJ...889..127Z}%
  \BibitemOpen
  \bibfield  {author} {\bibinfo {author} {\bibfnamefont {W.}~\bibnamefont
  {{Zhu}}}, \bibinfo {author} {\bibfnamefont {P.}~\bibnamefont {{Liu}}},
  \bibinfo {author} {\bibfnamefont {J.}~\bibnamefont {{Ruan}}},\ and\ \bibinfo
  {author} {\bibfnamefont {F.}~\bibnamefont {{Wang}}},\ }\bibfield  {title}
  {\bibinfo {title} {{\it Possible Evidence for the Gluon Condensation Effect in
  Cosmic Positron and Gamma-Ray Spectra}},\ }\href
  {https://doi.org/10.3847/1538-4357/ab6214} {\bibfield  {journal} {\bibinfo
  {journal} {ApJ}\ }\textbf {\bibinfo {volume} {889}},\ \bibinfo {eid} {127}
  (\bibinfo {year} {2020})},\ \Eprint {https://arxiv.org/abs/1912.12842}
  {arXiv:1912.12842 [astro-ph.HE]} \BibitemShut {NoStop}%
\bibitem [{\citenamefont {{Zhu}}\ \emph {et~al.}(2021)\citenamefont {{Zhu}},
  \citenamefont {{Zheng}}, \citenamefont {{Liu}}, \citenamefont {{Wan}},
  \citenamefont {{Ruan}},\ and\ \citenamefont {{Wang}}}]{2021JCAP...01..038Z}%
  \BibitemOpen
  \bibfield  {author} {\bibinfo {author} {\bibfnamefont {W.}~\bibnamefont
  {{Zhu}}}, \bibinfo {author} {\bibfnamefont {Z.}~\bibnamefont {{Zheng}}},
  \bibinfo {author} {\bibfnamefont {P.}~\bibnamefont {{Liu}}}, \bibinfo
  {author} {\bibfnamefont {L.}~\bibnamefont {{Wan}}}, \bibinfo {author}
  {\bibfnamefont {J.}~\bibnamefont {{Ruan}}},\ and\ \bibinfo {author}
  {\bibfnamefont {F.}~\bibnamefont {{Wang}}},\ }\bibfield  {title} {\bibinfo
  {title} {{\it Looking for the possible gluon condensation signature in sub-TeV
  gamma-ray spectra: from active galactic nuclei to gamma ray bursts}},\ }\href
  {https://doi.org/10.1088/1475-7516/2021/01/038} {\bibfield  {journal}
  {\bibinfo  {journal} {JCAP}\ }\textbf {\bibinfo {volume} {2021}}\bibfield
  {number} {\bibinfo  {number} { (1)},\ \bibinfo {eid} {038}},\ }\Eprint
  {https://arxiv.org/abs/2009.01984} {arXiv:2009.01984 [astro-ph.HE]}
  \BibitemShut {NoStop}%
\bibitem [{\citenamefont {{Ruan}}\ \emph {et~al.}(2021)\citenamefont {{Ruan}},
  \citenamefont {{Zheng}},\ and\ \citenamefont {{Zhu}}}]{2021JCAP...08..065R}%
  \BibitemOpen
  \bibfield  {author} {\bibinfo {author} {\bibfnamefont {J.}~\bibnamefont
  {{Ruan}}}, \bibinfo {author} {\bibfnamefont {Z.}~\bibnamefont {{Zheng}}},\
  and\ \bibinfo {author} {\bibfnamefont {W.}~\bibnamefont {{Zhu}}},\ }\bibfield
   {title} {\bibinfo {title} {{\it Exploring the possible gluon condensation
  signature in gamma-ray emission from pulsars}},\ }\href
  {https://doi.org/10.1088/1475-7516/2021/08/065} {\bibfield  {journal}
  {\bibinfo  {journal} {JCAP}\ }\textbf {\bibinfo {volume} {2021}}\bibfield
  {number} {\bibinfo  {number} { (8)},\ \bibinfo {eid} {065}},\ }\Eprint
  {https://arxiv.org/abs/2012.08767} {arXiv:2012.08767 [astro-ph.HE]}
  \BibitemShut {NoStop}%
\bibitem [{\citenamefont {{Abdo}}\ \emph {et~al.}(2013)\citenamefont {{Abdo}},
  \citenamefont {{Ajello}}, \citenamefont {{Allafort}}, \citenamefont
  {{Baldini}}, \citenamefont {{Ballet}}, \citenamefont {{Barbiellini}},
  \citenamefont {{Baring}}, \citenamefont {{Bastieri}}, \citenamefont
  {{Belfiore}}, \citenamefont {{Bellazzini}}, \citenamefont {{Bhattacharyya}},
  \citenamefont {{Bissaldi}}, \citenamefont {{Bloom}}, \citenamefont
  {{Bonamente}}, \citenamefont {{Bottacini}}, \citenamefont {{Brandt}},
  \citenamefont {{Bregeon}}, \citenamefont {{Brigida}}, \citenamefont
  {{Bruel}}, \citenamefont {{Buehler}}, \citenamefont {{Burgay}}, \citenamefont
  {{Burnett}}, \citenamefont {{Busetto}}, \citenamefont {{Buson}},
  \citenamefont {{Caliandro}}, \citenamefont {{Cameron}}, \citenamefont
  {{Camilo}}, \citenamefont {{Caraveo}}, \citenamefont {{Casandjian}},
  \citenamefont {{Cecchi}}, \citenamefont {{{\c{C}}elik}}, \citenamefont
  {{Charles}}, \citenamefont {{Chaty}}, \citenamefont {{Chaves}}, \citenamefont
  {{Chekhtman}}, \citenamefont {{Chen}}, \citenamefont {{Chiang}},
  \citenamefont {{Chiaro}}, \citenamefont {{Ciprini}}, \citenamefont {{Claus}},
  \citenamefont {{Cognard}}, \citenamefont {{Cohen-Tanugi}}, \citenamefont
  {{Cominsky}}, \citenamefont {{Conrad}}, \citenamefont {{Cutini}},
  \citenamefont {{D'Ammando}}, \citenamefont {{de Angelis}}, \citenamefont
  {{DeCesar}}, \citenamefont {{De Luca}}, \citenamefont {{den Hartog}},
  \citenamefont {{de Palma}}, \citenamefont {{Dermer}}, \citenamefont
  {{Desvignes}}, \citenamefont {{Digel}}, \citenamefont {{Di Venere}},
  \citenamefont {{Drell}}, \citenamefont {{Drlica-Wagner}}, \citenamefont
  {{Dubois}}, \citenamefont {{Dumora}}, \citenamefont {{Espinoza}},
  \citenamefont {{Falletti}}, \citenamefont {{Favuzzi}}, \citenamefont
  {{Ferrara}}, \citenamefont {{Focke}}, \citenamefont {{Franckowiak}},
  \citenamefont {{Freire}}, \citenamefont {{Funk}}, \citenamefont {{Fusco}},
  \citenamefont {{Gargano}}, \citenamefont {{Gasparrini}}, \citenamefont
  {{Germani}}, \citenamefont {{Giglietto}}, \citenamefont {{Giommi}},
  \citenamefont {{Giordano}}, \citenamefont {{Giroletti}}, \citenamefont
  {{Glanzman}}, \citenamefont {{Godfrey}}, \citenamefont {{Gotthelf}},
  \citenamefont {{Grenier}}, \citenamefont {{Grondin}}, \citenamefont
  {{Grove}}, \citenamefont {{Guillemot}}, \citenamefont {{Guiriec}},
  \citenamefont {{Hadasch}}, \citenamefont {{Hanabata}}, \citenamefont
  {{Harding}}, \citenamefont {{Hayashida}}, \citenamefont {{Hays}},
  \citenamefont {{Hessels}}, \citenamefont {{Hewitt}}, \citenamefont {{Hill}},
  \citenamefont {{Horan}}, \citenamefont {{Hou}}, \citenamefont {{Hughes}},
  \citenamefont {{Jackson}}, \citenamefont {{Janssen}}, \citenamefont
  {{Jogler}}, \citenamefont {{J{\'o}hannesson}}, \citenamefont {{Johnson}},
  \citenamefont {{Johnson}}, \citenamefont {{Johnson}}, \citenamefont
  {{Johnson}}, \citenamefont {{Johnston}}, \citenamefont {{Kamae}},
  \citenamefont {{Kataoka}}, \citenamefont {{Keith}}, \citenamefont {{Kerr}},
  \citenamefont {{Kn{\"o}dlseder}}, \citenamefont {{Kramer}}, \citenamefont
  {{Kuss}}, \citenamefont {{Lande}}, \citenamefont {{Larsson}}, \citenamefont
  {{Latronico}}, \citenamefont {{Lemoine-Goumard}}, \citenamefont {{Longo}},
  \citenamefont {{Loparco}}, \citenamefont {{Lovellette}}, \citenamefont
  {{Lubrano}}, \citenamefont {{Lyne}}, \citenamefont {{Manchester}},
  \citenamefont {{Marelli}}, \citenamefont {{Massaro}}, \citenamefont
  {{Mayer}}, \citenamefont {{Mazziotta}}, \citenamefont {{McEnery}},
  \citenamefont {{McLaughlin}}, \citenamefont {{Mehault}}, \citenamefont
  {{Michelson}}, \citenamefont {{Mignani}}, \citenamefont {{Mitthumsiri}},
  \citenamefont {{Mizuno}}, \citenamefont {{Moiseev}}, \citenamefont
  {{Monzani}}, \citenamefont {{Morselli}}, \citenamefont {{Moskalenko}},
  \citenamefont {{Murgia}}, \citenamefont {{Nakamori}}, \citenamefont
  {{Nemmen}}, \citenamefont {{Nuss}}, \citenamefont {{Ohno}}, \citenamefont
  {{Ohsugi}}, \citenamefont {{Orienti}}, \citenamefont {{Orlando}},
  \citenamefont {{Ormes}}, \citenamefont {{Paneque}}, \citenamefont
  {{Panetta}}, \citenamefont {{Parent}}, \citenamefont {{Perkins}},
  \citenamefont {{Pesce-Rollins}}, \citenamefont {{Pierbattista}},
  \citenamefont {{Piron}}, \citenamefont {{Pivato}}, \citenamefont {{Pletsch}},
  \citenamefont {{Porter}}, \citenamefont {{Possenti}}, \citenamefont
  {{Rain{\`o}}}, \citenamefont {{Rando}}, \citenamefont {{Ransom}},
  \citenamefont {{Ray}}, \citenamefont {{Razzano}}, \citenamefont {{Rea}},
  \citenamefont {{Reimer}}, \citenamefont {{Reimer}}, \citenamefont
  {{Renault}}, \citenamefont {{Reposeur}}, \citenamefont {{Ritz}},
  \citenamefont {{Romani}}, \citenamefont {{Roth}}, \citenamefont {{Rousseau}},
  \citenamefont {{Roy}}, \citenamefont {{Ruan}}, \citenamefont {{Sartori}},
  \citenamefont {{Saz Parkinson}}, \citenamefont {{Scargle}}, \citenamefont
  {{Schulz}}, \citenamefont {{Sgr{\`o}}}, \citenamefont {{Shannon}},
  \citenamefont {{Siskind}}, \citenamefont {{Smith}}, \citenamefont
  {{Spandre}}, \citenamefont {{Spinelli}}, \citenamefont {{Stappers}},
  \citenamefont {{Strong}}, \citenamefont {{Suson}}, \citenamefont
  {{Takahashi}}, \citenamefont {{Thayer}}, \citenamefont {{Thayer}},
  \citenamefont {{Theureau}}, \citenamefont {{Thompson}}, \citenamefont
  {{Thorsett}}, \citenamefont {{Tibaldo}}, \citenamefont {{Tibolla}},
  \citenamefont {{Tinivella}}, \citenamefont {{Torres}}, \citenamefont
  {{Tosti}}, \citenamefont {{Troja}}, \citenamefont {{Uchiyama}}, \citenamefont
  {{Usher}}, \citenamefont {{Vandenbroucke}}, \citenamefont {{Vasileiou}},
  \citenamefont {{Venter}}, \citenamefont {{Vianello}}, \citenamefont
  {{Vitale}}, \citenamefont {{Wang}}, \citenamefont {{Weltevrede}},
  \citenamefont {{Winer}}, \citenamefont {{Wolff}}, \citenamefont {{Wood}},
  \citenamefont {{Wood}}, \citenamefont {{Wood}},\ and\ \citenamefont
  {{Yang}}}]{2013ApJS..208...17A}%
  \BibitemOpen
  \bibfield  {author} {\bibinfo {author} {\bibfnamefont {A.~A.}\ \bibnamefont
  {{Abdo}}} \bibinfo {author} \emph{et al.},\ }\bibfield  {title} {\bibinfo
  {title} {{\it The Second Fermi Large Area Telescope Catalog of Gamma-Ray
  Pulsars}},\ }\href {https://doi.org/10.1088/0067-0049/208/2/17} {\bibfield
  {journal} {\bibinfo  {journal} {ApJS}\ }\textbf {\bibinfo {volume} {208}},\
  \bibinfo {eid} {17} (\bibinfo {year} {2013})},\ \Eprint
  {https://arxiv.org/abs/1305.4385} {arXiv:1305.4385 [astro-ph.HE]}
  \BibitemShut {NoStop}%
\bibitem [{\citenamefont {{Bykov}}\ \emph {et~al.}(2019)\citenamefont
  {{Bykov}}, \citenamefont {{Petrov}}, \citenamefont {{Krassilchtchikov}},
  \citenamefont {{Levenfish}}, \citenamefont {{Osipov}},\ and\ \citenamefont
  {{Pavlov}}}]{2019ApJ...876L...8B}%
  \BibitemOpen
  \bibfield  {author} {\bibinfo {author} {\bibfnamefont {A.~M.}\ \bibnamefont
  {{Bykov}}}, \bibinfo {author} {\bibfnamefont {A.~E.}\ \bibnamefont
  {{Petrov}}}, \bibinfo {author} {\bibfnamefont {A.~M.}\ \bibnamefont
  {{Krassilchtchikov}}}, \bibinfo {author} {\bibfnamefont {K.~P.}\ \bibnamefont
  {{Levenfish}}}, \bibinfo {author} {\bibfnamefont {S.~M.}\ \bibnamefont
  {{Osipov}}},\ and\ \bibinfo {author} {\bibfnamefont {G.~G.}\ \bibnamefont
  {{Pavlov}}},\ }\bibfield  {title} {\bibinfo {title} {{\it GeV-TeV Cosmic-Ray
  Leptons in the Solar System from the Bow Shock Wind Nebula of the Nearest
  Millisecond Pulsar J0437-4715}},\ }\href
  {https://doi.org/10.3847/2041-8213/ab1922} {\bibfield  {journal} {\bibinfo
  {journal} {ApJL}\ }\textbf {\bibinfo {volume} {876}},\ \bibinfo {eid} {L8}
  (\bibinfo {year} {2019})},\ \Eprint {https://arxiv.org/abs/1904.09430}
  {arXiv:1904.09430 [astro-ph.HE]} \BibitemShut {NoStop}%
\bibitem [{\citenamefont {{Torres}}\ \emph {et~al.}(2019)\citenamefont
  {{Torres}}, \citenamefont {{Vigan{\`o}}}, \citenamefont {{Coti Zelati}},\
  and\ \citenamefont {{Li}}}]{2019MNRAS.489.5494T}%
  \BibitemOpen
  \bibfield  {author} {\bibinfo {author} {\bibfnamefont {D.~F.}\ \bibnamefont
  {{Torres}}}, \bibinfo {author} {\bibfnamefont {D.}~\bibnamefont
  {{Vigan{\`o}}}}, \bibinfo {author} {\bibfnamefont {F.}~\bibnamefont {{Coti
  Zelati}}},\ and\ \bibinfo {author} {\bibfnamefont {J.}~\bibnamefont {{Li}}},\
  }\bibfield  {title} {\bibinfo {title} {{\it Synchrocurvature modelling of the
  multifrequency non-thermal emission of pulsars}},\ }\href
  {https://doi.org/10.1093/mnras/stz2403} {\bibfield  {journal} {\bibinfo
  {journal} {MNRAS}\ }\textbf {\bibinfo {volume} {489}},\ \bibinfo {pages}
  {5494} (\bibinfo {year} {2019})},\ \Eprint {https://arxiv.org/abs/1908.11574}
  {arXiv:1908.11574 [astro-ph.HE]} \BibitemShut {NoStop}%
\bibitem [{\citenamefont {{Mei}}\ \emph {et~al.}(2022)\citenamefont {{Mei}},
  \citenamefont {{Banerjee}}, \citenamefont {{Oganesyan}}, \citenamefont
  {{Sharan Salafia}}, \citenamefont {{Giarratana}}, \citenamefont
  {{Branchesi}}, \citenamefont {{D'Avanzo}}, \citenamefont {{Campana}},
  \citenamefont {{Ghirlanda}}, \citenamefont {{Ronchini}}, \citenamefont
  {{Shukla}},\ and\ \citenamefont {{Tiwari}}}]{2022arXiv220508566M}%
  \BibitemOpen
  \bibfield  {author} {\bibinfo {author} {\bibfnamefont {A.}~\bibnamefont
  {{Mei}}}, \bibinfo {author} {\bibfnamefont {B.}~\bibnamefont {{Banerjee}}},
  \bibinfo {author} {\bibfnamefont {G.}~\bibnamefont {{Oganesyan}}}, \bibinfo
  {author} {\bibfnamefont {O.}~\bibnamefont {{Sharan Salafia}}}, \bibinfo
  {author} {\bibfnamefont {S.}~\bibnamefont {{Giarratana}}}, \bibinfo {author}
  {\bibfnamefont {M.}~\bibnamefont {{Branchesi}}}, \bibinfo {author}
  {\bibfnamefont {P.}~\bibnamefont {{D'Avanzo}}}, \bibinfo {author}
  {\bibfnamefont {S.}~\bibnamefont {{Campana}}}, \bibinfo {author}
  {\bibfnamefont {G.}~\bibnamefont {{Ghirlanda}}}, \bibinfo {author}
  {\bibfnamefont {S.}~\bibnamefont {{Ronchini}}}, \bibinfo {author}
  {\bibfnamefont {A.}~\bibnamefont {{Shukla}}},\ and\ \bibinfo {author}
  {\bibfnamefont {P.}~\bibnamefont {{Tiwari}}},\ }\bibfield  {title} {\bibinfo
  {title} {{\it Gigaelectronvolt emission from a compact binary merger}},\ }\href
  {https://doi.org/10.1038/s41586-022-05404-7} {\bibfield  {journal} {\bibinfo
  {journal} {Nature}\ }\textbf {\bibinfo {volume} {612}},\ \bibinfo {eid} {236}
  (\bibinfo {year} {2022})},\ \Eprint {https://arxiv.org/abs/2205.08566}
  {arXiv:2205.08566 [astro-ph.HE]} \BibitemShut {NoStop}%
\bibitem [{\citenamefont {{Aharonian}}\ \emph {et~al.}(2006)\citenamefont
  {{Aharonian}}, \citenamefont {{Akhperjanian}}, \citenamefont {{Bazer-Bachi}},
  \citenamefont {{Beilicke}}, \citenamefont {{Benbow}}, \citenamefont
  {{Berge}}, \citenamefont {{Bernl{\"o}hr}}, \citenamefont {{Boisson}},
  \citenamefont {{Bolz}}, \citenamefont {{Borrel}}, \citenamefont {{Braun}},
  \citenamefont {{Brown}}, \citenamefont {{B{\"u}hler}}, \citenamefont
  {{B{\"u}sching}}, \citenamefont {{Carrigan}}, \citenamefont {{Chadwick}},
  \citenamefont {{Chounet}}, \citenamefont {{Cornils}}, \citenamefont
  {{Costamante}}, \citenamefont {{Degrange}}, \citenamefont {{Dickinson}},
  \citenamefont {{Djannati-Ata{\"\i}}}, \citenamefont {{O'C. Drury}},
  \citenamefont {{Dubus}}, \citenamefont {{Egberts}}, \citenamefont
  {{Emmanoulopoulos}}, \citenamefont {{Espigat}}, \citenamefont {{Feinstein}},
  \citenamefont {{Ferrero}}, \citenamefont {{Fiasson}}, \citenamefont
  {{Fontaine}}, \citenamefont {{Funk}}, \citenamefont {{Funk}}, \citenamefont
  {{F{\"u}{\ss}ling}}, \citenamefont {{Gallant}}, \citenamefont {{Giebels}},
  \citenamefont {{Glicenstein}}, \citenamefont {{Goret}}, \citenamefont
  {{Hadjichristidis}}, \citenamefont {{Hauser}}, \citenamefont {{Hauser}},
  \citenamefont {{Heinzelmann}}, \citenamefont {{Henri}}, \citenamefont
  {{Hermann}}, \citenamefont {{Hinton}}, \citenamefont {{Hoffmann}},
  \citenamefont {{Hofmann}}, \citenamefont {{Holleran}}, \citenamefont
  {{Horns}}, \citenamefont {{Jacholkowska}}, \citenamefont {{de Jager}},
  \citenamefont {{Kendziorra}}, \citenamefont {{Kh{\'e}lifi}}, \citenamefont
  {{Komin}}, \citenamefont {{Konopelko}}, \citenamefont {{Kosack}},
  \citenamefont {{Latham}}, \citenamefont {{Le Gallou}}, \citenamefont
  {{Lemi{\`e}re}}, \citenamefont {{Lemoine-Goumard}}, \citenamefont {{Lohse}},
  \citenamefont {{Martin}}, \citenamefont {{Martineau-Huynh}}, \citenamefont
  {{Marcowith}}, \citenamefont {{Masterson}}, \citenamefont {{Maurin}},
  \citenamefont {{McComb}}, \citenamefont {{de Naurois}}, \citenamefont
  {{Nedbal}}, \citenamefont {{Nolan}}, \citenamefont {{Noutsos}}, \citenamefont
  {{Orford}}, \citenamefont {{Osborne}}, \citenamefont {{Ouchrif}},
  \citenamefont {{Panter}}, \citenamefont {{Pelletier}}, \citenamefont
  {{Pita}}, \citenamefont {{P{\"u}hlhofer}}, \citenamefont {{Punch}},
  \citenamefont {{Raubenheimer}}, \citenamefont {{Raue}}, \citenamefont
  {{Rayner}}, \citenamefont {{Reimer}}, \citenamefont {{Reimer}}, \citenamefont
  {{Ripken}}, \citenamefont {{Rob}}, \citenamefont {{Rolland}}, \citenamefont
  {{Rowell}}, \citenamefont {{Sahakian}}, \citenamefont {{Santangelo}},
  \citenamefont {{Saug{\'e}}}, \citenamefont {{Schlenker}}, \citenamefont
  {{Schlickeiser}}, \citenamefont {{Schr{\"o}der}}, \citenamefont {{Schwanke}},
  \citenamefont {{Schwarzburg}}, \citenamefont {{Shalchi}}, \citenamefont
  {{Sol}}, \citenamefont {{Spangler}}, \citenamefont {{Spanier}}, \citenamefont
  {{Steenkamp}}, \citenamefont {{Stegmann}}, \citenamefont {{Superina}},
  \citenamefont {{Tavernet}}, \citenamefont {{Terrier}}, \citenamefont
  {{Th{\'e}oret}}, \citenamefont {{Tluczykont}}, \citenamefont {{van Eldik}},
  \citenamefont {{Vasileiadis}}, \citenamefont {{Venter}}, \citenamefont
  {{Vincent}}, \citenamefont {{V{\"o}lk}}, \citenamefont {{Wagner}},\ and\
  \citenamefont {{Ward}}}]{2006A&A...456..245A}%
  \BibitemOpen
  \bibfield  {author} {\bibinfo {author} {\bibfnamefont {F.}~\bibnamefont
  {{Aharonian}}} \bibinfo {author} \emph{et al.},\ }\bibfield  {title} {\bibinfo {title}
  {{\it Discovery of the two ``wings'' of the Kookaburra complex in VHE
  {\ensuremath{\gamma}}-rays with HESS}},\ }\href
  {https://doi.org/10.1051/0004-6361:20065511} {\bibfield  {journal} {\bibinfo
  {journal} {A\&A}\ }\textbf {\bibinfo {volume} {456}},\ \bibinfo {pages} {245}
  (\bibinfo {year} {2006})},\ \Eprint {https://arxiv.org/abs/astro-ph/0606311}
  {arXiv:astro-ph/0606311 [astro-ph]} \BibitemShut {NoStop}%
\bibitem [{\citenamefont {{Wang}}\ \emph {et~al.}(2020)\citenamefont {{Wang}},
  \citenamefont {{Lin}}, \citenamefont {{Dai}}, \citenamefont {{Takata}},
  \citenamefont {{Li}}, \citenamefont {{Hu}},\ and\ \citenamefont
  {{Hou}}}]{2020ApJ...902...96W}%
  \BibitemOpen
  \bibfield  {author} {\bibinfo {author} {\bibfnamefont {H.~H.}\ \bibnamefont
  {{Wang}}}, \bibinfo {author} {\bibfnamefont {L.~C.~C.}\ \bibnamefont
  {{Lin}}}, \bibinfo {author} {\bibfnamefont {S.}~\bibnamefont {{Dai}}},
  \bibinfo {author} {\bibfnamefont {J.}~\bibnamefont {{Takata}}}, \bibinfo
  {author} {\bibfnamefont {K.~L.}\ \bibnamefont {{Li}}}, \bibinfo {author}
  {\bibfnamefont {C.~P.}\ \bibnamefont {{Hu}}},\ and\ \bibinfo {author}
  {\bibfnamefont {X.}~\bibnamefont {{Hou}}},\ }\bibfield  {title} {\bibinfo
  {title} {{\it A Multiwavelength Study of PSR J1119-6127 after 2016 Outburst}},\
  }\href {https://doi.org/10.3847/1538-4357/abb3c4} {\bibfield  {journal}
  {\bibinfo  {journal} {ApJ}\ }\textbf {\bibinfo {volume} {902}},\ \bibinfo
  {eid} {96} (\bibinfo {year} {2020})},\ \Eprint
  {https://arxiv.org/abs/2008.12585} {arXiv:2008.12585 [astro-ph.HE]}
  \BibitemShut {NoStop}%
\bibitem [{\citenamefont {{Li}}\ \emph {et~al.}(2018)\citenamefont {{Li}},
  \citenamefont {{Torres}}, \citenamefont {{Lin}}, \citenamefont {{Grondin}},
  \citenamefont {{Kerr}}, \citenamefont {{Lemoine-Goumard}},\ and\
  \citenamefont {{de O{\~n}a Wilhelmi}}}]{2018ApJ...858...84L}%
  \BibitemOpen
  \bibfield  {author} {\bibinfo {author} {\bibfnamefont {J.}~\bibnamefont
  {{Li}}}, \bibinfo {author} {\bibfnamefont {D.~F.}\ \bibnamefont {{Torres}}},
  \bibinfo {author} {\bibfnamefont {T.~T.}\ \bibnamefont {{Lin}}}, \bibinfo
  {author} {\bibfnamefont {M.-H.}\ \bibnamefont {{Grondin}}}, \bibinfo {author}
  {\bibfnamefont {M.}~\bibnamefont {{Kerr}}}, \bibinfo {author} {\bibfnamefont
  {M.}~\bibnamefont {{Lemoine-Goumard}}},\ and\ \bibinfo {author}
  {\bibfnamefont {E.}~\bibnamefont {{de O{\~n}a Wilhelmi}}},\ }\bibfield
  {title} {\bibinfo {title} {{\it Observing and Modeling the Gamma-Ray Emission
  from Pulsar/Pulsar Wind Nebula Complex PSR J0205+6449/3C 58}},\ }\href
  {https://doi.org/10.3847/1538-4357/aabac9} {\bibfield  {journal} {\bibinfo
  {journal} {ApJ}\ }\textbf {\bibinfo {volume} {858}},\ \bibinfo {eid} {84}
  (\bibinfo {year} {2018})},\ \Eprint {https://arxiv.org/abs/1803.10863}
  {arXiv:1803.10863 [astro-ph.HE]} \BibitemShut {NoStop}%
\bibitem [{\citenamefont {{Cao}}\ \emph {et~al.}(2021)\citenamefont {{Cao}},
  \citenamefont {{Aharonian}}, \citenamefont {{An}}, \citenamefont {{Axikegu}},
  \citenamefont {{Bai}}, \citenamefont {{Bao}}, \citenamefont {{Bastieri}},
  \citenamefont {{Bi}}, \citenamefont {{Bi}}, \citenamefont {{Cai}},
  \citenamefont {{Cai}}, \citenamefont {{Cao}}, \citenamefont {{Chang}},
  \citenamefont {{Chang}}, \citenamefont {{Chang}}, \citenamefont {{Chen}},
  \citenamefont {{Chen}}, \citenamefont {{Chen}}, \citenamefont {{Chen}},
  \citenamefont {{Chen}}, \citenamefont {{Chen}}, \citenamefont {{Chen}},
  \citenamefont {{Chen}}, \citenamefont {{Chen}}, \citenamefont {{Chen}},
  \citenamefont {{Chen}}, \citenamefont {{Chen}}, \citenamefont {{Chen}},
  \citenamefont {{Cheng}}, \citenamefont {{Cheng}}, \citenamefont {{Cui}},
  \citenamefont {{Cui}}, \citenamefont {{Cui}}, \citenamefont {{Dai}},
  \citenamefont {{Dai}}, \citenamefont {{Dai}}, \citenamefont {{Danzengluobu}},
  \citenamefont {{della Volpe}}, \citenamefont {{D'Ettorre Piazzoli}},
  \citenamefont {{Dong}}, \citenamefont {{Fan}}, \citenamefont {{Fan}},
  \citenamefont {{Fan}}, \citenamefont {{Fang}}, \citenamefont {{Fang}},
  \citenamefont {{Feng}}, \citenamefont {{Feng}}, \citenamefont {{Feng}},
  \citenamefont {{Feng}}, \citenamefont {{Gao}}, \citenamefont {{Gao}},
  \citenamefont {{Gao}}, \citenamefont {{Gao}}, \citenamefont {{Ge}},
  \citenamefont {{Geng}}, \citenamefont {{Gong}}, \citenamefont {{Gou}},
  \citenamefont {{Gu}}, \citenamefont {{Guo}}, \citenamefont {{Guo}},
  \citenamefont {{Guo}}, \citenamefont {{Guo}}, \citenamefont {{Han}},
  \citenamefont {{He}}, \citenamefont {{He}}, \citenamefont {{He}},
  \citenamefont {{He}}, \citenamefont {{He}}, \citenamefont {{He}},
  \citenamefont {{Heller}}, \citenamefont {{Hor}}, \citenamefont {{Hou}},
  \citenamefont {{Hou}}, \citenamefont {{Hu}}, \citenamefont {{Hu}},
  \citenamefont {{Hu}}, \citenamefont {{Hu}}, \citenamefont {{Huang}},
  \citenamefont {{Huang}}, \citenamefont {{Huang}}, \citenamefont {{Huang}},
  \citenamefont {{Huang}}, \citenamefont {{Ji}}, \citenamefont {{Ji}},
  \citenamefont {{Jia}}, \citenamefont {{Jiang}}, \citenamefont {{Jiang}},
  \citenamefont {{Jin}}, \citenamefont {{Kuleshov}}, \citenamefont
  {{Levochkin}}, \citenamefont {{Li}}, \citenamefont {{Li}}, \citenamefont
  {{Li}}, \citenamefont {{Li}}, \citenamefont {{Li}}, \citenamefont {{Li}},
  \citenamefont {{Li}}, \citenamefont {{Li}}, \citenamefont {{Li}},
  \citenamefont {{Li}}, \citenamefont {{Li}}, \citenamefont {{Li}},
  \citenamefont {{Li}}, \citenamefont {{Li}}, \citenamefont {{Li}},
  \citenamefont {{Li}}, \citenamefont {{Li}}, \citenamefont {{Liang}},
  \citenamefont {{Liang}}, \citenamefont {{Lin}}, \citenamefont {{Liu}},
  \citenamefont {{Liu}}, \citenamefont {{Liu}}, \citenamefont {{Liu}},
  \citenamefont {{Liu}}, \citenamefont {{Liu}}, \citenamefont {{Liu}},
  \citenamefont {{Liu}}, \citenamefont {{Liu}}, \citenamefont {{Liu}},
  \citenamefont {{Liu}}, \citenamefont {{Liu}}, \citenamefont {{Liu}},
  \citenamefont {{Liu}}, \citenamefont {{Liu}}, \citenamefont {{Long}},
  \citenamefont {{Lu}}, \citenamefont {{Lv}}, \citenamefont {{Ma}},
  \citenamefont {{Ma}}, \citenamefont {{Ma}}, \citenamefont {{Mao}},
  \citenamefont {{Masood}}, \citenamefont {{Mitthumsiri}}, \citenamefont
  {{Montaruli}}, \citenamefont {{Nan}}, \citenamefont {{Pang}}, \citenamefont
  {{Pattarakijwanich}}, \citenamefont {{Pei}}, \citenamefont {{Qi}},
  \citenamefont {{Ruffolo}}, \citenamefont {{Rulev}}, \citenamefont
  {{S{\'a}iz}}, \citenamefont {{Shao}}, \citenamefont {{Shchegolev}},
  \citenamefont {{Sheng}}, \citenamefont {{Shi}}, \citenamefont {{Song}},
  \citenamefont {{Stenkin}}, \citenamefont {{Stepanov}}, \citenamefont {{Sun}},
  \citenamefont {{Sun}}, \citenamefont {{Sun}}, \citenamefont {{Tam}},
  \citenamefont {{Tang}}, \citenamefont {{Tian}}, \citenamefont {{Wang}},
  \citenamefont {{Wang}}, \citenamefont {{Wang}}, \citenamefont {{Wang}},
  \citenamefont {{Wang}}, \citenamefont {{Wang}}, \citenamefont {{Wang}},
  \citenamefont {{Wang}}, \citenamefont {{Wang}}, \citenamefont {{Wang}},
  \citenamefont {{Wang}}, \citenamefont {{Wang}}, \citenamefont {{Wang}},
  \citenamefont {{Wang}}, \citenamefont {{Wang}}, \citenamefont {{Wang}},
  \citenamefont {{Wang}}, \citenamefont {{Wang}}, \citenamefont {{Wang}},
  \citenamefont {{Wang}}, \citenamefont {{Wang}}, \citenamefont {{Wei}},
  \citenamefont {{Wei}}, \citenamefont {{Wei}}, \citenamefont {{Wen}},
  \citenamefont {{Wu}}, \citenamefont {{Wu}}, \citenamefont {{Wu}},
  \citenamefont {{Wu}}, \citenamefont {{Wu}}, \citenamefont {{Xi}},
  \citenamefont {{Xia}}, \citenamefont {{Xia}}, \citenamefont {{Xiang}},
  \citenamefont {{Xiao}}, \citenamefont {{Xiao}}, \citenamefont {{Xin}},
  \citenamefont {{Xin}}, \citenamefont {{Xing}}, \citenamefont {{Xu}},
  \citenamefont {{Xu}}, \citenamefont {{Xue}}, \citenamefont {{Yan}},
  \citenamefont {{Yang}}, \citenamefont {{Yang}}, \citenamefont {{Yang}},
  \citenamefont {{Yang}}, \citenamefont {{Yang}}, \citenamefont {{Yang}},
  \citenamefont {{Yang}}, \citenamefont {{Yao}}, \citenamefont {{Yao}},
  \citenamefont {{Ye}}, \citenamefont {{Yin}}, \citenamefont {{Yin}},
  \citenamefont {{You}}, \citenamefont {{You}}, \citenamefont {{Yu}},
  \citenamefont {{Yuan}}, \citenamefont {{Zeng}}, \citenamefont {{Zeng}},
  \citenamefont {{Zeng}}, \citenamefont {{Zeng}}, \citenamefont {{Zha}},
  \citenamefont {{Zhai}}, \citenamefont {{Zhang}}, \citenamefont {{Zhang}},
  \citenamefont {{Zhang}}, \citenamefont {{Zhang}}, \citenamefont {{Zhang}},
  \citenamefont {{Zhang}}, \citenamefont {{Zhang}}, \citenamefont {{Zhang}},
  \citenamefont {{Zhang}}, \citenamefont {{Zhang}}, \citenamefont {{Zhang}},
  \citenamefont {{Zhang}}, \citenamefont {{Zhang}}, \citenamefont {{Zhang}},
  \citenamefont {{Zhang}}, \citenamefont {{Zhang}}, \citenamefont {{Zhang}},
  \citenamefont {{Zhang}}, \citenamefont {{Zhang}}, \citenamefont {{Zhao}},
  \citenamefont {{Zhao}}, \citenamefont {{Zhao}}, \citenamefont {{Zhao}},
  \citenamefont {{Zhao}}, \citenamefont {{Zheng}}, \citenamefont {{Zheng}},
  \citenamefont {{Zhou}}, \citenamefont {{Zhou}}, \citenamefont {{Zhou}},
  \citenamefont {{Zhou}}, \citenamefont {{Zhou}}, \citenamefont {{Zhou}},
  \citenamefont {{Zhu}}, \citenamefont {{Zhu}}, \citenamefont {{Zhu}},
  \citenamefont {{Zhu}},\ and\ \citenamefont {{Zuo}}}]{2021Natur.594...33C}%
  \BibitemOpen
  \bibfield  {author} {\bibinfo {author} {\bibfnamefont {Z.}~\bibnamefont
  {{Cao}}} \bibinfo {author} \emph{et al.},\ }\bibfield  {title} {\bibinfo {title}
  {{\it Ultrahigh-energy photons up to 1.4 petaelectronvolts from 12
  {\ensuremath{\gamma}}-ray Galactic sources}},\ }\href
  {https://doi.org/10.1038/s41586-021-03498-z} {\bibfield  {journal} {\bibinfo
  {journal} {Nature}\ }\textbf {\bibinfo {volume} {594}},\ \bibinfo {pages}
  {33} (\bibinfo {year} {2021})}\BibitemShut {NoStop}%
\bibitem [{\citenamefont {{Tibet AS{\ensuremath{\gamma}} Collaboration}}\ \emph
  {et~al.}(2021)\citenamefont {{Tibet AS{\ensuremath{\gamma}} Collaboration}},
  \citenamefont {{Amenomori}}, \citenamefont {{Bao}}, \citenamefont {{Bi}},
  \citenamefont {{Chen}}, \citenamefont {{Chen}}, \citenamefont {{Chen}},
  \citenamefont {{Chen}}, \citenamefont {{Chen}}, \citenamefont {{Cirennima}},
  \citenamefont {{Danzengluobu}}, \citenamefont {{Fang}}, \citenamefont
  {{Fang}}, \citenamefont {{Feng}}, \citenamefont {{Feng}}, \citenamefont
  {{Feng}}, \citenamefont {{Gao}}, \citenamefont {{Gou}}, \citenamefont
  {{Guo}}, \citenamefont {{Guo}}, \citenamefont {{He}}, \citenamefont {{He}},
  \citenamefont {{Hibino}}, \citenamefont {{Hotta}}, \citenamefont {{Hu}},
  \citenamefont {{Hu}}, \citenamefont {{Huang}}, \citenamefont {{Jia}},
  \citenamefont {{Jiang}}, \citenamefont {{Jin}}, \citenamefont {{Kasahara}},
  \citenamefont {{Katayose}}, \citenamefont {{Kato}}, \citenamefont {{Kato}},
  \citenamefont {{Kawata}}, \citenamefont {{Kihara}}, \citenamefont {{Ko}},
  \citenamefont {{Kozai}}, \citenamefont {{Labaciren}}, \citenamefont {{Li}},
  \citenamefont {{Li}}, \citenamefont {{Li}}, \citenamefont {{Lin}},
  \citenamefont {{Liu}}, \citenamefont {{Liu}}, \citenamefont {{Liu}},
  \citenamefont {{Liu}}, \citenamefont {{Liu}}, \citenamefont {{Lou}},
  \citenamefont {{Lu}}, \citenamefont {{Meng}}, \citenamefont {{Munakata}},
  \citenamefont {{Nakada}}, \citenamefont {{Nakamura}}, \citenamefont
  {{Nanjo}}, \citenamefont {{Nishizawa}}, \citenamefont {{Ohnishi}},
  \citenamefont {{Ohura}}, \citenamefont {{Ozawa}}, \citenamefont {{Qian}},
  \citenamefont {{Qu}}, \citenamefont {{Saito}}, \citenamefont {{Sakata}},
  \citenamefont {{Sako}}, \citenamefont {{Shao}}, \citenamefont {{Shibata}},
  \citenamefont {{Shiomi}}, \citenamefont {{Sugimoto}}, \citenamefont
  {{Takano}}, \citenamefont {{Takita}}, \citenamefont {{Tan}}, \citenamefont
  {{Tateyama}}, \citenamefont {{Torii}}, \citenamefont {{Tsuchiya}},
  \citenamefont {{Udo}}, \citenamefont {{Wang}}, \citenamefont {{Wu}},
  \citenamefont {{Xue}}, \citenamefont {{Yamamoto}}, \citenamefont {{Yang}},
  \citenamefont {{Yokoe}}, \citenamefont {{Yuan}}, \citenamefont {{Zhai}},
  \citenamefont {{Zhang}}, \citenamefont {{Zhang}}, \citenamefont {{Zhang}},
  \citenamefont {{Zhang}}, \citenamefont {{Zhang}}, \citenamefont {{Zhang}},
  \citenamefont {{Zhang}}, \citenamefont {{Zhao}}, \citenamefont
  {{Zhaxisangzhu}},\ and\ \citenamefont {{Zhou}}}]{2021NatAs...5..460T}%
  \BibitemOpen
  \bibfield  {author} {\bibinfo {author} {\bibfnamefont
  {M.}~\bibnamefont {{Amenomori}}} \bibinfo {author} \emph{et al.} \bibinfo 
  {author} {\bibnamefont {{(Tibet AS{\ensuremath{\gamma}} Collaboration)}}},\
  }\bibfield  {title} {\bibinfo {title} {{\it Potential PeVatron supernova remnant
  G106.3+2.7 seen in the highest-energy gamma rays}},\ }\href
  {https://doi.org/10.1038/s41550-020-01294-9} {\bibfield  {journal} {\bibinfo
  {journal} {Nat. Astron.}\ }\textbf {\bibinfo {volume} {5}},\ \bibinfo
  {pages} {460} (\bibinfo {year} {2021})},\ \Eprint
  {https://arxiv.org/abs/2109.02898} {arXiv:2109.02898 [astro-ph.HE]}
  \BibitemShut {NoStop}%
\bibitem [{\citenamefont {{Lhaaso Collaboration}}\ \emph
  {et~al.}(2021)\citenamefont {{Lhaaso Collaboration}}, \citenamefont {{Cao}},
  \citenamefont {{Aharonian}}, \citenamefont {{An}}, \citenamefont {{Axikegu}},
  \citenamefont {{Bai}}, \citenamefont {{Bai}}, \citenamefont {{Bao}},
  \citenamefont {{Bastieri}}, \citenamefont {{Bi}}, \citenamefont {{Bi}},
  \citenamefont {{Cai}}, \citenamefont {{Cai}}, \citenamefont {{Cao}},
  \citenamefont {{Chang}}, \citenamefont {{Chang}}, \citenamefont {{Chen}},
  \citenamefont {{Chen}}, \citenamefont {{Chen}}, \citenamefont {{Chen}},
  \citenamefont {{Chen}}, \citenamefont {{Chen}}, \citenamefont {{Chen}},
  \citenamefont {{Chen}}, \citenamefont {{Chen}}, \citenamefont {{Chen}},
  \citenamefont {{Chen}}, \citenamefont {{Chen}}, \citenamefont {{Chen}},
  \citenamefont {{Chen}}, \citenamefont {{Cheng}}, \citenamefont {{Cheng}},
  \citenamefont {{Cui}}, \citenamefont {{Cui}}, \citenamefont {{Cui}},
  \citenamefont {{D'Ettorre Piazzoli}}, \citenamefont {{Dai}}, \citenamefont
  {{Dai}}, \citenamefont {{Dai}}, \citenamefont {{Danzengluobu}}, \citenamefont
  {{Della Volpe}}, \citenamefont {{Dong}}, \citenamefont {{Duan}},
  \citenamefont {{Fan}}, \citenamefont {{Fan}}, \citenamefont {{Fan}},
  \citenamefont {{Fang}}, \citenamefont {{Fang}}, \citenamefont {{Feng}},
  \citenamefont {{Feng}}, \citenamefont {{Feng}}, \citenamefont {{Feng}},
  \citenamefont {{Gao}}, \citenamefont {{Gao}}, \citenamefont {{Gao}},
  \citenamefont {{Gao}}, \citenamefont {{Gao}}, \citenamefont {{Ge}},
  \citenamefont {{Geng}}, \citenamefont {{Gong}}, \citenamefont {{Gou}},
  \citenamefont {{Gu}}, \citenamefont {{Guo}}, \citenamefont {{Guo}},
  \citenamefont {{Guo}}, \citenamefont {{Guo}}, \citenamefont {{Guo}},
  \citenamefont {{Han}}, \citenamefont {{He}}, \citenamefont {{He}},
  \citenamefont {{He}}, \citenamefont {{He}}, \citenamefont {{He}},
  \citenamefont {{He}}, \citenamefont {{Heller}}, \citenamefont {{Hor}},
  \citenamefont {{Hou}}, \citenamefont {{Hou}}, \citenamefont {{Hu}},
  \citenamefont {{Hu}}, \citenamefont {{Hu}}, \citenamefont {{Hu}},
  \citenamefont {{Huang}}, \citenamefont {{Huang}}, \citenamefont {{Huang}},
  \citenamefont {{Huang}}, \citenamefont {{Huang}}, \citenamefont {{Huang}},
  \citenamefont {{Ji}}, \citenamefont {{Ji}}, \citenamefont {{Jia}},
  \citenamefont {{Jiang}}, \citenamefont {{Jiang}}, \citenamefont {{Jin}},
  \citenamefont {{Ke}}, \citenamefont {{Kuleshov}}, \citenamefont
  {{Levochkin}}, \citenamefont {{Li}}, \citenamefont {{Li}}, \citenamefont
  {{Li}}, \citenamefont {{Li}}, \citenamefont {{Li}}, \citenamefont {{Li}},
  \citenamefont {{Li}}, \citenamefont {{Li}}, \citenamefont {{Li}},
  \citenamefont {{Li}}, \citenamefont {{Li}}, \citenamefont {{Li}},
  \citenamefont {{Li}}, \citenamefont {{Li}}, \citenamefont {{Li}},
  \citenamefont {{Li}}, \citenamefont {{Li}}, \citenamefont {{Li}},
  \citenamefont {{Liang}}, \citenamefont {{Liang}}, \citenamefont {{Lin}},
  \citenamefont {{Liu}}, \citenamefont {{Liu}}, \citenamefont {{Liu}},
  \citenamefont {{Liu}}, \citenamefont {{Liu}}, \citenamefont {{Liu}},
  \citenamefont {{Liu}}, \citenamefont {{Liu}}, \citenamefont {{Liu}},
  \citenamefont {{Liu}}, \citenamefont {{Liu}}, \citenamefont {{Liu}},
  \citenamefont {{Liu}}, \citenamefont {{Liu}}, \citenamefont {{Liu}},
  \citenamefont {{Liu}}, \citenamefont {{Long}}, \citenamefont {{Lu}},
  \citenamefont {{Lv}}, \citenamefont {{Ma}}, \citenamefont {{Ma}},
  \citenamefont {{Ma}}, \citenamefont {{Mao}}, \citenamefont {{Masood}},
  \citenamefont {{Min}}, \citenamefont {{Mitthumsiri}}, \citenamefont
  {{Montaruli}}, \citenamefont {{Nan}}, \citenamefont {{Pang}}, \citenamefont
  {{Pattarakijwanich}}, \citenamefont {{Pei}}, \citenamefont {{Qi}},
  \citenamefont {{Qi}}, \citenamefont {{Qiao}}, \citenamefont {{Qin}},
  \citenamefont {{Ruffolo}}, \citenamefont {{Rulev}}, \citenamefont {{Saiz}},
  \citenamefont {{Shao}}, \citenamefont {{Shchegolev}}, \citenamefont
  {{Sheng}}, \citenamefont {{Shi}}, \citenamefont {{Song}}, \citenamefont
  {{Stenkin}}, \citenamefont {{Stepanov}}, \citenamefont {{Su}}, \citenamefont
  {{Sun}}, \citenamefont {{Sun}}, \citenamefont {{Sun}}, \citenamefont {{Tam}},
  \citenamefont {{Tang}}, \citenamefont {{Tian}}, \citenamefont {{Wang}},
  \citenamefont {{Wang}}, \citenamefont {{Wang}}, \citenamefont {{Wang}},
  \citenamefont {{Wang}}, \citenamefont {{Wang}}, \citenamefont {{Wang}},
  \citenamefont {{Wang}}, \citenamefont {{Wang}}, \citenamefont {{Wang}},
  \citenamefont {{Wang}}, \citenamefont {{Wang}}, \citenamefont {{Wang}},
  \citenamefont {{Wang}}, \citenamefont {{Wang}}, \citenamefont {{Wang}},
  \citenamefont {{Wang}}, \citenamefont {{Wang}}, \citenamefont {{Wang}},
  \citenamefont {{Wang}}, \citenamefont {{Wang}}, \citenamefont {{Wang}},
  \citenamefont {{Wei}}, \citenamefont {{Wei}}, \citenamefont {{Wei}},
  \citenamefont {{Wen}}, \citenamefont {{Wu}}, \citenamefont {{Wu}},
  \citenamefont {{Wu}}, \citenamefont {{Wu}}, \citenamefont {{Wu}},
  \citenamefont {{Xi}}, \citenamefont {{Xia}}, \citenamefont {{Xia}},
  \citenamefont {{Xiang}}, \citenamefont {{Xiao}}, \citenamefont {{Xiao}},
  \citenamefont {{Xiao}}, \citenamefont {{Xin}}, \citenamefont {{Xin}},
  \citenamefont {{Xing}}, \citenamefont {{Xu}}, \citenamefont {{Xu}},
  \citenamefont {{Xue}}, \citenamefont {{Yan}}, \citenamefont {{Yan}},
  \citenamefont {{Yang}}, \citenamefont {{Yang}}, \citenamefont {{Yang}},
  \citenamefont {{Yang}}, \citenamefont {{Yang}}, \citenamefont {{Yang}},
  \citenamefont {{Yang}}, \citenamefont {{Yao}}, \citenamefont {{Yao}},
  \citenamefont {{Ye}}, \citenamefont {{Yin}}, \citenamefont {{Yin}},
  \citenamefont {{You}}, \citenamefont {{You}}, \citenamefont {{Yu}},
  \citenamefont {{Yuan}}, \citenamefont {{Zeng}}, \citenamefont {{Zeng}},
  \citenamefont {{Zeng}}, \citenamefont {{Zeng}}, \citenamefont {{Zha}},
  \citenamefont {{Zhai}}, \citenamefont {{Zhang}}, \citenamefont {{Zhang}},
  \citenamefont {{Zhang}}, \citenamefont {{Zhang}}, \citenamefont {{Zhang}},
  \citenamefont {{Zhang}}, \citenamefont {{Zhang}}, \citenamefont {{Zhang}},
  \citenamefont {{Zhang}}, \citenamefont {{Zhang}}, \citenamefont {{Zhang}},
  \citenamefont {{Zhang}}, \citenamefont {{Zhang}}, \citenamefont {{Zhang}},
  \citenamefont {{Zhang}}, \citenamefont {{Zhang}}, \citenamefont {{Zhang}},
  \citenamefont {{Zhang}}, \citenamefont {{Zhang}}, \citenamefont {{Zhao}},
  \citenamefont {{Zhao}}, \citenamefont {{Zhao}}, \citenamefont {{Zhao}},
  \citenamefont {{Zhao}}, \citenamefont {{Zheng}}, \citenamefont {{Zheng}},
  \citenamefont {{Zhou}}, \citenamefont {{Zhou}}, \citenamefont {{Zhou}},
  \citenamefont {{Zhou}}, \citenamefont {{Zhou}}, \citenamefont {{Zhou}},
  \citenamefont {{Zhu}}, \citenamefont {{Zhu}}, \citenamefont {{Zhu}},
  \citenamefont {{Zhu}},\ and\ \citenamefont {{Zuo}}}]{2021Sci...373..425L}%
  \BibitemOpen
  \bibfield  {author} {\bibinfo {author} {\bibfnamefont {Z.}~\bibnamefont
  {{Cao}}} \bibinfo {author} \emph{et al.} \bibinfo {author} {\bibnamefont 
  {{(Lhaaso Collaboration)}}}, \ }\bibfield
  {title} {\bibinfo {title} {{\it Peta-electron volt gamma-ray emission from the
  Crab Nebula}},\ }\href {https://doi.org/10.1126/science.abg5137} {\bibfield
  {journal} {\bibinfo  {journal} {Science}\ }\textbf {\bibinfo {volume}
  {373}},\ \bibinfo {pages} {425} (\bibinfo {year} {2021})},\ \Eprint
  {https://arxiv.org/abs/2111.06545} {arXiv:2111.06545 [astro-ph.HE]}
  \BibitemShut {NoStop}%
\bibitem [{\citenamefont {{Ruderman}}(1972)}]{1972ARA&A..10..427R}%
  \BibitemOpen
  \bibfield  {author} {\bibinfo {author} {\bibfnamefont {M.}~\bibnamefont
  {{Ruderman}}},\ }\bibfield  {title} {\bibinfo {title} {{\it Pulsars: Structure
  and Dynamics}},\ }\href {https://doi.org/10.1146/annurev.aa.10.090172.002235}
  {\bibfield  {journal} {\bibinfo  {journal} {ARA\&A}\ }\textbf {\bibinfo
  {volume} {10}},\ \bibinfo {pages} {427} (\bibinfo {year} {1972})}\BibitemShut
  {NoStop}%
\bibitem [{\citenamefont {{Tsytovich}}(1973)}]{1973ARA&A..11..363T}%
  \BibitemOpen
  \bibfield  {author} {\bibinfo {author} {\bibfnamefont {V.~N.}\ \bibnamefont
  {{Tsytovich}}},\ }\bibfield  {title} {\bibinfo {title} {{\it Interaction of Fast
  Particles with Waves in Cosmic Magnetoactive Plasma}},\ }\href
  {https://doi.org/10.1146/annurev.aa.11.090173.002051} {\bibfield  {journal}
  {\bibinfo  {journal} {ARA\&A}\ }\textbf {\bibinfo {volume} {11}},\ \bibinfo
  {pages} {363} (\bibinfo {year} {1973})}\BibitemShut {NoStop}%
\bibitem [{\citenamefont {Gaisser}\ \emph {et~al.}(2016)\citenamefont
  {Gaisser}, \citenamefont {Engel},\ and\ \citenamefont
  {Resconi}}]{1990cup..book.....G}%
  \BibitemOpen
  \bibfield  {author} {\bibinfo {author} {\bibfnamefont {T.~K.}\ \bibnamefont
  {Gaisser}}, \bibinfo {author} {\bibfnamefont {R.}~\bibnamefont {Engel}},\
  and\ \bibinfo {author} {\bibfnamefont {E.}~\bibnamefont {Resconi}},\ }\href
  {https://doi.org/10.1017/CBO9781139192194} {\emph {\bibinfo {title} {{\it Cosmic
  Rays and Particle Physics}}}}\ (\bibinfo  {publisher} {{Cambridge University
  Press}},\ \bibinfo {address} {{Cambridge}},\ \bibinfo {year}
  {2016})\BibitemShut {NoStop}%
\bibitem [{\citenamefont {You}(1998)}]{1998scichina..book.....Y}%
  \BibitemOpen
  \bibfield  {author} {\bibinfo {author} {\bibfnamefont {J.-H.}\ \bibnamefont
  {You}},\ }\href@noop {} {\emph {\bibinfo {title} {{\it Radiation Mechanisms in
  Astrophysics}}}}\ (\bibinfo  {publisher} {{China Science Publishing}},\
  \bibinfo {address} {{Beijing}},\ \bibinfo {year} {1998})\BibitemShut
  {NoStop}%
\bibitem [{\citenamefont {{Abdollahi}}\ \emph {et~al.}(2020)\citenamefont
  {{Abdollahi}}, \citenamefont {{Ballet}}, \citenamefont {{Fukazawa}},
  \citenamefont {{Katagiri}},\ and\ \citenamefont
  {{Condon}}}]{2020ApJ...896...76A}%
  \BibitemOpen
  \bibfield  {author} {\bibinfo {author} {\bibfnamefont {S.}~\bibnamefont
  {{Abdollahi}}}, \bibinfo {author} {\bibfnamefont {J.}~\bibnamefont
  {{Ballet}}}, \bibinfo {author} {\bibfnamefont {Y.}~\bibnamefont
  {{Fukazawa}}}, \bibinfo {author} {\bibfnamefont {H.}~\bibnamefont
  {{Katagiri}}},\ and\ \bibinfo {author} {\bibfnamefont {B.}~\bibnamefont
  {{Condon}}},\ }\bibfield  {title} {\bibinfo {title} {{\it On the Origin of the
  Gamma-Ray Emission toward SNR CTB 37A with Fermi-LAT}},\ }\href
  {https://doi.org/10.3847/1538-4357/ab91b3} {\bibfield  {journal} {\bibinfo
  {journal} {ApJ}\ }\textbf {\bibinfo {volume} {896}},\ \bibinfo {eid} {76}
  (\bibinfo {year} {2020})},\ \Eprint {https://arxiv.org/abs/2006.05731}
  {arXiv:2006.05731 [astro-ph.HE]} \BibitemShut {NoStop}%
\bibitem [{\citenamefont {{Arneodo}}(1994)}]{1994PhR...240..301A}%
  \BibitemOpen
  \bibfield  {author} {\bibinfo {author} {\bibfnamefont {M.}~\bibnamefont
  {{Arneodo}}},\ }\bibfield  {title} {\bibinfo {title} {{\it Nuclear effects in
  structure functions}},\ }\href {https://doi.org/10.1016/0370-1573(94)90048-5}
  {\bibfield  {journal} {\bibinfo  {journal} {Physics Reports}\ }\textbf
  {\bibinfo {volume} {240}},\ \bibinfo {pages} {301} (\bibinfo {year}
  {1994})}\BibitemShut {NoStop}%
\end{thebibliography}
\end{document}